\documentclass{aa}  

\usepackage{hyperref}
\hypersetup{colorlinks,citecolor=blue}
\usepackage{graphicx}
\usepackage{natbib}
\usepackage{threeparttable}
\usepackage{txfonts}
\usepackage{float}

\def\msun {$M_{\odot}$}
\def\lsun {$L_{\odot}$}

\def\sori {$\sigma$-Ori}
\def\Vp {$V_{\rm peak}$}
\def\sorionis {$\sigma$-Orionis}
\newcommand{\degree}{\ensuremath{^\circ}}

\newcommand{\Lacc}{{$L_{\rm acc}$}}
\newcommand{\Macc}{{$\dot{M}_{\rm acc}$}}
\newcommand{\OI}{$\rm [OI]\,\lambda$6300}
\newcommand{\NII}{$\rm [NII]\,\lambda$6583}
\newcommand{\SII}{$\rm [SII]\,\lambda$6731}
\newcommand{\SIIb}{$\rm [SII]\,\lambda$6716}

\usepackage{xcolor}

\begin{document} 

   \title{A New Look at Disk Winds and External Photoevaporation \\in the \sorionis\,Cluster \thanks{Based on observations collected at the European Southern Observatory under ESO programmes 0104.C-0454(A) and 108.22CB.001.}}


   \author{K. Maucó\inst{\ref{instESO}}
          \and
          C.~F. Manara\inst{\ref{instESO}}
          \and
          A. Bayo\inst{\ref{instESO}}
          \and
          J. Hernández\inst{\ref{UNAM-E}}
          \and
          J. Campbell-White\inst{\ref{instESO}}
          \and
          N. Calvet\inst{\ref{UMich}}
          \and
          G. Ballabio\inst{\ref{Imperial}}
          \and
          M.~L. Aru\inst{\ref{instESO}}
          \and
          J. M. Alcalá\inst{\ref{instNA}}
          \and
          M. Ansdell\inst{\ref{instNASA}}
          \and
          C. Brice\~no\inst{\ref{instNOIRLab}}
          \and
          S. Facchini\inst{\ref{instUniMI}}
          \and
          T.~J. Haworth\inst{\ref{instQM}}
          \and
          M. McClure\inst{\ref{instLeiden}}
          \and
          J.~P. Williams\inst{\ref{instHW}}}

   \institute{ European Southern Observatory, Karl-Schwarzschild-Strasse 2, 85748 Garching bei M\"unchen, Germany\label{instESO}\\
              \email{kmaucoco@eso.org}
         \and
             Instituto de Astronomía, Universidad Nacional Autónoma de México, Ensenada, B.C., México\label{UNAM-E}
        \and
             University of Michigan, UMICH, Ann Arbor, MI 48109, USA\label{UMich}
        \and
            Imperial Astrophysics, Imperial College London, Blackett Laboratory, Prince Consort Road, London SW7 2AZ, UK\label{Imperial}
        \and
            INAF -- Osservatorio Astronomico di Capodimonte, via Moiariello 16, 80131 Napoli, Italy\label{instNA}   
        \and
            Astronomy Unit, School of Physics and Astronomy, Queen Mary University of London, London E1 4NS, UK\label{instQM}
        \and
            Dipartimento di Fisica, Universit\'a degli Studi di Milano, Via Giovanni Celoria 16, 20133 Milano, Italy\label{instUniMI}
        \and
            NASA Headquarters, 300 E Street SW, Washington, DC 20546, USA\label{instNASA}
        \and
            Institute for Astronomy, University of Hawai‘i at Mānoa, Honolulu, HI, USA\label{instHW}
        \and
            SOAR Telescope/NSF NOIRLab, Casilla 603, La Serena, Chile\label{instNOIRLab}
        \and
            Leiden Observatory, Leiden University, PO Box 9513, 2300 RA Leiden, The Netherlands\label{instLeiden}
        }


 
  \abstract
   {Disk winds play a crucial role in the evolution of protoplanetary disks. Typical conditions for star and planet formation are in regions with intermediate or strong UV radiation fields produced by massive stars. In these environments, internally or externally driven winds can occur. The \sorionis\,cluster is the ideal site to study disk winds under these conditions; its outer parts, exposed only to mild UV fields, can be used to study disk evolution, while its innermost regions to study the effect of external irradiation.}
   {Our goal is to study disk winds in the \sorionis\,cluster by looking at the properties of optical forbidden lines, and comparing them with other star-forming regions at different ages, to search for potential signatures of disk evolution and external photoevaporation.}
   {We analyze the \OI, \NII, and \SII,$\lambda$6716 lines using high-resolution MIKE spectra for a sample of 27 classical T Tauri stars and complemented by intermediate-resolution X-shooter data. We decompose the line profiles into multiple Gaussian components. We calculated luminosities, line ratios, and kinematic properties of these components.}
   {We found that the \OI\,line luminosity and kinematic properties for our \sorionis\,sample are similar to those found in low-mass star-forming regions. The frequency of single-component \OI\,line profiles reflects the expected evolutionary stage given the intermediate age of \sorionis\,($\sim$3-5 Myrs). This points to internal processes contributing to the line emission. However, the highly irradiated disks in the cluster do not follow the accretion luminosity-\OI\, line luminosity relation found in low-mass star-forming regions, and all exhibit single-component line profiles. Line ratios of highly ionized species of [NII] and [SII] show higher ratios than typical values found in sources in low-mass star-forming regions. These are interpreted as signatures of external photoevaporation.}
   {We show the potential of using multiple forbidden emission lines to study both internally 
   and externally driven disk winds. In the case of \sorionis, the innermost regions are 
   clearly affected by external irradiation, evidenced by the lack of correlation in the accretion - [OI] luminosity relation. The broad line widths of close-in sources, however, indicate a possible contribution from internal processes, such as magnetohydrodynamical winds and/or internal photoevaporation. This suggests a coevolution of internal and external winds in the \sorionis\,disks, while pointing towards a new way to disentangle these processes.}
    
   \keywords{Stars: winds, outflows -- protoplanetary disks -- stars:pre-main sequence -- stars:variables:TTauri}

   \maketitle
%

\section{Introduction}

Protoplanetary disks, made of gas and dust, govern the growth of stars and are the birthplace of planets. Several processes contribute to their evolution, mainly mediated by the accretion of matter onto the central star and mass loss through winds \citep[e.g.,][]{Hartmann2016,pascucci2022,Coleman2022,Coleman2024,Serna2024}. 
These winds can have an origin internal to the star-disk system, associated with the host star and produced in the inner disk \citep[e.g.,][]{Ercolano2017}, or due to external causes linked to the environment \citep[e.g.,][]{Winter2022}. A comprehensive evaluation of the physics of disk evolution and planet formation, therefore, requires understanding the intricate relationships between mass accretion and mass loss through internal and external winds.

Optical forbidden lines, particularly the \OI\,line, are well-known wind tracers accessible with medium to high-resolution spectrographs and have been widely used to characterize internal winds in disks located in different low-mass star-forming regions (SFRs) in the last decades \citep[e.g.,][]{Hartigan95,Rigliaco2013,Natta2014,Simon2016,Fang2018,McGinnis2018,Banzatti2019,Gangi2022}. 
The lines are composed of multiple components attributed to different launching regions in the disk \citep{Edwards87}. A high-velocity component (HVC) peaking at hundreds of km/s is frequently observed in younger regions, and it is known to be associated with fast collimated jets as a result of the accretion process \citep[e.g.,][]{Hartigan95,Nisini2018}. A low-velocity component (LVC) peaking between $\pm$30  km/s is also commonly found among T Tauri stars, and may be itself decomposed in a broad (BC) and a narrow component \citep[NC, e.g.,][]{Rigliaco2013,Simon2016}. It is well established that the BC of the LVC is linked to a magnetohydrodynamical (MHD) wind in the inner disk \citep[e.g.,][]{Simon2016,Whelan2021,Nisini2024}. \citet{Banzatti2019} further argued that the entire LVC traces a radially extended MHD wind based on the findings that the kinematic properties of the BC correlate with those of the NC. However, \citet{weber2020} pointed out that these observed correlations do not necessarily imply a common origin in an extended MHD wind, and concluded that the forbidden line profiles are best reproduced by the combination of an inner MHD wind (producing the HVC, and BC) and a photoevaporative wind associated with the NC. Therefore, the origin of the NC is still under debate, but it is possibly related to slow, photoevaporative winds. 

The morphology of the \OI\,line seems to also evolve with age, losing first the HVC, followed by the broad LVC as the inner disk dissipates, and ending up with a single-line profile for older regions and disks with central cavities \citep{Ercolano2017,McGinnis2018,Banzatti2019,Fang2023}, reinforcing the value of this line as a proxy for disk evolution. Recently, a peculiarly bright \OI\,line was detected by \citet{Campbell-White23} in the planet-host young star PDS70, pointing towards a still-to-be-understood relation between planet formation and the presence of winds in the inner regions of disks.

Since it is now widely established that the majority of exo-planet host stars form in dense and massive stellar clusters \citep[e.g.,][]{Winter2022,Gautam2024}, modern theoretical models \citep[e.g.,][]{Fatuzzo2008,Facchini2016,Haworth2016,Winter2020a,Winter2020b,ballabio2023,Hallatt2024} and growing observational evidence \citep[e.g.,][]{Mann2014, Guarcello2016,Kim2016,Ansdell2017,vanTerwisga2023,Haworth2023,Ballering2023, Mauco2023,Aru2024,Boyden2024} point to external photoevaporation by OB stars playing a significant role in the evolution of planet-forming disks in massive clusters. In these environments, massive OB stars externally irradiate nearby protoplanetary disks, heating the gas to escape velocity, and creating an externally driven photoevaporative wind that rapidly removes material from the outer part of the disk, strongly impacting the
size, mass, and survival timescale of protoplanetary disks \citep[e.g.,][]{Clarke2007}.
However, all previous works studying the \OI\,and other forbidden lines have mainly focused on low UV irradiated environments with no massive stars, hence tracing internal winds, either photoevaporative or MHD in origin, through the thermally excited \OI\,emission.
In contrast, in more massive SFRs, the line is not thermally excited but instead results from the dissociation of OH near the H-H2 transition in external winds \citep{Storzer1999}. 

Since both internally and externally driven winds contribute to the \OI\,line, several efforts have been made to find observables that can distinguish between the origin of the emission contributing to the line profiles \citep[e.g.,][]{ballabio2023}. \citet{Rigliaco2009} reported the discovery of an externally photoevaporated disk in the \sorionis \,cluster by detecting asymmetric line profiles in several forbidden optical lines. Furthermore, the line ratios they measured differed from the ones measured in low-mass SFRs, and mainly resembled ratios observed in proplyds in the Orion Nebula Clusters. Such pioneering work still lacks proper follow-up to date.

The way forward to understand disk evolution in more representative environments (i.e., similar birth environments of the Sun) is to characterize wind tracers in star clusters containing OB stars and compare them with what is known in low-mass SFRs. 
The \sorionis\, cluster is an ideal target for this goal.
Its massive multiple system \sori\, has impacted the evolution of disks during the $\sim$3-5 Myr of life of the cluster \citep{Oliveira2004}, but at a level of UV-radiation $\sim 10^4-10^5\, G_0$\footnote{expressed in terms of the Habing unit $G_0$ ($G_0 = \rm 1.6 \times 10^{-3}\, erg\, cm^{-2}\, s^{-1}$, over 912-2400 \AA, \citealt{Habing1968})} at most \citep{Mauco2023,Mauco2016}, lower than levels found in proplyds in the Orion Nebular Cluster \citep[$\sim 10^6-10^7\, G_0$,][]{Aru2024}.
Far-ultraviolet (FUV) radiation fields from massive stars externally illuminate the disks in the inner regions of the cluster. Indeed, \citet{Mauco2016} found evidence of external photoevaporation within the first 0.6 pc from \sori\,by the presence of [NII] lines for a small sample of disks observed at low resolution. This was later confirmed by ALMA surveys of this region founding a dearth of massive ($> 3 M_{\oplus}$) disks close ($<$ 0.5 pc) to the central OB stars \citep{Ansdell2017,Mauco2023}. This points to external photoevaporation acting at the innermost regions of the cluster in line with previous findings, which suggested that even moderate FUV fields ($\lesssim$ 2$\times 10^3\, G_0$) can drive significant disk mass loss \citep{Kim2016,vanTerwisga2023}. 
 Therefore, disk dissipation due to external photoevaporation can be observed on long enough timescales to see its full impact.  
Given its intermediate age, \sorionis\, also allows us to study disk evolution in the outer parts of the cluster where only low FUV fields ($\lesssim 10^3\, G_0$) permeate the disks, and internal processes mainly drive the evolution. 

In this paper, we characterize disk winds diagnostic in the \sorionis\,cluster by looking at forbidden lines in high-to-intermediate resolution optical spectra. After describing the sample in Sect.~\ref{sec:sample} and the observations and data reduction in Sect.~\ref{sec:obs}, we present our analysis of the line profiles in Sect.~\ref{sec:analysis}. The results of the luminosities, line ratios, and kinematic properties of the forbidden lines are described in Sect.~\ref{sec:results}. We discuss the implication of these results in light of disk evolution and external photoevaporation on Sect.~\ref{sec:discussion}. Our conclusions are presented in Sect.~\ref{sec:conclusions}.


\section{Sample}\label{sec:sample}

Low-mass members of the \sorionis\, cluster were first reported by \citet{Walter1997} who found over 80 X-ray sources and spectroscopically identified more than 100 low-mass, pre-main-sequence (PMS) members lying within 1$^{\circ}$ from the star $\sigma$ Ori, which is a massive quintuplet system \citep{caballero2014}. The brightest stars of the system, the $\sigma$ Ori Aa, Ab (O9.5, B0.5) pair and $\sigma$ Ori B ($\sim$B0-B2) with a total mass of $\sim$44 $M_{\odot}$ \citep{Simon2015}, form a large, low-density HII region and the bright photo-dissociation region known as the Horsehead nebula \citep{habart2005,Abergel2003}. \citet{caballero2019} carried out an exhaustive membership classification of the cluster and found that 10\% of the whole $\sigma$ Orionis stellar population (mostly in the intermediate mass regime) were non-clustered members. Several brown dwarfs have also been identified as part of the cluster \citep{damian2023a,damian2023b}. 

As part of the Orion OB1 variability survey, a sample of $\sim$350 T Tauri stars (TTS), mostly class II sources, belonging to the \sorionis\, cluster were observed and their circumstellar disks were first characterized with Spitzer IRAC and MIPS photometry \citep{Hernandez2007, Luhman2008}, then followed with Herschel PACS \citep{Mauco2016}, and more recently, imaged with ALMA \citep{Ansdell2017,Mauco2023}. We selected stars that have been confirmed as members of the \sorionis\, cluster based on their spectral types, H$\alpha$ emission, and the presence of Li I ($\lambda$6707 \AA) in absorption \citep{Hernandez2014}.
We focus on sources with Herschel PACS detections at 70 $\mu$m and 160 $\mu$m, especially those modeled in \citet{Mauco2016}, and with ALMA observations \citep{Ansdell2017, Mauco2023}. We followed these sources with optical, high-resolution spectroscopy using the Magellan
Inamori Kyocera Echelle (MIKE) instrument \citep{Bernstein2003}. 

The MIKE sample consists of 27 Classical T-Tauri stars (CTTS). We complemented this sample with the X-shooter sample (50 sources) reported in \citet{Mauco2023}. Of the MIKE sample, 22 stars also have X-shooter spectra. This leaves 5 stars with only MIKE observations and 28 CTTS with only X-shooter spectra. The stars in our sample are located at different projected distances from the cluster center. 
Our MIKE sample includes the three \sorionis\, stars (SO518, SO583, SO1153) studied in the PENELLOPE Large Program \citep{Manara2021,Gangi2023}. 
Using the disk census performed by \citet{Hernandez2007}, we evaluate the completeness of the studied sample, which includes the sample with Herschel PACS photometry \citep{Mauco2016}. Our sample includes 83\% (25/30) of the systems classified as full or transitional disks by \citet{Hernandez2007} with J$<$12.5 that fall within the PACS's FOV. Using the 4 Myr isochrones from \citet{Siess2000}, our sample includes most stars with stellar masses of 0.5 \msun\,or larger. Additionally, excluding the star SO514, which is anomalously faint in the J band, the limiting magnitude in our work is J$\sim$13, corresponding to 0.33 \msun.

The stellar parameters are those estimated in \citet{Mauco2023} for the 22 stars with X-shooter spectra, while for the rest of the MIKE sample, are those reported in \citet{Mauco2016}, that used the J and V photometry, the visual extinctions $A_V$, and the spectral types from \citet{Hernandez2014}, and their updated GAIA EDR3 distances \citep{Gaia2021} to estimate the stellar luminosity. The bolometric corrections and effective temperatures were obtained from the standard table for 5–30 Myr old PMS stars from
\citet{Pecaut2013}, and used to estimate stellar radii. 
Their stellar mass and age were estimated using the stellar isochrones from \citet{Siess2000}. Table ~\ref{tab:targets_info} shows the stellar parameters and additional information of our sample (MIKE and X-shooter). 

\section{Observations and Data Reduction} \label{sec:obs}

The observations were carried out during several nights in November-December 2016, 2017, and 2020 using the MIKE instrument \citep{Bernstein2003} on the Magellan Clay 6.5 m telescope at Las Campanas Observatory in Chile. 
The instrument is a high-resolution
spectrograph with a wavelength coverage of 3350--9500 \AA\,with 34 echelle orders,
but in this work, we focus only on the red arm (4900 to 9500 \AA), which covers the forbidden emission lines of \OI, \NII, and \SII, $\lambda$6716. Given our exposure times, the blue arm of MIKE (3350--5000 \AA) does not have enough sensitivity or signal-to-noise to detect emission lines of interest.
We used the 0.75"$\times$5" slit which implies a resolution of R$\sim$28,000 ($\Delta v$ = 10.7 km/s near the \OI\,line).   
The data were reduced using the 
MIKE pipeline in the Carnegie Observatories' \texttt{CarPy} package \citep{Kelson2000,Kelson2003}, which applies flat fields, removes scattered light, and subtracts the sky background.
Order by order, the pipeline extracts the stellar spectrum and
applies a wavelength calibration based on Th–Ar lamp
exposures taken before and after each science spectrum. To remove telluric lines, we used the \texttt{telluric} task of the \texttt{Image Reduction and Analysis Facility} package \citep[\texttt{IRAF,}][]{Davis1993}, which removes Earth's atmospheric features based on a featureless spectrum of a hot, rapidly rotating star. In our case, we used the telluric standard star HR 8998 for this purpose which was observed as part of our MIKE campaign. 

Details on the X-shooter observations (R$\sim$18,400 in the visible arm) and data reduction can be found in \citet{Mauco2023}. For the X-shooter sample, the telluric removal was done using the \texttt{molecfit} package \citep{Smette2015}, which models the telluric absorption lines on the observed spectra using information on the atmospheric conditions of the night. 

\section{Analysis}\label{sec:analysis}

\subsection{Corrected line Profiles}

Before analyzing the optical forbidden line profiles, we need to remove any telluric and photospheric absorption contaminating the region of interest in the spectra. The telluric lines were removed during the reduction of the spectra (Sect.~\ref{sec:obs}). 
Photospheric absorption feature removal, on the other hand, was carried out using the automated routines within the \texttt{STAR-MELT PYTHON} package \citep[][Campbell-White, in prep.]{Campbell-white2021}. For the emission line regions of interest, the best fitting photospheric removal was found after applying the standard steps of matching absorption features between the target and template star, i.e., continuum normalization, radial velocity shifting, instrument and rotational broadening, and veiling. To obtain the radial velocity of each target, we use the \texttt{rvsao} package from \texttt{IRAF} for the MIKE sample and the dedicated module in the \texttt{STAR-MELT PYTHON} package for the X-shooter sample, which calculates the radial velocity of targets using template spectra
broadened to the resolution of the instrument. The radial velocities of our sample are listed in Tab~\ref{tab:targets_info}. These steps were applied iteratively for several template standard stars (both main sequence and Class III, Campbell-White, in prep.), within 2 spectral types of the target stars. An effective $\chi^2$ measure was employed to determine the parameters for the best-fitting subtraction, as well as the overall best-fitting template from those available. 

\begin{figure}
    \centering
	\includegraphics[width=0.5\textwidth]{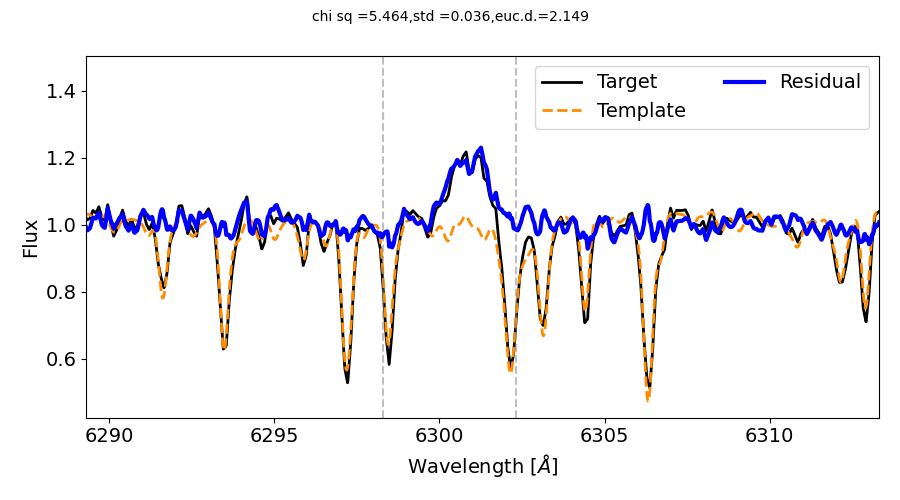}
    \caption{MIKE spectra of SO774 showing the photospheric removal around the \OI\,line using the \texttt{STAR-MELT PYTHON} package \citep{Campbell-white2021}.} 
    \label{fig:photo_removal}
\end{figure}

Figure~\ref{fig:photo_removal} shows an example of the photospheric removal around the \OI\,line for one star in our sample. The final output is a telluric-photospheric- and radial velocity-corrected spectrum of each star in our sample.

\subsection{Gaussian Fitting} \label{sec:Gauss_fit}

\begin{figure*}
    \centering
	\includegraphics[width=0.8\textwidth]{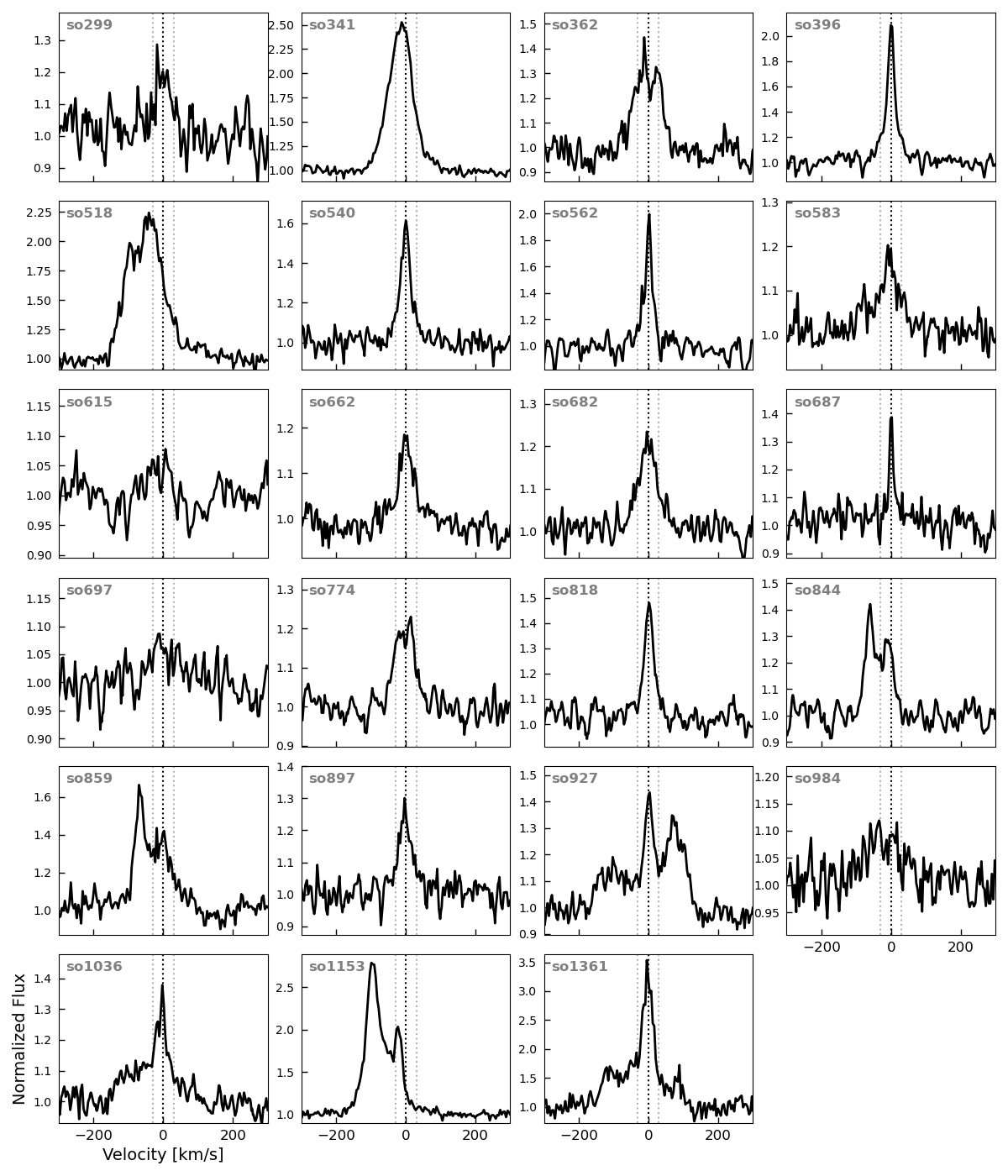}
    \caption{Telluric- photospheric- and radial velocity corrected \OI\,lines profiles for all the MIKE sources that exhibit the line. The black dotted line is located at 0 km/s, while the gray dotted lines indicate the velocity threshold for the high/low-velocity component ($\pm$30 km/s).}
    \label{fig:OI6300_profiles}
\end{figure*}

The corrected spectrum around the optical forbidden lines shows structured profiles composite of multiple peaks, as shown in Fig.~\ref{fig:OI6300_profiles} for the \OI\,line observed in the MIKE spectra. Motivated by this and following previous works \citep[e.g.,][]{Rigliaco2013,Simon2016,McGinnis2018,Gangi2023,Fang2018,Fang2023}, we decomposed the line profiles into different Gaussian components (GC) using the \texttt{STAR-MELT} package.  \texttt{STAR-MELT} uses the PYTHON package
\texttt{LMFIT9} \citep{Newville2014}, which carries out non-linear least-squares model fitting. The fitting procedure provides measurements and uncertainties of velocity centroids, amplitudes, and line widths for each velocity component. 
To discriminate between the HVC and the LVC in the line profiles, we adopted the velocity threshold of $\pm$30 km/s following \citet{Banzatti2019}. 
The LVC can also be subdivided into a narrow and broad component. This is motivated by the fact that some LVC profiles needed two Gaussian components, one significantly broader than the other. The width threshold between the NC and BC was taken as 40 km/s following \citet{Banzatti2019}. This paper focuses on the LVC since it is thought to be related to slow winds, and not to jets. 

A function with multiple Gaussian components provides good fits in most spectra in our sample. 
For the best fit, the number of Gaussian components used to match the line profiles was determined by $\chi^2$ minimization, where a new component was added only if it improves the reduced $\chi^2$ by more than 20\% as done in \citet{Banzatti2019}. To ensure the best-fit was reliable, we measured the rms of the residual spectrum (Gaussians minus original spectrum) and verified it was within 2$\sigma$ of the rms of the original spectrum (next to the line of interest) as done in \citet{Fang2018,Fang2023}. Since in Sect.~\ref{sec:age_evo} we compare our sample with those from \citet{Banzatti2019} and \citet{Fang2023}, following this approach makes the comparison more appropriate. Through this procedure, we were able to robustly fit all but one source, SO341, which needed an additional component to fulfill the residual rms constraints. In contrast, to ensure we are not over fitting, we also verified that the rms constraints of the residual spectrum were satisfied one component at a time. Only 3 sources have an added component (all broad components) with which the residual rms lies very close to the threshold. This is the case of SO518, SO774, and SO1036, which, according to the measured fitted parameters produced by \texttt{STAR-MELT} (see Tab.~\ref{tab:OIfit}), also have the highest uncertainties in the kinematics properties of this component, which highlights the reliability of our results.

\begin{figure*}
    \centering
	\includegraphics[width=0.9\textwidth]{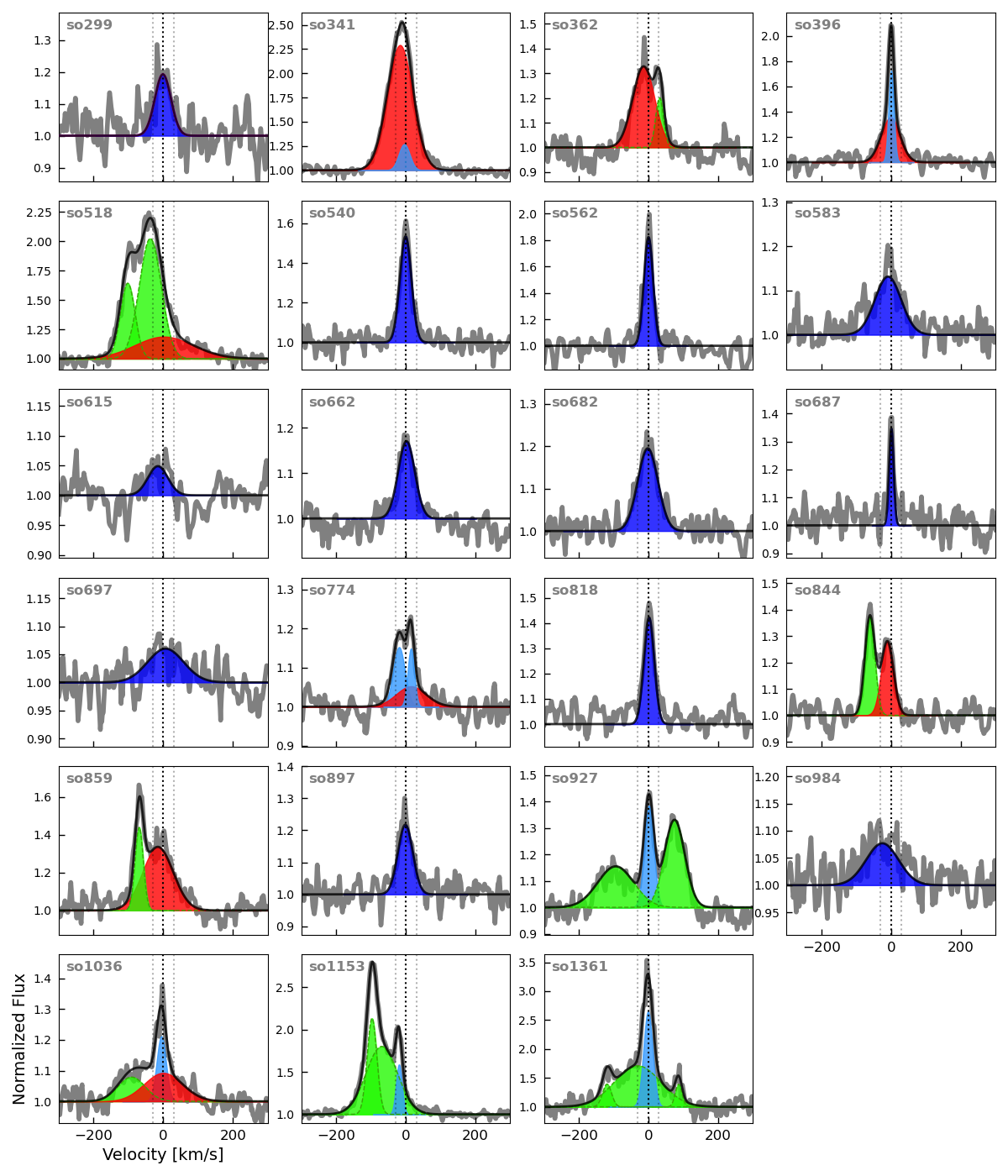}
    \caption{\OI\,lines fit profiles for the MIKE spectra. The colors indicate the type of Gaussian component as described in sec~\ref{sec:Gauss_fit}: single component (dark blue), low-velocity narrow component (light blue), low-velocity broad component (red), high-velocity component (green). The black dotted line is located at 0 km/s, while the gray dotted lines indicate the velocity threshold for the high/low-velocity component ($\pm$30 km/s).}
    \label{fig:OI6300_fit}
\end{figure*}

\section{Results}\label{sec:results}

In this section, we discuss the main results from fitting the optical forbidden lines of \OI, \NII, and \SII, $\lambda$6716 with multiple Gaussian components, with a particular focus on the brightest \OI\,line. These lines are particularly interesting for our study since they are usually bright and have been widely used for characterizing disk winds, produced by either the central star or by environmental factors, in different low-mass SFRs \citep[e.g.,][]{Rigliaco2009,Rigliaco2013, Natta2014,Simon2016,Banzatti2019, McGinnis2018,Fang2018,Fang2023}.
We exclude the other commonly studied $\rm [OI]$ line at 5577 \AA\,because the part of the spectrum containing this line in the MIKE spectra (the blue arm) has significantly lower signal-to-noise, which does not enable us to perform any meaningful analysis.

\subsection{Optical Forbidden Emission Lines Profiles}

We detected the \OI\,line transition in 85\% (23/27) of the stars with MIKE spectra. Figure~\ref{fig:OI6300_fit} shows the fitted \OI\,line profiles for the MIKE sample. The different fit components are indicated with filled colors depending on the classification described in~\ref{sec:Gauss_fit} as follows: single component (dark blue), low-velocity narrow component (light blue), low-velocity broad component (red), and high-velocity component (green).   

The most common morphology for the \OI\,line LVC is that of a single Gaussian component centered around $\sim$0 km/s, found in 12 out of 23 sources ($\sim$52\%). The rest of the sample needed multiple Gaussian components. 
For sources with two components, at least one of them is consistent with a HVC (either red or blue-shifted) except for SO396. 
In contrast, six sources (SO518, SO774, SO927, SO1036, SO1153, and SO1361) needed three or more components to fit their line profiles. All of these, with the exception of SO774, exhibit HVC in their spectra, with the particular cases of SO927 and SO1361 showing, in addition to the blueshifted HVC, also a redshifted HVC. SO518 and SO1153 are peculiar sources. SO518 seems to be a highly inclined system ($>$70\degree), while SO1153 is a class I star, which in the \citet{Lada2006} classification points to a strong IR excess rather than to an embedded object, and is identified as a high accretor \citep[log\Macc = -7.3 \msun/yr,][]{Mauco2023}. These sources, with more than 3 components, also present the largest blue shifts ($\sim$100 km/s) probably related to collimated jets, and at least one very broad ($>$50 km/s) component most likely linked to a MHD wind coming from the inner regions very close to the central star. 
Table~\ref{tab:OIfit} lists the \OI\,line fit parameters found for the MIKE sample.

For the X-shooter sample, we found a similar detection rate of 80\% (40/50) of the stars. The fitted \OI\,line profiles and the corresponding Gaussian components fit results for X-shooter sources can be found in Appendix~\ref{ap:OI_XS}. In general, the Gaussian decomposition between the MIKE and the X-shooter sample agrees fairly well. Most of the stars are fitted with the same number and type of components (SO299, SO341, SO518, SO540, SO562, SO583, SO662, SO697, SO818, SO859, SO897), others are similar but with a narrow component in MIKE spectra seen as a broad component in X-shooter (SO927, SO1036, SO1153, SO1361), as expected given the lower resolution. Only 3 sources with multiple components in MIKE spectra are identified as a single-component (SC) line in X-shooter (SO362, SO774, SO844), again expected given the lower resolution. Finally, for SO687, the HVC observed in X-shooter, and absent in MIKE, is probably not real but the result of residuals during the photospheric lines removal, which was particularly challenging for this source. This source is the closest to the massive system \sori. Figure~\ref{fig:MIKE_vs_XS} shows the comparison between MIKE and X-shooter spectra for the overlapping sample.

\begin{table*}
\begin{center}
\begin{threeparttable}[b]
\caption{\label{tab:OIfit} Parameters for the \OI\,LVC for the MIKE Sample}
\begin{tabular}{lccccccc}
\hline\hline   
Name & logL$_{\rm [OI], LVC}$\tnote{1} & $v_{\rm peak}$  & FWHM    & EW    &  $F_{\rm cont}$   & logL$_{\rm [OI],GC}$\tnote{2} & Class\tnote{3} \\
     & [\lsun] & [km/s] & [km/s]  & [\AA] &   [erg $\rm s^{-1}\,nm^{-1}\,cm^{-2}$ ]   &   [\lsun]   &       \\
\hline
SO299 &     -6.0 $\pm$     -0.6 &    -0.98 $\pm$      0.8  &     54.4 $\pm$      0.2 &     0.23 & 1.08e-14 &     -6.0 & LVC-SC \\ [0.5ex]
SO341 &    -3.98 $\pm$     -0.4 &   -14.73 $\pm$      2.1  &     81.6 $\pm$      3.4 &     2.72 & 7.41e-14 &    -3.98 & LVC-SC \\ [0.5ex]
SO362 &    -4.82 $\pm$    -0.48 &   -13.81 $\pm$      1.1  &     74.2 $\pm$      0.1 &     0.54 & 5.44e-14 &    -4.82 & LVC-BC \\ [0.5ex]
\hline
SO396 &     -5.89 $\pm$ -0.46   &    -1.15 $\pm$      1.0  &     63.0 $\pm$      0.1 &      0.5 & 3.27e-15 &    -6.09 & LVC-BC  \\ [0.5ex]
SO396 &                         &     0.47 $\pm$      1.0  &     17.8 $\pm$      0.1 &      0.29 & 3.27e-15&    -6.33 & LVC-NC  \\ [0.5ex]
\hline
SO518 &    -4.31 $\pm$    -0.43 &     1.72 $\pm$      0.8  &    198.7 $\pm$      0.1 &     0.84 & 1.17e-13 &    -4.31 & LVC-BC \\ [0.5ex]
SO540 &    -4.65 $\pm$    -0.47 &    -0.65 $\pm$      1.2  &     39.0 $\pm$      1.8 &     0.47 & 9.22e-14 &    -4.65 & LVC-SC \\ [0.5ex]
SO562 &    -5.44 $\pm$    -0.54 &     0.29 $\pm$      2.0  &     31.4 $\pm$      2.2 &      0.6 & 1.22e-14 &    -5.44 & LVC-SC \\ [0.5ex]
SO583 &    -3.96 $\pm$     -0.4 &    -8.51 $\pm$      8.4  &     88.6 $\pm$     13.9 &     0.26 & 8.46e-13 &    -3.96 & LVC-SC \\ [0.5ex]
SO615 &    -5.82 $\pm$    -0.50 &    -15.41$\pm$      1.9  &     67.3 $\pm$     4.8  &     0.07 & 3.92e-14 &    -5.82 & LVC-SC  \\ [0.5ex]
SO662 &    -5.04 $\pm$     -0.5 &     1.91 $\pm$      2.9  &     56.7 $\pm$      4.7 &     0.22 &  8.2e-14 &    -5.04 & LVC-SC \\ [0.5ex]
SO682 &     -4.9 $\pm$    -0.49 &    -1.72 $\pm$      2.7  &     68.5 $\pm$      4.4 &      0.3 & 8.04e-14 &     -4.9 & LVC-SC \\ [0.5ex]
SO687 &    -5.41 $\pm$    -0.54 &     1.06 $\pm$      1.1  &     15.6 $\pm$      1.1 &     0.12 & 6.17e-14 &    -5.41 & LVC-SC \\ [0.5ex]
SO697 &    -4.88 $\pm$    -0.49 &      6.7 $\pm$     18.9  &    121.6 $\pm$     31.4 &     0.16 & 1.57e-13 &    -4.88 & LVC-SC \\ [0.5ex]
\hline
SO774 &    -5.33 $\pm$    -0.53 &   -19.46 $\pm$      2.1  &     39.9 $\pm$      3.4 &     0.14 & 2.76e-14 &    -5.72 & LVC-NC \\ [0.5ex]
SO774 &                         &   15.1   $\pm$      15.1 &     101.3$\pm$      25.1&     0.12 & 2.76e-14 &    -5.78 & LVC-BC  \\ [0.5ex]
SO774 &                         &   14.78  $\pm$      1.2  &     22.4 $\pm$      2.0 &     0.07 & 2.76e-14 &    -5.98 & LVC-NC  \\ [0.5ex]
\hline
SO818 &    -5.26 $\pm$    -0.53 &      1.9 $\pm$      1.2  &     35.0 $\pm$      1.5 &     0.32 & 3.29e-14 &    -5.26 & LVC-SC \\ [0.5ex]
SO844 &    -5.59 $\pm$    -0.56 &   -10.41 $\pm$      1.2  &     41.5 $\pm$      1.5 &     0.26 & 1.82e-14 &    -5.59 & LVC-BC \\ [0.5ex]
SO859 &    -5.16 $\pm$    -0.52 &   -15.64 $\pm$      1.8  &    103.8 $\pm$      0.1 &     0.78 &  1.7e-14 &    -5.16 & LVC-BC \\ [0.5ex]
SO897 &    -4.77 $\pm$    -0.48 &     -1.4 $\pm$      3.2  &     50.3 $\pm$      5.2 &     0.24 & 1.39e-13 &    -4.77 & LVC-SC \\ [0.5ex]
SO927 &    -5.27 $\pm$    -0.53 &     1.63 $\pm$      0.8  &     32.4 $\pm$      0.1 &     0.28 & 3.57e-14 &    -5.27 & LVC-NC \\ [0.5ex]
SO984 &    -5.09 $\pm$    -0.51 &   -24.93 $\pm$     15.1  &    109.5 $\pm$     24.9 &     0.19 & 8.32e-14 &    -5.09 & LVC-SC \\ [0.5ex]
\hline
SO1036 &    -4.75 $\pm$    -0.52 &    -5.49 $\pm$     1.6  &     31.8 $\pm$      2.6 &     0.15 & 8.73e-14 &    -5.19 & LVC-NC \\ [0.5ex]
SO1036 &                         &    0.0   $\pm$     14.6 &     129.9$\pm$      24.3&     0.27 & 8.73e-14 &    -4.94 & LVC-BC  \\ [0.5ex]
\hline
SO1153 &    -4.61 $\pm$    -0.46 &   -19.67 $\pm$      0.8 &     20.4 $\pm$      0.1 &     0.27 & 1.86e-13 &    -4.61 & LVC-NC \\ [0.5ex]
SO1361 &    -4.24 $\pm$    -0.42 &    -1.11 $\pm$      3.2 &     31.2 $\pm$      5.2 &     1.15 & 9.66e-14 &    -4.24 & LVC-NC \\ [0.5ex]
\hline
\end{tabular}
\begin{tablenotes}
    \item [1] Total luminosity of the low-velocity component.
    \item [2] Luminosity of each Gaussian component.
    \item [3] LVC: Low-velocity component. SC: single-component. NC: narrow-component. BC: broad-component.
\end{tablenotes}
\end{threeparttable}
\end{center}
\end{table*}

For the other optical forbidden lines of \NII\,and \SII,$\lambda$6716, considering both samples (MIKE+X-shooter), we found lower detection rates between $\sim$$20-60$\% in agreement with previous results \citep[e.g.,][]{McGinnis2018}. The [SII] lines show preferentially HVC (see Fig.~\ref{fig:SII6730_fit}, \ref{fig:SII6716_fit}, and \ref{fig:SII6730_fit_XS}) most likely related to collimated jets as also seen in low-mass SFRs \citep{Hartigan95}. 
Since the \NII\,lines can have a contribution from nebular emission \citep{Gangi2023},  we verified that none of our detections were contaminated by nebular emission by inspecting the 2D spectral images.
Table~\ref{tab:line_detect_rates} lists the detection rates found in both MIKE and X-shooter samples for all the forbidden lines studied in this work. The profile shapes are consistent in the two instruments, with differences only due to resolution, as shown in Fig.~\ref{fig:MIKE_vs_XS} in Appendix~\ref{ap:MIKE_vs_XS}. For the overlapping sample, if a line is detected in both data sets, the higher resolution MIKE observations are favored over the X-shooter data throughout this work. The fitted line profiles and the results from the line fits can be found in Appendix~\ref{ap:other_lines}. 

\begin{table}
\begin{center}
\caption{\label{tab:line_detect_rates} Optical Forbidden Lines LVC Detection Rates}
\begin{tabular}{lccc}
\hline\hline
Line    &   MIKE    &   X-shooter \\
        &   (\%)    &   (\%)     \\
\hline
\OI     &    85     &   80       \\ [0.5ex]
\hline
\NII    &    63     &   38        \\ [0.5ex]
\hline
\SII    &    19     &   22        \\ [0.5ex] 
\hline
\end{tabular}
\end{center}
\end{table}

\subsection{Kinematics of the \OI\,line}

\begin{figure}
    \centering
	\includegraphics[width=0.5\textwidth]{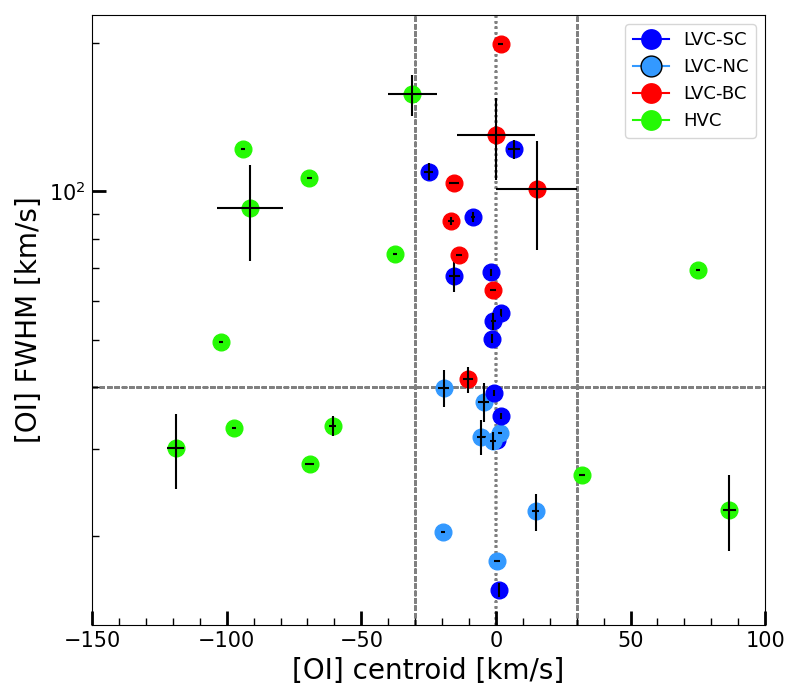}
    \caption{Kinematics of the \OI: FWHM vs \Vp\, for \sorionis\, sources for the MIKE sample. Colors indicate the type of Gaussian component as described in Fig.~\ref{fig:OI6300_fit}. Vertical lines indicate the systemic stellar velocity (0 km/s) and the velocity limit between low and high-velocity components ($\pm$30 km/s). The horizontal line indicates the width threshold assumed for the broad and narrow components (40 km/s).}
    \label{fig:OI_FWHM_Vp}
\end{figure}

Figure~\ref{fig:OI_FWHM_Vp} shows the kinematic properties of the \OI\,line in the FWHM - peak velocity plane for the MIKE sample. Vertical lines indicate the systemic stellar velocity (0 km/s) and the velocity limit between low and high-velocity components ($\pm$30 km/s). The horizontal line indicates the width threshold assumed for the broad and narrow components instead (40 km/s). 
The peak velocities range from -119 to +86 km/s, with a higher number of blueshifted lines than redshifted. Note the high concentration of peak velocities around 0 km/s. HVC are present in 8 out of 27 sources ($\sim$30\%), with 5 sources showing HVC with blueshifts larger than -100 km/s, indicative of collimated jets. The lesser blueshifted components may be linked to micro-jets instead. Only 3 stars (SO362, SO927, SO1361) show a red-shifted HVC.

As shown in Fig.~\ref{fig:OI_FWHM_Vp}, most of the single-line LVC have peak velocities around 0 km/s, with four targets having |$v_{\rm peak}$| $>$ 5 km/s, and most of them are blue-shifted, indicative of winds. However, they mostly exhibit broad (FWHM $>$ 40 km/s) profiles as well. 
The BC and NC line profiles are more or less equally distributed in velocity, with a preference for blue-shifted emission over red-shifted components. The frequency of single-component over NC and BC for \sorionis\, sources in our MIKE sample is 65\%/35\%. 

Similar results are found for the X-shooter sample as shown in Fig.~\ref{fig:OI_FWHM_Vp_XS}. There is a greater prevalence of blueshifted over redshifted components with a larger frequency of broad, single-line components, expected from the lower resolution. 

\subsection{\OI\,LVC Line Luminosities} \label{sec:results_OIlum}

We measured equivalent widths of the \OI\,LVC line profile for our sample, and derived their line fluxes as $F_{\rm line} \approx$ EW(line) $\times\, F_{\rm cont}$, where EW(line) and $F_{\rm cont}$ are the equivalent width and the continuum flux around the line, respectively.  
We used X-shooter flux-calibrated spectra from \citet{Mauco2023} to estimate $F_{\rm cont}$ for the MIKE spectra of the overlapping sample. For those stars without X-shooter spectra, we calculated the $F_{\rm cont}$ from the extinction-corrected $R$-band magnitude of each source reported in \citet{Hernandez2007}. Once the line fluxes are estimated, the line luminosities are given by $L_{\rm line} = 4\pi d^2 F_{\rm line}$, where $d$ is the distance to each star from the \textit{Gaia} EDR3 \citep{Gaia2021} and reported in \citet{Mauco2023}. These values are reported in Table~\ref{tab:OIfit} for the MIKE sample and are listed in Table~\ref{tab:OIfit_XS} for sources with X-shooter spectra.    

\begin{figure}
	\includegraphics[width=0.5\textwidth]{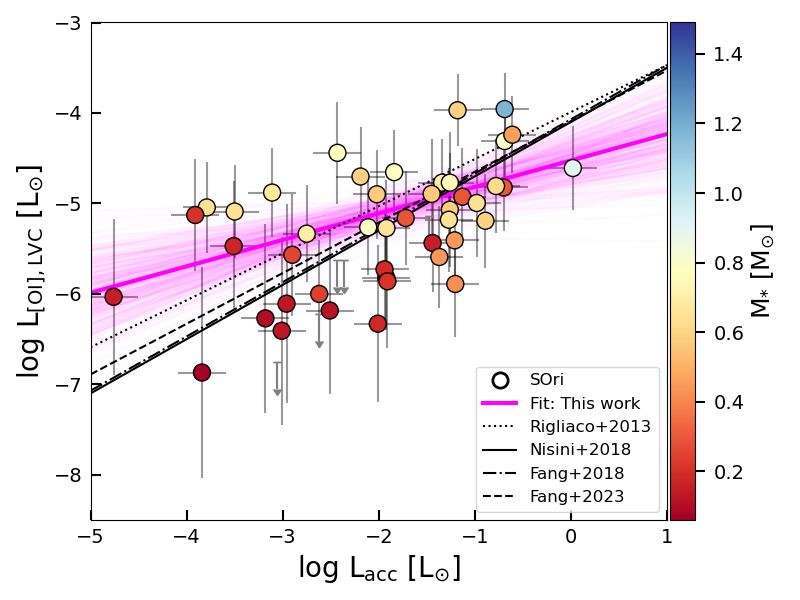}
        \includegraphics[width=0.438\textwidth]{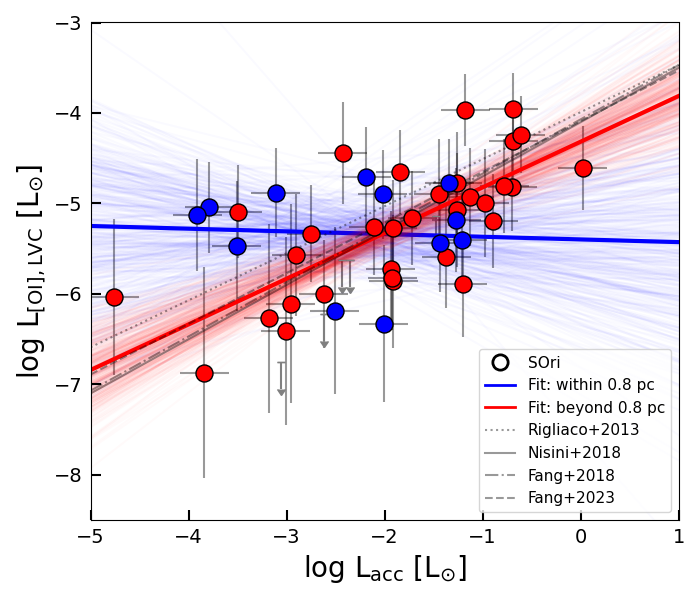}
    \caption{\OI\,line luminosity of the LVC as a function of $L_{\rm acc}$ of \sorionis\, sources in both data sets (MIKE and X-shooter). \textit{Top}: A positive correlation is found with a r=0.79 (pink line). Correlations found in other low-mass SFRs are also shown for comparison \citep{McGinnis2018,Fang2018,Fang2023,Nisini2018,Rigliaco2013}.
    The distribution of stellar masses is indicated by the color gradient.  \textit{Bottom}: Correlations for the highly irradiated disks within 0.8 pc from \sori\,(blue), and for the less irradiated disks located further out (red).}
    \label{fig:OI_Lacc}
\end{figure}

It is well-known that the \OI\,line luminosity positively correlates with accretion luminosity for protoplanetary disks in low-mass star-forming regions \citep[e.g.,][]{McGinnis2018,Fang2018,Fang2023,Nisini2018,Rigliaco2013}. 
Fig~\ref{fig:OI_Lacc} shows the \OI\,line luminosity of the LVC (log$\rm L_{OI,LVC}$) as a function of the accretion luminosity (\Lacc), taken from \citet{Mauco2023}, for all the sources with measured log$\rm L_{OI,LVC}$ in both data sets (MIKE and X-shooter). We found a positive correlation (pink line) with a correlation coefficient r = 0.81. We compute the linear regression coefficients using the \texttt{linmix} package \citep{Kelly2007}, which allows us to include uncertainties on both axes and upper limits in the fitting procedure. We adopt the median of the results of the chains as best-fit values. The best fit obtained with this method has a slope $\beta$ = 0.29$\pm$0.10, a square of the standard deviation $\sigma^2$ = 0.05$^{+0.07}_{-0.04}$, and r = 0.81$^{+0.12}_{-0.21}$. 
We also added the stellar mass of our sources as a color gradient in the top panel. 

The bottom panel shows the fitting for sources within and beyond 0.8 pc, which is the distance within which we believe external photoevaporation is most significant. This is based on: 1) the fact that, as discussed in Sect.~\ref{sec:OI_shape}, all sources within 0.8 pc show a particular \OI\,line morphology (single-component) probably due to the effect of the external radiation field, and 2) this range of projected distances is equal to the range of FUV field strengths that have been shown to be significant in producing the \OI\,line in external photoevaporation models \citep[$10^3-10^6 G_0$,][see Fig.~\ref{fig:OI_lum}]{ballabio2023}.
We found a similar correlation to that of the entire sample for sources located beyond 0.8 pc (red line), with values of $\beta$ = 0.51$^{+0.09}_{-0.10}$, $\sigma^2$ = 0.22$^{+0.10}_{-0.06}$, r = 0.77$^{+0.08}_{-0.11}$.
In contrast, we found no correlation for sources within 0.8 pc (blue line) with values of $\beta$ = -0.02$\pm$0.24, $\sigma^2$ = 0.45$^{+0.37}_{-0.18}$, r = -0.04$\pm$0.37, as expected from external photoevaporation. Appendix~\ref{ap:corner} shows the corner plots with the posterior analysis results.

The flat slope observed for sources within 0.8 pc is most probably independent of the range of stellar masses or accretion luminosities within that sample. K-S tests performed on the distribution of stellar masses and accretion luminosities for both samples (beyond and within 0.8 pc) returned high probabilities that both samples are drawn from the same parent population with p-values equal to 0.57, and 0.20, respectively. Moreover, using a bootstrap sampling method with $10^4$ iterations \citep[as done in,][]{Aru2024,McLeod2021}, we confirmed the positive correlation for sources beyond 0.8 pc and the lack of correlation for sources within 0.8 pc. The results of this method are also shown in Appendix~\ref{ap:corner}.  
The implications of these findings are discussed later in Sect.~\ref{sec:OI_lacc}.

\subsection{Line Ratios}

We estimated line ratios of [OI]6300/([SII]6716+6731) and [NII]6583/([SII]6716+6731) of our \sorionis\,sources in both datasets (MIKE and X-shooter) following \citet{Rigliaco2009}. As in the case of the \OI\,line, we only considered the LVC of each line in our estimates, and the [NII] and [SII] line fluxes were estimated following the same methodology as in sec.~\ref{sec:results_OIlum}. In cases where the $\rm [SII]$ lines were not detected, we estimated a 3-$\sigma$ upper limit as 3$\times$ RMS $\times\,\Delta\lambda$, with RMS as the local flux noise and $\Delta\lambda$ as the expected line width, following \citet{Gangi2023}. The line width was estimated from the other detected [SII] lines which have an average value of $\sim$1 \AA.
Table~\ref{tab:line_ratios} lists these line ratios for our \sorionis\,sample. In section~\ref{sec:line_ratio_discussion} we discuss these results in the context of external photoevaporation in the cluster. 

\begin{table}
\begin{center}
\caption{\label{tab:line_ratios} Line Ratios}
\begin{tabular}{lccc}
\hline\hline
Name    &   $\rm \frac{[OI]6300}{[SII]6716+6731}$    &   $\rm \frac{[NII]6583}{[SII]6716+6731}$  \\
\hline
*MIKE &  &\\ [0.5ex]
\hline
SO396	&	0.61 	    &  3.61 \\ [0.5ex]
SO341	&	3.68		&  0.16 \\ [0.5ex]
SO662	&	2.32		&  4.38 \\ [0.5ex]
SO1036	&	1.70		&  4.42 \\ [0.5ex]
\hline
 *X-shooter &  &\\ [0.5ex]
\hline
SO299   &   >1.48       &  --     \\ [0.5ex]
SO362   &   >31.47      &  --     \\ [0.5ex]
SO467   &   >4.67       &  --     \\ [0.5ex]
SO490   &   >2.84       &  --     \\ [0.5ex]
SO500	&	4.07		&  --     \\ [0.5ex]
SO520	&	>1.19		&  --     \\ [0.5ex]
SO540	&	>8.35		&  --     \\ [0.5ex]
SO562	&	>6.09		&  --     \\ [0.5ex]
SO563	&	2.24		&  --     \\ [0.5ex]
SO583	&	3.80		&  1.95   \\ [0.5ex]
SO587	&	0.41		&  1.13   \\ [0.5ex]
SO646	&	>7.43		&  --     \\ [0.5ex]
SO662	&	2.16		&  4.49   \\ [0.5ex]
SO682   &   --          &  >2.49  \\ [0.5ex] 
SO687	&	>0.99		&  >16.05 \\ [0.5ex]
SO694	&	>2.58		&  >21.92 \\ [0.5ex]
SO697	&	2.33		&  --     \\ [0.5ex]
SO774	&	>3.87		&  --     \\ [0.5ex]
SO818	&	>6.65	    &  --     \\ [0.5ex]
SO823	&	>40.27	    &  >2.23  \\ [0.5ex]
SO848	&	0.96		&  1.61   \\ [0.5ex]
SO897	&	>5.76 	    &  >3.37  \\ [0.5ex]
SO1036	&	2.77		&  5.97   \\ [0.5ex]
SO1152	&	>3.57 	    &  --     \\ [0.5ex]
SO1156	&	>4.53 	    &  --     \\ [0.5ex]
SO1260	&	4.50		&  1.27   \\ [0.5ex]
SO1266	&	3.11		&  --     \\ [0.5ex]
SO1267  &   --          &  >0.82  \\ [0.5ex]
SO1274	&	>1.89 	    &  --     \\ [0.5ex]
SO1327	&	>2.30 	    &  --     \\ [0.5ex]
SO1361	&	>2.82 	    &  --     \\ [0.5ex]
SO1369	&	>1.58 	    &  --     \\ [0.5ex]
\hline
\end{tabular}
\end{center}
\end{table}

\section{Discussion}\label{sec:discussion}

\begin{figure}
	\includegraphics[width=0.5\textwidth]{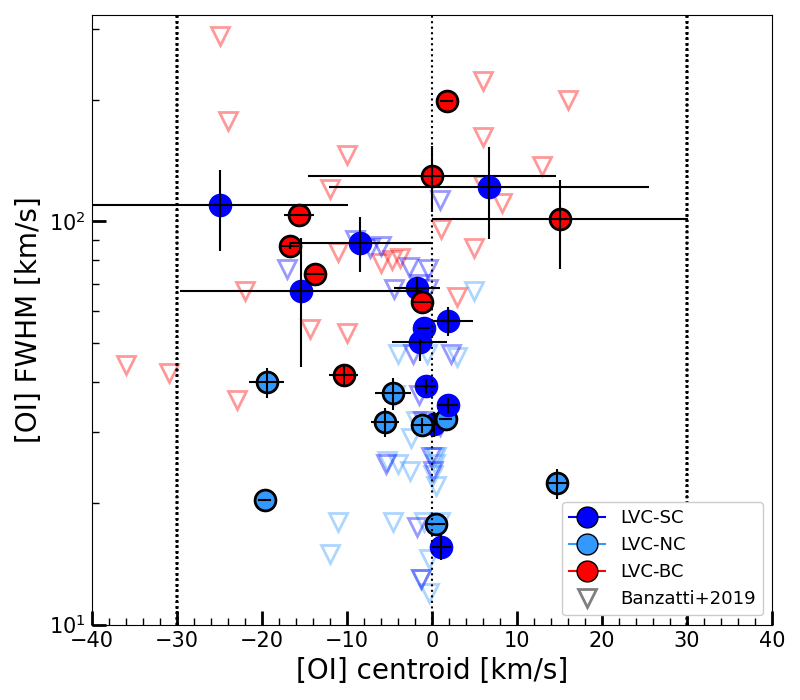}
        \includegraphics[width=0.5\textwidth]{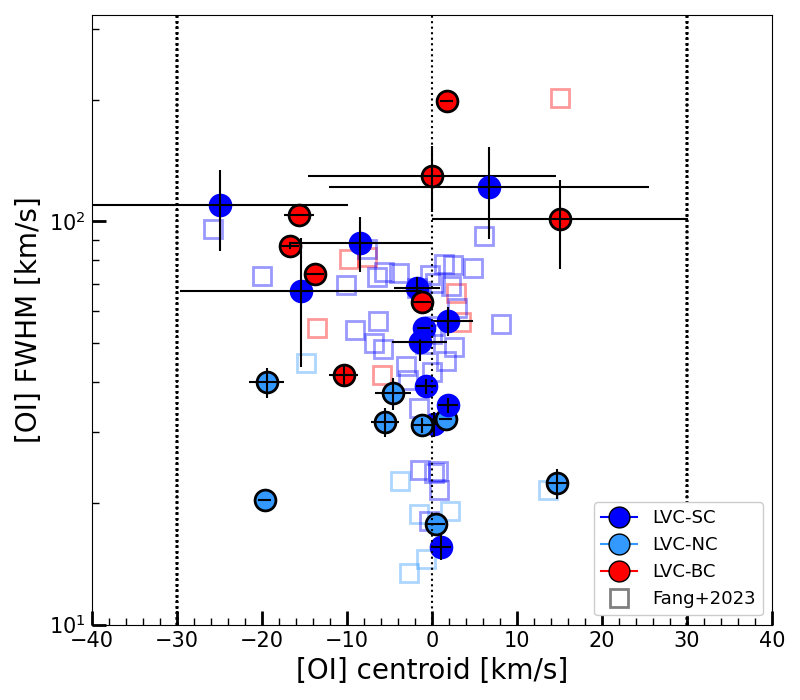}
    \caption{\OI\,LVC line kinematics for \sorionis\,sources in the MIKE sample (circles) compared to others low-mass SFRs. \textit{Top}: 1-3 Myr younger regions of Lupus, Taurus, and Ophiuchus from \citet{Banzatti2019}. \textit{Bottom}: 5-10 Myr older region of Upper Sco from \citet{Fang2023}.}
    \label{fig:OI6300_SFR_comparison}
\end{figure}

\subsection{Tracing Disk Evolution through the \OI\,line}\label{sec:age_evo}

The \OI\,line is particularly useful for disk evolution studies. On the one hand, it is the brightest line among the optical forbidden lines, making it the most commonly observed and most studied line in the literature, with well-characterized line profiles for several stars in different low-mass SFRs. This makes it ideal for survey studies. On the other hand, it is also a proxy for dust evolution: the line changes from multiple components for full disks to a single, centrally peak line for disks with central cavities \citep{Ercolano2017, Banzatti2019, pascucci2022}. Moreover, the line profile shows signs of evolving with age as the disk disperses, with the HVC and broad LVC disappearing for older regions as the inner disk becomes optically thin, pointing to inside-out gas dissipation in the inner disk \citep{McGinnis2018, Fang2018, Fang2023}. As a result, a single component \OI\,line profile becomes the dominant mode in older regions \citep{Nisini2024}.

Given the intermediate age of \sorionis, and the fact that we can use the outermost part of the cluster for disk evolution studies, it is interesting to compare the kinematic properties ($v_{\rm peak}$, FWHM) of the \OI\,line of our \sorionis\,sample with those found in other SFRs with different ages. For that, we used the young ($\sim$1-3 Myr) sample of \citet{Banzatti2019}, which studied the \OI\,line in Taurus, Lupus, and Ophiuchus, and the old (5-10 Myr) sample of \citet{Fang2023} in Upper Sco. We only considered the MIKE sample for this analysis given its higher resolution, similar to these works. Because observations were carried out at slightly different spectral resolutions, we deconvolved all observed FWHMs from the instrumental broadening by taking as Gaussian FWHM that of the respective sample resolution following \citet{McGinnis2018,Fang2023}.

In Figure~\ref{fig:OI6300_SFR_comparison}, we show the FWHM-$v_{\rm peak}$, plane for the \OI\,line LVC of our MIKE \sorionis\, sources compared to the younger regions from \citet{Banzatti2019} on the top panel, and with the older Upper Sco region from \citet{Fang2023} on the bottom panel.
As shown in the figure, the kinematic properties of the \OI\,line LVC for \sorionis\, sources seem to agree relatively well with those found in younger regions, they populate more or less similar areas in the FWHM-$v_{\rm peak}$, plane. The K-S test performed among the different samples returns high probabilities that the centroid velocities and FWHMs are drawn from the same parent population when considering the complete sample of each region ($p$ = 0.24-0.80 and 0.36-0.40, respectively) as well as the sub-samples divided by Gaussian components ($p_{\rm SC}$ = 0.57-0.77 and 0.63-0.72, $p_{\rm NC}$ = 0.45-0.96 and 0.57-0.55, $p_{\rm BC}$ = 0.76-96 and 0.57-0.99).

This is quantified on Tab.~\ref{tab:kine_regions} where we summarized the median FWHM and centroid velocities for each region. The distribution of centroid velocities and FWHM for the SC and NC are in agreement, within the uncertainties, among the different regions. This is also the case for the BC between the younger regions and our sample, with \sorionis\,having a slightly larger FWHM median. The only notable difference is seen on the BC between Upper Sco and the other regions, showing BC with smaller FWHM but similar centroid velocities (given the uncertainties). 
More importantly, Upper Sco stands out as the region with more single-line \OI\,profiles, as reported in previous results and as expected for later evolutionary stages \citep{McGinnis2018,Fang2023,Nisini2024}. 

\begin{figure}
    \centering
    \includegraphics[width=0.5\textwidth]{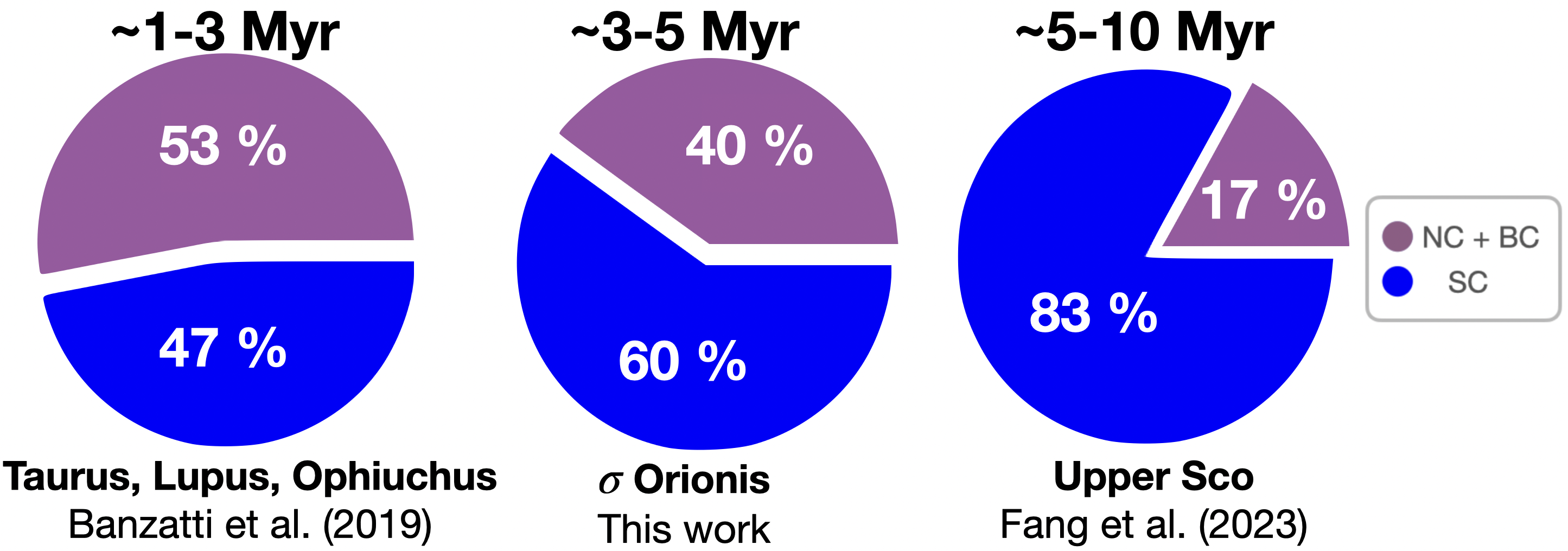}
    \caption{Frequency of the single-component to broad/narrow component of the \OI\,line among SFRs with different ages: $\sim$1-3 Myr old Lupus, Taurus, and Ophiuchus from \citet{Banzatti2019}, $\sim$3-5 Myr old \sorionis\,from this work (MIKE sample), and $\sim$5-10 Myr older Upper Sco from \citet{Fang2023}.}
    \label{fig:piecharts}
\end{figure}

This is better visualized in Fig.~\ref{fig:piecharts}, which shows as piecharts the frequency of single-line (SC) to broad/narrow (NC+BC) components of the \OI\,line LVC among these regions. The younger regions studied in \citet{Banzatti2019} show a more or less balance frequency of 47\% (SC) to 53\% (NC+BC), for the \sorionis\,cluster, we found frequencies of 60\% (SC) to 40\% (NC+BC), while for the older, Upper Sco region the numbers are 83\% (SC) to 17\% (NC+BC). The occurrence of single-peak \OI\,line profiles is found to increase with age, in agreement with previous results \citep{McGinnis2018, Fang2018, Fang2023}, with the \sorionis\, cluster indeed revealing an intermediate stage of evolution.
A caveat to this result would be the fact that, although these numbers remain the same for the younger regions when considering only sources within our completeness (0.5\msun) and/or our magnitude limit (0.33\msun, see Sect.~\ref{sec:sample}), they do change for the older Upper Sco region with frequencies dropping to values around $\sim$60\%. This is mainly because the Upper Sco sample has a significant fraction of lower-mass stars ($M_{*} < 0.33$\msun). If we do the calculations including the wider mass limit covered by our X-shooter sample ($\sim$0.1\msun), the frequencies remain the same as those shown in Figure~\ref{fig:piecharts}.

Although \sorionis\,sources in our sample seem to be evolving similarly as SFRs without massive stars, based on the frequency of its SC \OI\,lines, we show in the next section, robust evidence of external photoevaporation affecting the \OI\,line profiles for sources in close proximity to \sori\,($d_p < 0.8 $ pc). This rather suggests a co-evolution of external as well as internal processes affecting the disks, with the former being less prominent for stars located at projected separations from \sori\, $\gtrsim$ 0.8 pc, where only mild ($\sim 10^3-10^2\, G_0$) FUV fields are present \citep{Mauco2023}. If we exclude the most highly irradiated sources (within 0.8 pc from \sori, see Sect.~\ref{sec:OI_shape}) the SC to NC+BC frequency for our sample\,is 43\% to 57\%, which remains consistent with the expected trend with age.

\begin{table}
\begin{center}
\caption{\label{tab:kine_regions} Median FWHM (km/s) and Centroids velocities (km/s) of the \OI\,LVC}
\begin{tabular}{lccc}
\hline\hline
[OI] LVC                       &   SC      &    NC    &    BC     \\
\hline
Young SFRs & & & \\[0.2ex]
  $v_{\rm peak}$               &    -1.5 $\pm$ 2.2   &   -0.9 $\pm$ 2.2   &    -6.0 $\pm$ 4.0   \\ [0.5ex]
  FWHM                         &    57.1 $\pm$ 2.6   &   24.0 $\pm$ 1.5   &   85.7 $\pm$ 8.1     \\ [0.5ex]
\hline
\sorionis\, & & & \\[0.2ex]
  $v_{\rm peak}$              &    -0.1 $\pm$ 5.6    & -1.1 $\pm$ 1.2     &   -1.2 $\pm$ 5.2  \\ [0.5ex]
  FWHM                        &    55.7 $\pm$ 8.9    &  29.3 $\pm$ 1.3    &   100.7 $\pm$ 7.5  \\ [0.5ex]
\hline
Upper Sco & & & \\[0.2ex]
  $v_{\rm peak}$              &    -0.6 $\pm$ 1.3    & -1.5 $\pm$ 1.3     &   -5.9 $\pm$ 1.3  \\ [0.5ex] 
  FWHM                        &    55.5 $\pm$ 1.5    &  17.0 $\pm$ 1.5    &   65.7 $\pm$ 1.5  \\ [0.5ex] 

\hline
\end{tabular}
\end{center}
\end{table}


\begin{figure}
    \centering
	\includegraphics[width=0.5\textwidth]{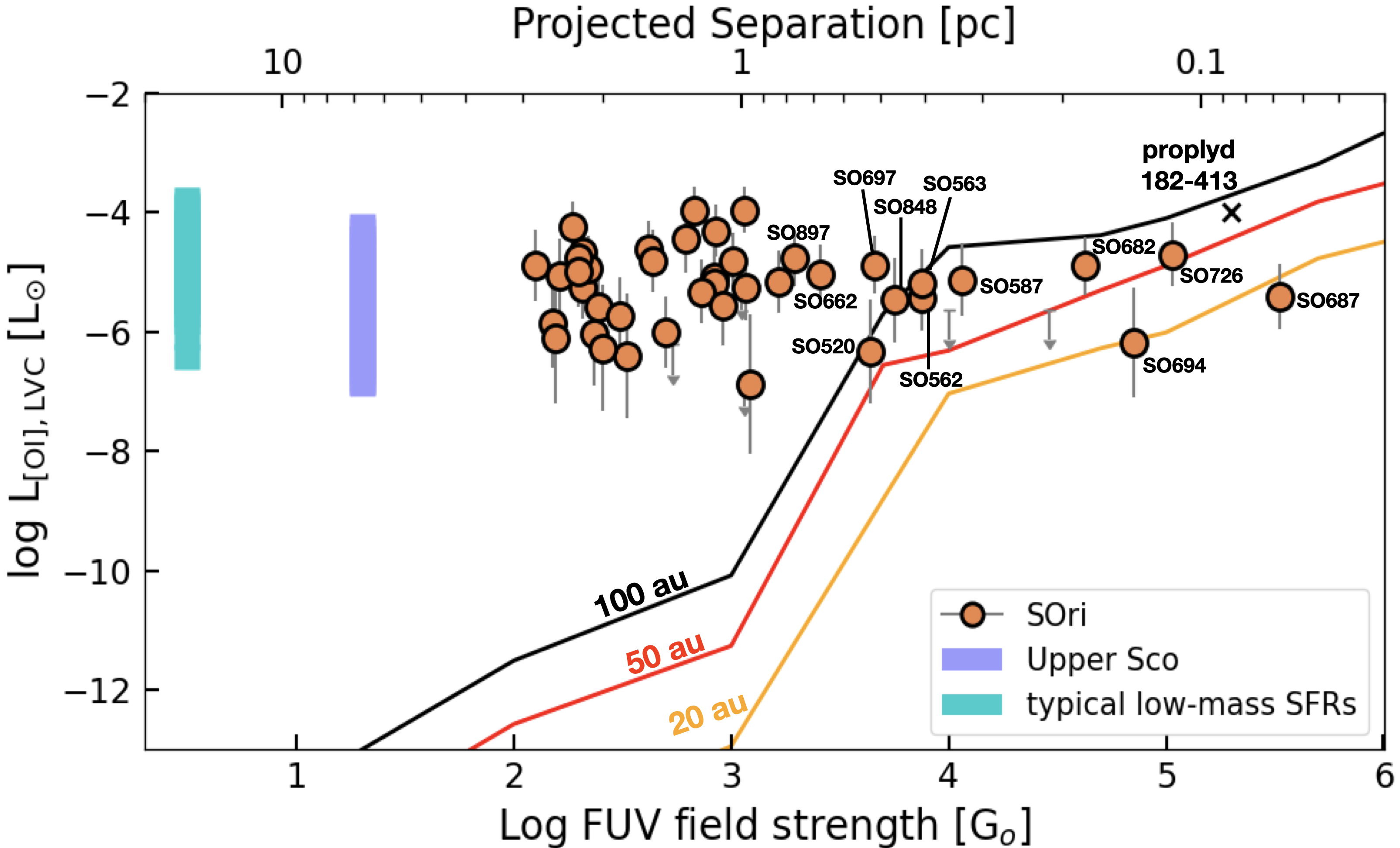}
    \caption{\OI\,LVC luminosities of \sorionis\,sources in both datasets (MIKE and X-shooter) as a function of FUV field strength. Expected luminosities from external photoevaporation models \citep{ballabio2023} for disks with different initial sizes (solid lines) are shown. Disks exposed to FUV field strength $> 10^3\,G_0$ (within 0.8 pc from \sori) are labeled. The range of luminosities found in other low-mass SFRs is also shown for comparison \citep[][]{Fang2018, Fang2023}. The location of proplyd 182-413 is taken from \citep{ballabio2023} which assumes an accretion luminosity estimated using the empirical relations from \citet{Nisini2018} for the stellar mass of the object.}
    \label{fig:OI_lum}
\end{figure}

\subsection{External Photoevaporation in the \sorionis\,Cluster}

The \sorionis\, cluster is an ideal region to study external photoevaporation. This has been highlighted in the past where one externally irradiated disk was proposed based on the presence of blue-shifted, asymmetric line profiles of high-ionization species of [NII] and [SII] in high-resolution spectra \citep[][SO587]{Rigliaco2009}, as well as through the very low dust disk masses found for sources within the first 0.5 pc from the center of the cluster \citep{Ansdell2017,Mauco2023}. Following these studies, we analyzed wind diagnostics for a sample of \sorionis\, sources using high resolution ($\Delta v \sim$10 km/s) optical spectra using as our main tracer the \OI\,line. The premise of this work is that, if environmental effects, such as external photoevaporation due to the FUV field produced by \sori, are indeed affecting the disks in the cluster, then we expect to see a correlation of the properties of the \OI\,line with the distance to the massive system. This is what we investigate in this section.

\subsubsection{Implications from the \OI\,Line Luminosities}\label{sec:OI_lacc}

Fig.~\ref{fig:OI_lum} shows the measured luminosity of the \OI\,line LVC as a function of the FUV field strength for our complete sample (MIKE + X-shooter). We also compared with luminosities observed in other low mass-star forming regions \citep[e.g., Lupus, Taurus, Ophiuchus,][]{Fang2018} and the older Upper Sco region \citep{Fang2023}. The \OI\,line luminosities for \sorionis\, sources are mainly constant regardless of the strength of the FUV field produced by \sori\, and are comparable to those found in other low-mass SFRs. 

The fact that we found a constant distribution of $[\rm OI]$ line luminosities for the LVC regardless of the FUV field strength and at similar levels than in low-mass SFRs, indicates that internal processes may also play a role in generating the \OI\,line in \sorionis\,sources, particularly at FUV field strengths lower than $10^3\,G_0$, where the luminosity of the line due to external irradiation is expected to be low and has been related to internal winds \citep{ballabio2023}. Indeed, models of the collisionally excited \OI\,line show averaged line luminosities of $\sim10^{-5}$ \lsun\, \citep{Ercolano2010,Ercolano2016}, which is consistent with the luminosities we found in our sample. 
Similarly, based on theoretical predictions of externally photoevaporating models \citep{ballabio2023}, shown in Fig.~\ref{fig:OI_lum} for disks with different initial radii (solid lines), it is evident that at lower $G_0$ ($<10^3$) the models indicate a sharp drop in the \OI\,luminosity where external irradiation is not effective in producing the line. Sources exposed to FUV field strengths larger than $2\times 10^3\, G_0$, on the other hand, are mostly consistent with disks with small radii ($\lesssim$100 au), pointing to the effect of external irradiation. 

To investigate to what extent the [OI] line luminosity we observed is indeed related to internal processes, we analyzed how the [OI] luminosity correlates with accretion from the host star in Fig.~\ref{fig:OI_Lacc}. The \sorionis\, sources do exhibit a positive $\rm L_{[OI]} - L_{acc}$ correlation (pink line top panel), in agreement with other low-mass SFRs (gray lines) but with a less steep relationship.  The difference may indicate the effect of external photoevaporation, as in this scenario, we do not expect to find any correlation between the luminosity of the [OI] line and the luminosity of accretion, since the heating is dominated by the external radiation field \citep{ballabio2023}. We found this for the highly irradiated disks within 0.8 pc from \sori\, (blue line in the bottom panel). For these sources, the $\rm L_{[OI]}$ is independent of $\rm L_{acc}$ (constant). Moreover, the $\rm L_{[OI]} - L_{acc}$ correlation gets steeper, and more similar to other low-mass SFRs, for less irradiated sources beyond 0.8 pc (red line in the bottom panel). This reinforces the claim that sources near the massive system \sori\, have been affected by external irradiation and this process is responsible for the \OI\,line emission we observe, or at least is the dominant factor. Figure~\ref{fig:OI_Lacc} also shows that the \OI\,luminosity also depends on the mass of the central object, in that the more massive stars are responsible for the brightest lines, but this could be related to higher-mass stars having higher accretion rates. 

\begin{figure}
    \centering
	\includegraphics[width=0.5\textwidth]{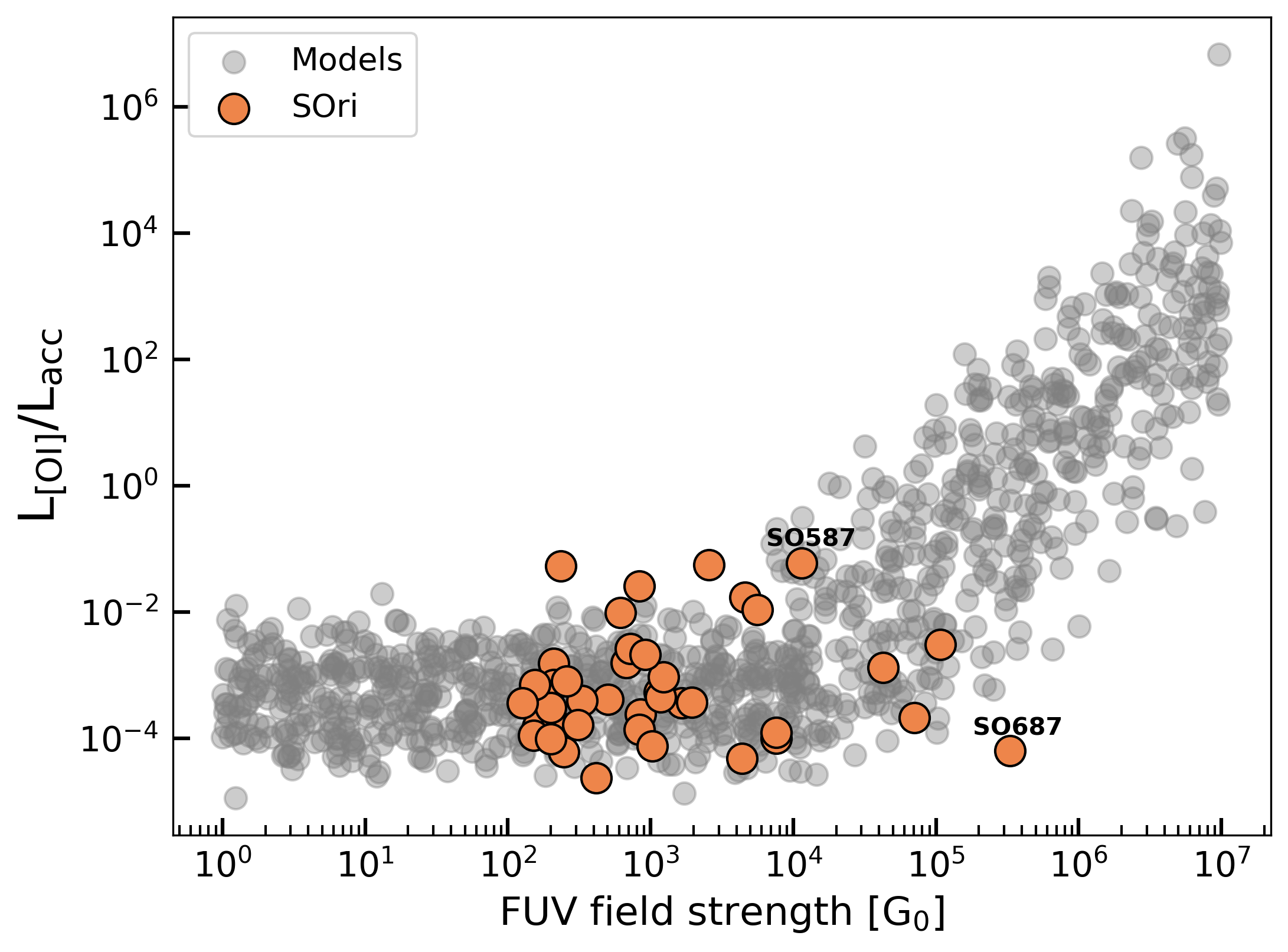}
    \caption{Ratio of the total \OI\,LVC line luminosity to accretion luminosity for \sorionis\, disks in both datasets (MIKE and X-shooter). Model predictions from \citet{ballabio2023} are also shown for comparison.}
    \label{fig:LOI_Lacc_Go}
\end{figure}

\citet{ballabio2023} proposed the $\rm L_{[OI]}/L_{acc}$ ratio as a diagnostic for identifying external photoevaporation as it potentially distinguishes bright internal contributions associated with high accretion and external contributions to the line luminosity. The models predict a significant increase in the $\rm L_{[OI]}/L_{acc}$ ratio for disks exposed to FUV fields larger than $10^4\,G_0$. Figure~\ref{fig:LOI_Lacc_Go} shows this ratio for \sorionis\,sources compared to the expected values. The \sorionis\, sources agree fairly well with this prediction within the range of FUV field strengths investigated by our observations. SO587, the externally photoevaporative disk reported by \cite{Rigliaco2009}, is among the sources with the highest L$_{\rm [OI]}$/L$_{\rm acc}$, as expected. In contrast, SO687, which is the closest disk to \sori, appears as an outlier on the figure with one of the lowest L$_{\rm [OI]}$/L$_{\rm acc}$. As mentioned before, this highly irradiated source is probably very small in size (radii $<$20 au, see Fig.~\ref{fig:OI_lum}) given its low L$_{\rm [OI]}$. However, it still exhibits a significant accretion luminosity (logL$_{\rm acc}$ = -1.21 \lsun), lowering its L$_{\rm [OI]}$/L$_{\rm acc}$. 

The K-S test performed between the observed $\rm L_{[OI]}/L_{acc}$ ratio (excluding SO687) and those from the models (within the observed FUV field strength range) shows that both samples are most probably drawn from the same parent distribution with a p-value equal to 0.14. If we include SO687 instead, the p-value drops to 0.034, rejecting the null hypothesis. 
Moreover, to test how strong the distribution of $\rm L_{[OI]}/L_{acc}$ is a monotonic function of the FUV field strength, we estimated the Spearman’s correlation coefficient ($r_{\rm s}$) of the observations (excluding the outlier SO687) and the models (with $G_0$ values within the observed FUV field strength range). The test showed that both, models and observations, are represented by only a weak monotonic relationship with Spearman’s coefficients, $r_{\rm s}$, of 0.38 and 0.23, respectively.
Unfortunately, there is a lack of sources at higher $G_0$ values to be able to test the theoretical trend with this sample beyond $10^{4}-10^{5}\,G_0$. More data probing highly irradiated environments is needed to confirm this trend.

\subsubsection{Implications from the morphology of the \OI\,line profile}\label{sec:OI_shape}

\begin{figure}
    \centering
	\includegraphics[width=1\columnwidth]{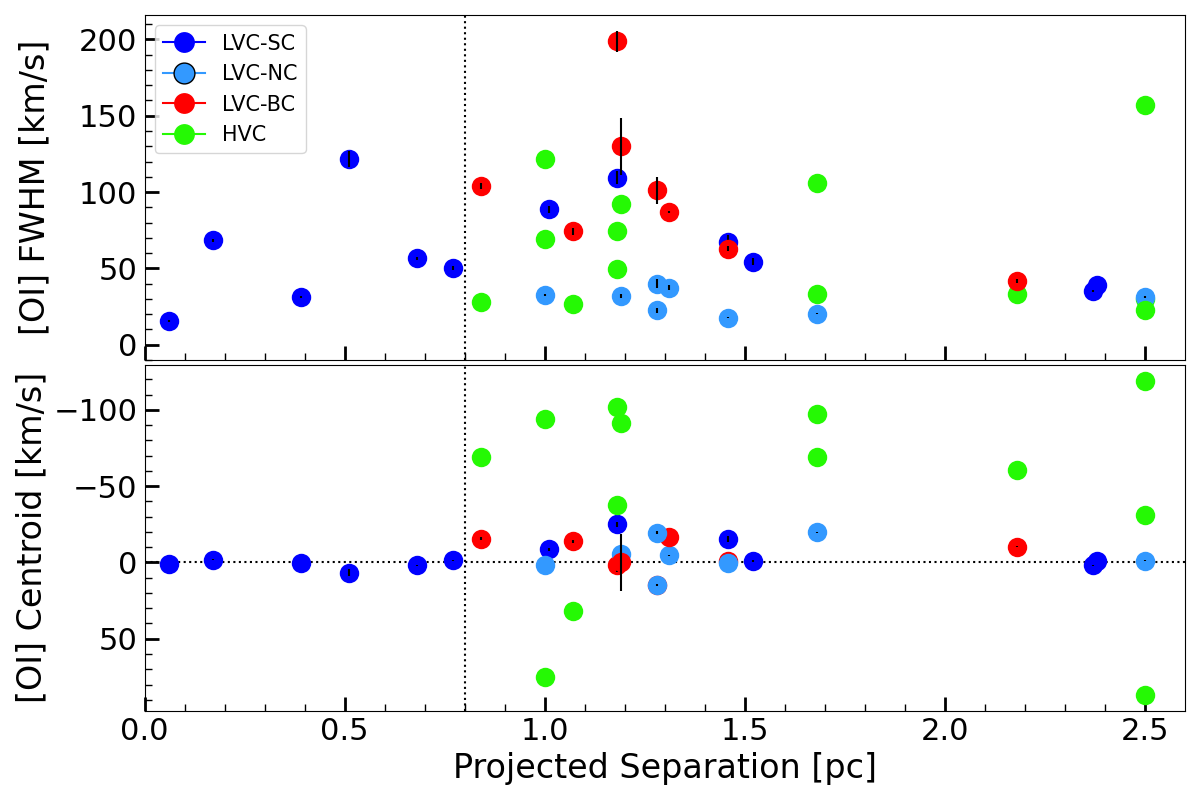}
    \caption{\OI\,lines FWHM (top) and peak velocity (bottom) as a function of projected separation from \sori\, for MIKE sources. The colors represent the type of Gaussian components as explained in sec~\ref{sec:Gauss_fit}. The vertical dotted line is positioned at 0.8 pc (see text for details). The horizontal line indicates the systemic stellar velocity (0 km/s).}
    \label{fig:OI6300_vs_dp}
\end{figure}

If external photoevaporation is indeed active in the inner regions of the cluster, then sources exposed to higher FUV fields (hence closer to \sori) are prime targets to investigate if the shape of their line profiles differs from those of more distant stars. 
In Fig~\ref{fig:OI6300_vs_dp}, we show the kinematic properties of the \OI\,line for \sorionis\,sources as a function of projected separation from the massive system \sori. As seen, the \OI\,line profiles for sources within the first 0.8 pc from \sori\, clearly deviate from stars further out in that they all exhibit only single-component line profiles. This result could point to external irradiation significantly impacting the \OI\,line profiles for stars in more massive SFRs. 

As discussed in Sect.~\ref{sec:age_evo}, this SC line profile is also the dominant mode in older regions and has been related to inner disk dust dissipation, given that the SC \OI\,profile becomes narrower as the infrared spectral index (an indicator of inner disk clearing) increases \citep{Banzatti2019,Fang2023}. However, the line widths are also found to be too broad ($>$ 40 km/s) in most stars to be compatible with thermal/photoevaporative winds and thus have been related to internal MHD winds \citep{Fang2023}. Currently,  as stated in \citet{Fang2023}, the transition of the LVC from BC+NC to SC is not yet understood and likely holds clues about the properties of the disk wind as the system evolves and the accretion rate decreases.

In the context of external photoevaporation in the \sorionis\,cluster, it appears that the rapid mass loss from the outer parts of the disks of sources close to \sori, evidenced by their low dust disk masses \citep{Ansdell2017,Mauco2023}, induces a "premature" late evolutionary stage similar to that expected in older regions.
This is consistent with the fact that the HVC in low-mass SFRs is only found in optically thick inner disks, which in our case are all found in sources beyond 0.8 pc from \sori. Thus, as the disks located closer to \sori\,deplete at a faster rate than those further away, a higher proportion of single-component lines may be expected there, which is what we see in Fig.~\ref{fig:OI6300_vs_dp}. The low centroid velocities for these close-in sources (some of them blueshifted) indicate that the SC \OI\,line is indeed probing slow winds, likely driven by external irradiation \citep{ballabio2023}. The rather broader lines (theoretical expectations predict FWHM $<$ 6 km/s, \citealt{ballabio2023}), then point to a possible contribution to the line emission from internal processes such as MHD winds, but not as the dominant factor given the lack of correlation of the \OI\,luminosity with L$_{\rm acc}$ found for inward sources in Sect.~\ref{sec:OI_lacc}, and/or internal photoevaporative winds, given the similar range of \OI\,luminosities observed across the sample and similar [OI]/[SII] line ratios compared to low-mass star-forming regions (Fig.~\ref{fig:Line_ratios}).

Higher spectral resolution observations ($R >80,000$) are needed to confirm if SC \OI\,line profiles are ubiquitous in externally irradiated disks. Given the resolution of our data, the profile we see may be a combination of a very narrow \OI\,line component, due to external photoevaporation, superimposed to a broader component (FWHM $>$ 30 km/s) produced by inner winds close to the central star.

\subsubsection{Implications from line ratios of highly ionized species}\label{sec:line_ratio_discussion}

Externally photoevaporating disks are expected to have stronger high-ionization emission lines, e.g. [NII] 6583 \AA\, and [SII] 6731 \AA, compared to disks in low-mass SFRs without massive OB stars, as these lines are formed in the outer part of the outflow, which is ionized, heated, and shocked by the external radiation impinging on the disks \citep{Rigliaco2009}. This was demonstrated using spatially resolved MUSE observations of Orion proplyds \citep{Aru2024}, showing that the \NII\,line mostly comes from the bright cusps near the ionization front while the \SII\,lines appear as the brightest lines at the tail of the proplyds. The low-ionization line of \OI\,in contrast, better traces the disk surface and part of the ionization front. The \OI\,line is known to be common and bright in low-mass SFRs and related to internal processes. Since the \OI\,luminosity in \sorionis\,appears to be constant regardless of the strength of the ambient FUV field (sec.~\ref{sec:results_OIlum}), using line ratios of high-ionized species can be a useful diagnostic to study external photoevaporation in the cluster. 

As shown in \citet{Rigliaco2009}, the [OI]6300/([SII]6716+6731) and [NII]6583/([SII]6716+6731) line ratios for SO587 (an externally irradiated disks in \sorionis), clearly deviates from those found in Taurus \citep{Hartigan95}, the former ratio being lower while the later larger compared to Taurus disks.   
Therefore, we expect a similar behavior for these line ratios in our sample: sources exposed to higher FUV fields (closer to \sori) should exhibit lower [OI]6300/([SII]6716+6731) ratios along with higher [NII]6583/([SII]6716+6731) ratios than in typical low-mass SFRs. Disks located in the outer parts of the cluster (beyond 0.8 pc), and exposed to mild FUV fields, should have values more similar to low-mass SFRs instead.

\begin{figure*}
    \centering
	\includegraphics[width=0.9\textwidth]{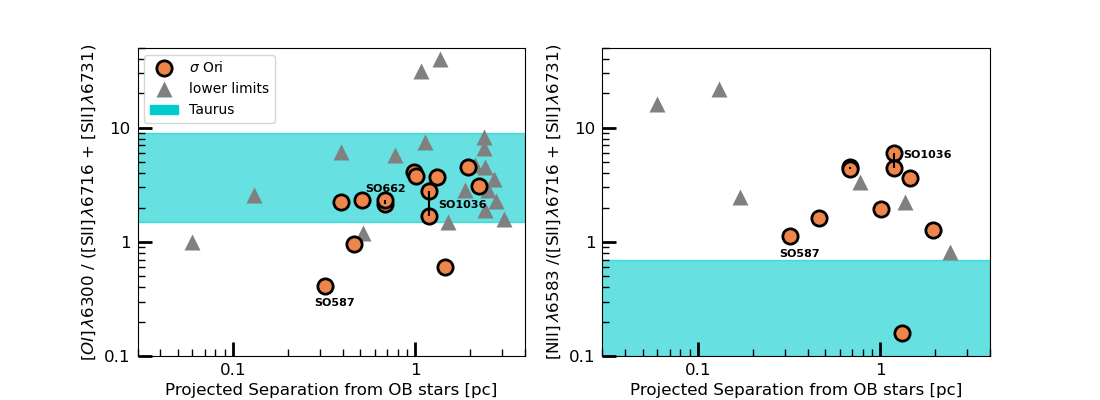}
    \caption{High-ionization emission lines ratios for \sorionis\,sources in both datasets (MIKE and X-shooter).
    \textit{Left:} [OI]6300/([SII]6716+[SII]6731) line ratio. \textit{Rigth:} [NII]6583/([SII]6716+[SII]6731) line ratio. Highly irradiated disks are expected to have a lower [OI]6300/([SII]6716+[SII]6731) ratio and higher [NII]6583/([SII]6716+[SII]6731) ratio compared to low-mass SFRs like Taurus. We only considered the LVC of each line in our estimate.}
    \label{fig:Line_ratios}
\end{figure*}

Figure~\ref{fig:Line_ratios} shows these ratios for \sorionis\,disks as a function of projected separation from the \sori\,massive system. Also shown is the typical range of values for Taurus sources as reported in \citet[][Fig.2]{Rigliaco2009}.
The expected behavior is found for disks in \sorionis\,for the highly ionized species: most of our objects (including upper limits) deviate from the typical values found in Taurus (light blue shaded regions) in the expected way, as seen in the right panel. These brighter lines of $[\rm NII]$ and $[\rm SII]$ (compared to regions without massive stars like Taurus) are consistent with the emission observed in the Orion proplyds by \citet{Aru2024}. In contrast, the [OI]6300/([SII]6716+6731) ratio is found to be similar as those observed in Taurus, pointing to a "contamination" of the \OI\,line luminosity from internal rather than external processes, as suggested in Sect.~\ref{sec:OI_lacc}. It seems that [NII]6583/([SII]6716+6731) ratio is a better diagnostic for identifying external photoevaporation than [OI]6300/([SII]6716+6731).

\section{Conclusions}\label{sec:conclusions}
We analyzed high-resolution ($\Delta v \sim10$ km/s) optical spectra for a sample of 27 CTTS in the $\sim$3-5 Myr-old \sorionis\,cluster, along with complementary, intermediate-resolution X-shooter spectra from \citet{Mauco2023},  to search for potential signatures of disk evolution and external photoevaporation. The \sorionis\,data set enables us to characterize the \OI\,line emission, as a function of age and external FUV field strength, and line ratios of highly ionized species.
Our main findings are as follows:

\begin{itemize}
    \item The \OI\,line LVC is detected in 23 out of 27 MIKE sources, and 40 out of 50 X-shooter sources, showing detection rates between $\sim$80-85\%. 
    The \OI\,luminosity for \sorionis\,sources shows a constant distribution throughout the FUV field strengths produced by the massive system \sori. The \OI\,luminosities have similar values as those found in low-mass SFRs. This indicates that internal processes may also play a role in generating the \OI\,line, particularly at FUV field strengths lower than $10^3\,G_0$, where the contribution due to external irradiation is expected to be low. 
    
    \item The \sorionis\,sample follows the positive $\rm L_{[OI]_{LVC}} - L_{acc}$ relation found in other low-mass SFRs but with a less steep relation. This is due to external photoevaporation, where the \OI\,luminosity of the  LVC for highly irradiated disks (within 0.8 pc) is independent of accretion luminosity, as expected from external photoevaporation models.

    \item The frequency of the \OI\,SC over NC+BC in \sorionis\, agrees with the expected trend found in other low-mass SFRs given its intermediate age ($\sim$3-5 Myrs). This result also holds when considering only sources exposed to mild FUV fields ($<10^3\,G_0$) and thus without a significant influence from external irradiation.
    
    \item Highly irradiated disks in \sorionis\,exhibit centrally-peaked, broad single-component \OI\,line profiles. Their line profiles resemble those preferentially found in older (more evolved) regions, probably because these disks are losing mass at a faster rate than disks further out in the cluster. However, the single-component lines are broader than expected, pointing to a contribution from inner winds. Higher resolution observations are needed to confirm if single-component line profiles are ubiquitous in externally irradiated disks.   

    \item Line ratios of highly ionized species (e.g., \NII\,and \SII) are consistent with externally irradiated disks, showing higher [NII]6583/([SII]6716+[SII]6731) ratios than stars in Taurus and similar to SO587 (a photoevaporating disk in the cluster). In contrast, the [OI]6300/([SII]6716+[SII]6731) ratio is found to be similar to stars in Taurus, pointing to a contamination of the \OI\,line luminosity from internal winds. The bright lines of $\rm [NII]$ and $\rm [SII]$ in \sorionis\,disks are consistent with the emission observed in the Orion proplyds. 
\end{itemize}

This work has shown that the study of multiple forbidden emission lines is a powerful tool to study both internally and externally driven disk winds. In particular, the results obtained on the peculiar kinematics and the line ratios in the inner regions of the $\sigma$-Orionis cluster show the potential to better constrain the external photoevaporation process. Theoretical modelling of these lines in the context of externally irradiated disks is needed, in order to further constrain the physical conditions of the disk winds. 

\begin{acknowledgements}
      We thank the anonymous referee for the helpful review of our work that improved the presented study.
      Funded by the European Union (ERC, WANDA, 101039452). Views and opinions expressed are however those of the author(s) only and do not necessarily reflect those of the European Union or the European Research Council Executive Agency. Neither the European Union nor the granting authority can be held responsible for them. TJH acknowledges funding from a Royal Society Dorothy Hodgkin Fellowship and UKRI guaranteed funding for a Horizon Europe ERC consolidator grant (EP/Y024710/1). S.F. is funded by the European Union (ERC, UNVEIL, 101076613), and acknowledges financial contribution from PRIN-MUR 2022YP5ACE. K.M. \& J.H. acknowledge support from the National Council of Humanities, Sciences, and Technologies (CONAHCyT) project No. 86372, the PAPIIT UNAM project  IG101723 and the UNAM Postdoctoral Fellowship Program. GB is supported by the European Research Council (ERC) under the European Union’s Horizon 2020 research and innovation programme (Grant agreement No. 853022, PEVAP). JMA acknowledges financial support from the Large Grant INAF 2022 “YSOs Outflows, Disks and Accretion: towards a global framework for the evolution of planet forming systems (YODA)”, from PRIN-MUR 2022 20228JPA3A “The path to star and planet formation in the JWST era (PATH)” funded by NextGeneration EU, and from INAF-GoG 2022 “NIR-dark Accretion Outbursts in Massive Young stellar objects (NAOMY)”.

\end{acknowledgements}

%
%

\bibliographystyle{aa}
\bibliography{bibliography}

\begin{appendix} 

\section{\OI\,LVC line profiles and luminosities for the X-shooter sample}\label{ap:OI_XS}

\subsection{Line profiles and luminosities}
\begin{figure*}[ht]
    \centering
	\includegraphics[width=0.9\textwidth]{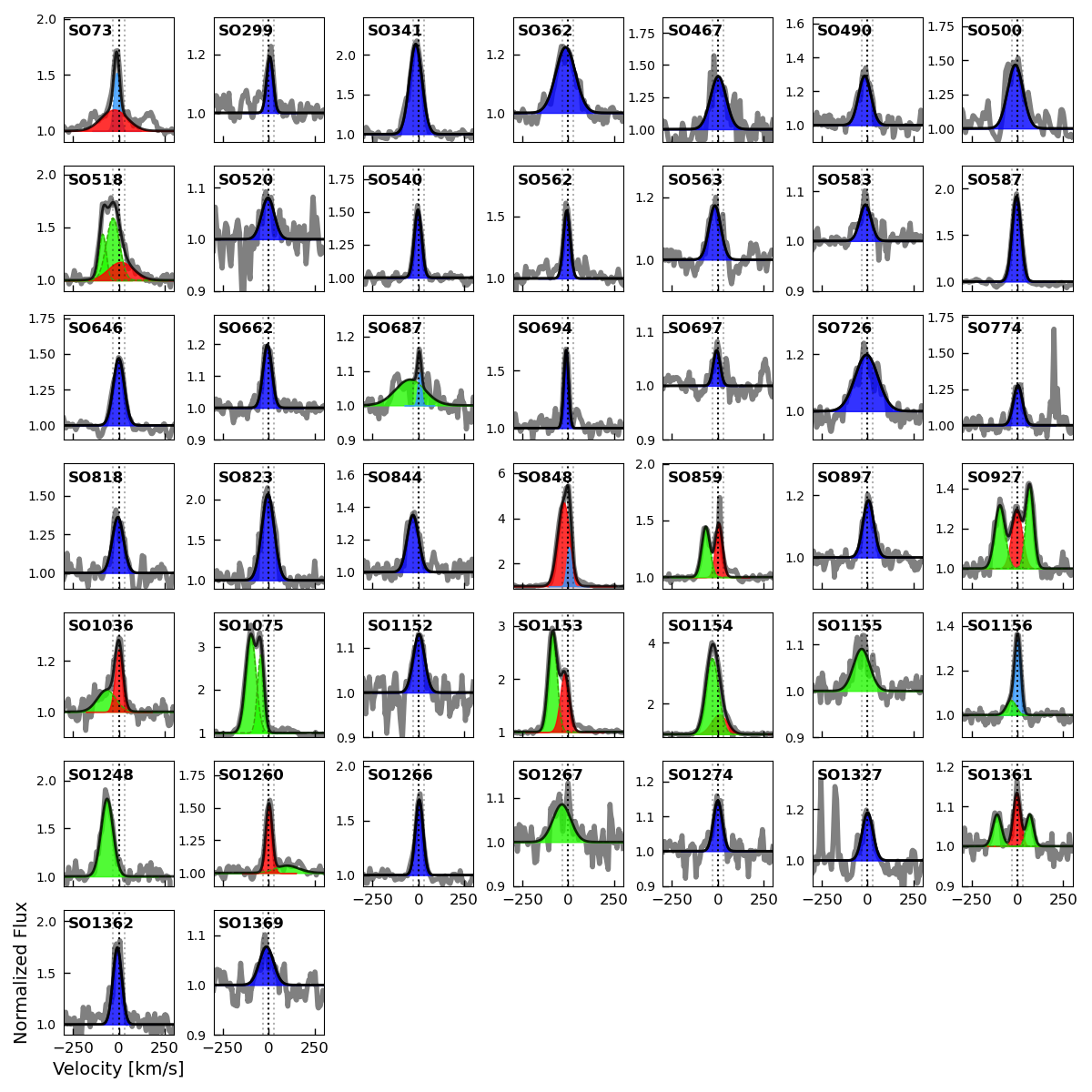}
    \caption{\OI\,lines fit for the X-shooter sample. The colors indicate the type of Gaussian component as described in Fig.~\ref{fig:OI6300_fit}. The black dotted line is located at 0 km/s, while the gray dotted lines indicate the velocity threshold for the high/low-velocity component ($\pm$30 km/s).}
    \label{fig:OI_Gauss_fit_XS}
\end{figure*}

\begin{figure}[ht]
    \centering
	\includegraphics[width=0.5\textwidth]{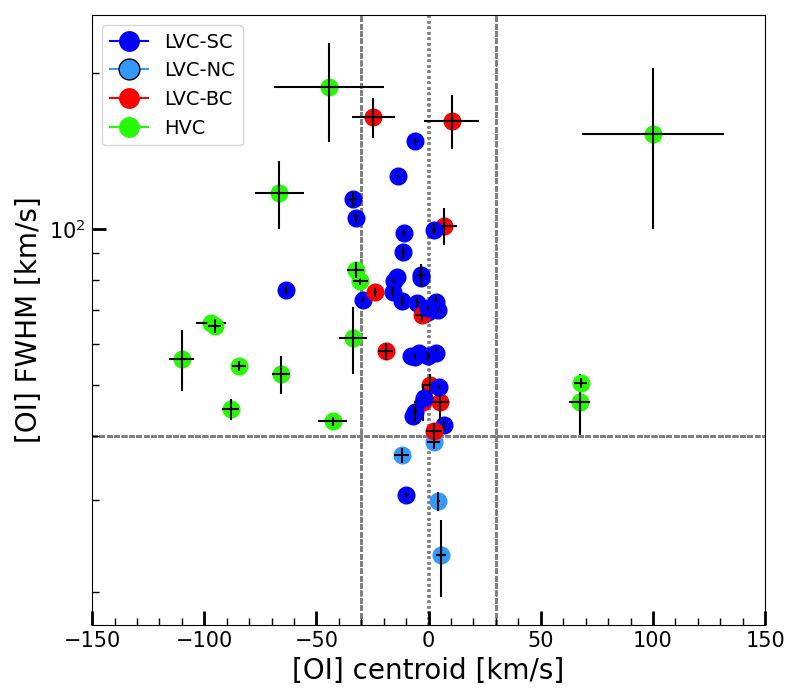}
    \caption{Kinematics of the \OI: FWHM vs $v_{\rm peak}$, for \sorionis\, sources for the X-shooter sample. Colors indicate the type of Gaussian component as described in Fig.~\ref{fig:OI6300_fit}. Vertical lines indicate the systemic stellar velocity (0 km/s) and the velocity limit between low and high-velocity components ($\pm$30 km/s). The horizontal line indicates the width threshold assumed for the broad and narrow components (40 km/s).}
    \label{fig:OI_FWHM_Vp_XS}
\end{figure}

\begin{figure*}[ht]
    \centering
	\includegraphics[width=0.9\textwidth]{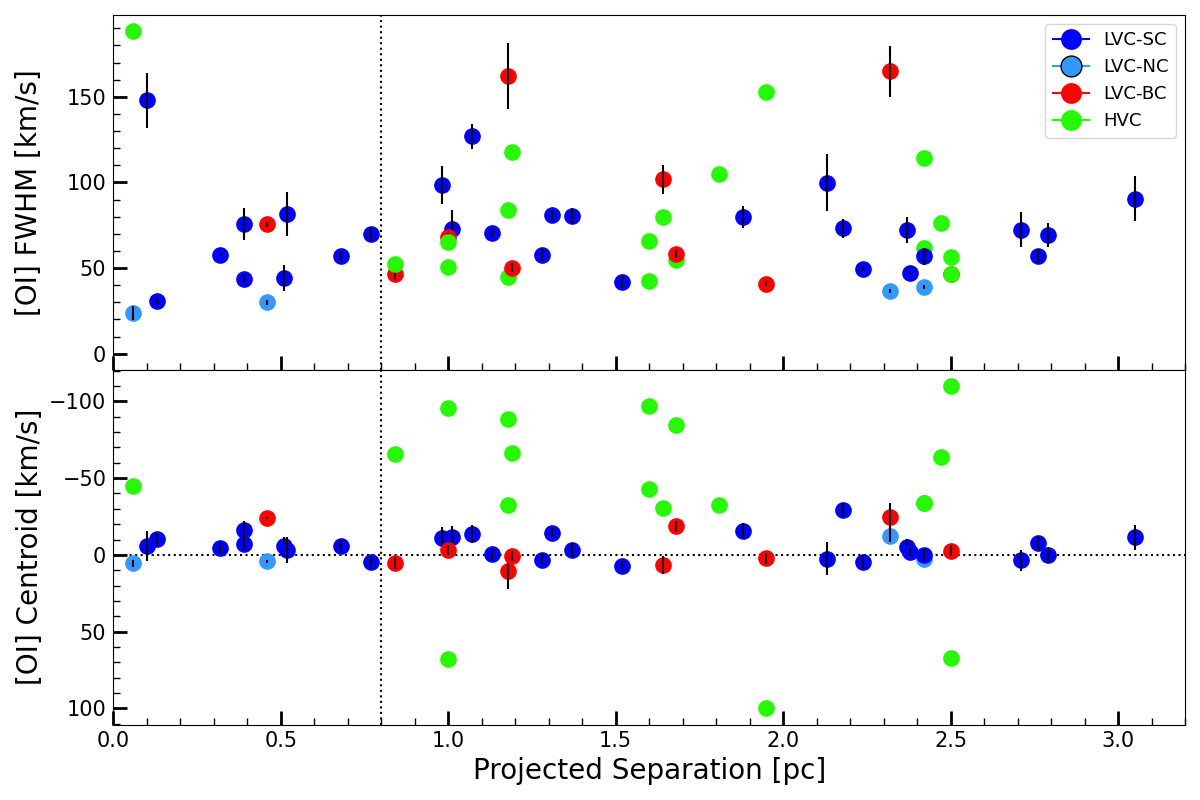}
    \caption{\OI\,lines FWHM (top) and peak velocity (bottom) as a function of projected separation from \sori\,for the X-shooter sample. The colors represent the type of Gaussian component as explained in sec~\ref{sec:Gauss_fit}. The vertical dotted line is positioned at 0.8 pc. The horizontal line indicates the systemic stellar velocity (0 km/s).}
    \label{fig:OI6300_dp_XS}
\end{figure*}

Table~\ref{tab:OIfit_XS} lists the \OI\,LVC fit parameters for the X-shooter sample. 
Figure~\ref{fig:OI_Gauss_fit_XS} shows the fitted \OI\,line profiles for the X-shooter sample. 
Figure~\ref{fig:OI_FWHM_Vp_XS} shows the kinematic properties of the \OI\,line for the X-shooter sample. Figure~\ref{fig:OI6300_dp_XS} shows the kinematic properties as a function of projected separation from \sori.

\begin{table*}
\begin{center}
\begin{threeparttable}[b]
\caption{\label{tab:OIfit_XS} Parameters for the \OI\,LVC for the X-shooter sample}
\begin{tabular}{lccccccc}
\hline\hline   
Name & log$L_{\rm OI, LVC}$\tnote{1} & FWHM\tnote{2}   & $v_{\rm peak}$\tnote{2} & EW    &  $F_{\rm cont}$   & log$L_{\rm OI,GC}$\tnote{3} & Class\tnote{4}\\
     & [\lsun] & [km/s] & [km/s]  & [\AA] &   [erg $\rm s^{-1}\,nm^{-1}\,cm^{-2}$ ]   &   [\lsun]   &       \\
\hline

SO73 &    -4.93 &     36.6 &   -12.03 &     0.42 & 2.66e-14 &    -5.34 & LVC-NC  \\ [0.5ex]
SO73 &          &    164.8 &   -24.61 &     0.71 & 2.66e-14 &    -5.12 & LVC-BC  \\ [0.5ex]
\hline
SO299 &    -6.12 &     41.9 &     6.96 &     0.18 & 1.08e-14 &    -6.12 & LVC-SC  \\ [0.5ex]
SO341 &     -4.1 &     80.9 &   -14.23 &     2.06 & 7.41e-14 &     -4.1 & LVC-SC  \\ [0.5ex]
SO362 &    -4.75 &    126.8 &   -13.51 &     0.64 & 5.44e-14 &    -4.75 & LVC-SC  \\ [0.5ex]   
SO467 &    -6.27 &     99.9 &     2.44 &     0.93 &  1.20e-15 &    -6.27 & LVC-SC  \\ [0.5ex]
SO490 &    -6.41 &     79.5 &   -15.57 &     0.52 &  1.50e-15 &    -6.41 & LVC-SC  \\ [0.5ex]
SO500 &    -6.87 &     98.4 &   -11.09 &     0.99 &  3.00e-16 &    -6.87 & LVC-SC  \\ [0.5ex]
SO518 &    -4.43 &    162.2 &    10.24 &     0.63 & 1.17e-13 &    -4.43 & LVC-BC  \\ [0.5ex]
SO520 &    -6.33 &     81.6 &     -3.3 &     0.15 &  6.30e-15 &    -6.33 & LVC-SC  \\ [0.5ex]
SO540 &    -4.58 &     47.3 &    -2.01 &     0.55 & 9.22e-14 &    -4.58 & LVC-SC  \\ [0.5ex]
SO562 &    -5.49 &     43.6 &    -7.16 &     0.53 & 1.22e-14 &    -5.49 & LVC-SC  \\ [0.5ex]
SO563 &    -5.18 &     75.9 &   -15.93 &     0.29 & 4.48e-14 &    -5.18 & LVC-SC  \\ [0.5ex]
SO583 &     -4.3 &     72.8 &   -11.82 &     0.12 & 8.46e-13 &     -4.3 & LVC-SC  \\ [0.5ex]
SO587 &    -5.13 &     57.7 &    -4.33 &      1.2 & 1.23e-14 &    -5.13 & LVC-SC  \\ [0.5ex]
SO646 &    -5.57 &     70.6 &    -0.39 &     0.75 &  7.10e-15 &    -5.57 & LVC-SC  \\ [0.5ex]
SO662 &    -4.98 &     56.8 &    -6.07 &     0.25 &  8.20e-14 &    -4.98 & LVC-SC  \\ [0.5ex]
SO687 &    -5.78 &     23.5 &     5.49 &     0.05 & 6.17e-14 &    -5.78 & LVC-NC  \\ [0.5ex]
SO694 &    -6.18 &     30.7 &   -10.27 &     0.46 &  3.00e-15 &    -6.18 & LVC-SC  \\ [0.5ex]
SO697 &    -5.28 &     44.3 &    -6.12 &     0.07 & 1.57e-13 &    -5.28 & LVC-SC  \\ [0.5ex]
SO726 &     -4.7 &    147.9 &     -6.0 &     0.66 & 5.88e-14 &     -4.7 & LVC-SC  \\ [0.5ex]
SO774 &     -5.3 &     57.7 &     3.13 &     0.36 & 2.76e-14 &     -5.3 & LVC-SC  \\ [0.5ex]
SO818 &     -5.0 &     72.2 &    -4.98 &     0.59 & 3.29e-14 &     -5.0 & LVC-SC  \\ [0.5ex]
SO823 &    -4.44 &     80.4 &    -3.36 &      1.9 & 3.79e-14 &    -4.44 & LVC-SC  \\ [0.5ex]
SO844 &    -5.25 &     73.2 &   -29.26 &     0.57 & 1.82e-14 &    -5.25 & LVC-SC  \\ [0.5ex]
\hline
SO848 &    -5.47 &     75.7 &   -24.02 &     6.27 &  1.10e-15 &    -5.55 & LVC-BC  \\ [0.5ex]
SO848 &          &     29.9 &     4.27 &     1.16 &  1.10e-15 &    -6.29 & LVC-NC  \\ [0.5ex]
\hline
SO859 &    -5.36 &     46.4 &     5.19 &     0.49 &  1.70e-14 &    -5.36 & LVC-BC  \\ [0.5ex]
SO897 &    -4.69 &     69.9 &     4.32 &     0.29 & 1.39e-13 &    -4.69 & LVC-SC  \\ [0.5ex]
SO927 &    -5.07 &     68.4 &    -3.12 &     0.45 & 3.57e-14 &    -5.07 & LVC-BC  \\ [0.5ex]
SO1036 &    -4.93 &     50.1 &     0.71 &     0.28 & 8.73e-14 &    -4.93 & LVC-BC  \\ [0.5ex]
SO1152 &    -5.07 &     72.4 &      3.4 &     0.21 & 8.03e-14 &    -5.07 & LVC-SC  \\ [0.5ex]
SO1153 &    -3.88 &     58.2 &   -19.09 &     1.44 & 1.86e-13 &    -3.88 & LVC-BC  \\ [0.5ex]
SO1154 &    -4.81 &    101.7 &     6.74 &     1.48 & 2.09e-14 &    -4.81 & LVC-BC  \\ [0.5ex]
SO1156 &    -4.77 &     38.9 &     2.38 &      0.3 & 1.13e-13 &    -4.77 & LVC-NC  \\ [0.5ex]
SO1260 &    -5.73 &     40.8 &     2.35 &     0.47 &  8.50e-15 &    -5.73 & LVC-BC  \\ [0.5ex]
SO1266 &    -6.03 &     49.7 &     4.73 &     0.77 &  2.40e-15 &    -6.03 & LVC-SC  \\ [0.5ex]
SO1274 &    -4.99 &     56.9 &    -0.05 &     0.18 & 1.05e-13 &    -4.99 & LVC-SC  \\ [0.5ex]
SO1327 &    -5.86 &     69.1 &    -0.07 &     0.29 &  9.50e-15 &    -5.86 & LVC-SC  \\ [0.5ex]
SO1361 &    -5.16 &     46.5 &    -2.66 &     0.14 & 9.66e-14 &    -5.16 & LVC-BC  \\ [0.5ex]
SO1362 &    -6.11 &     56.9 &    -7.88 &     0.93 &  1.70e-15 &    -6.11 & LVC-SC  \\ [0.5ex]
SO1369 &    -4.89 &     90.4 &   -11.34 &     0.16 & 1.61e-13 &    -4.89 & LVC-SC  \\ [0.5ex]
\hline
\end{tabular}
\begin{tablenotes}
    \item [1] Total LVC luminosity.
    \item [2] Median uncertainties. SC: $\delta_{\rm v_{peak}}$ = 0.6, $\delta_{\rm FWHM}$ = 1.4. NC: $\delta_{\rm v_{peak}}$ = 0.3 , $\delta_{\rm FWHM}$ = 0.9. BC: $\delta_{\rm v_{peak}}$ = 4.3, $\delta_{\rm FWHM}$ = 3.3.
    \item [3] Luminosity of each Gaussian component.
    \item [4] LVC: Low-velocity component. SC: single-component. NC: narrow-component. BC: broad-component.
\end{tablenotes}
\end{threeparttable}
\end{center}
\end{table*}

\subsection{Comparison with MIKE Observations}\label{ap:MIKE_vs_XS}

\begin{figure*}[ht]
    \centering
	\includegraphics[width=0.9\textwidth]{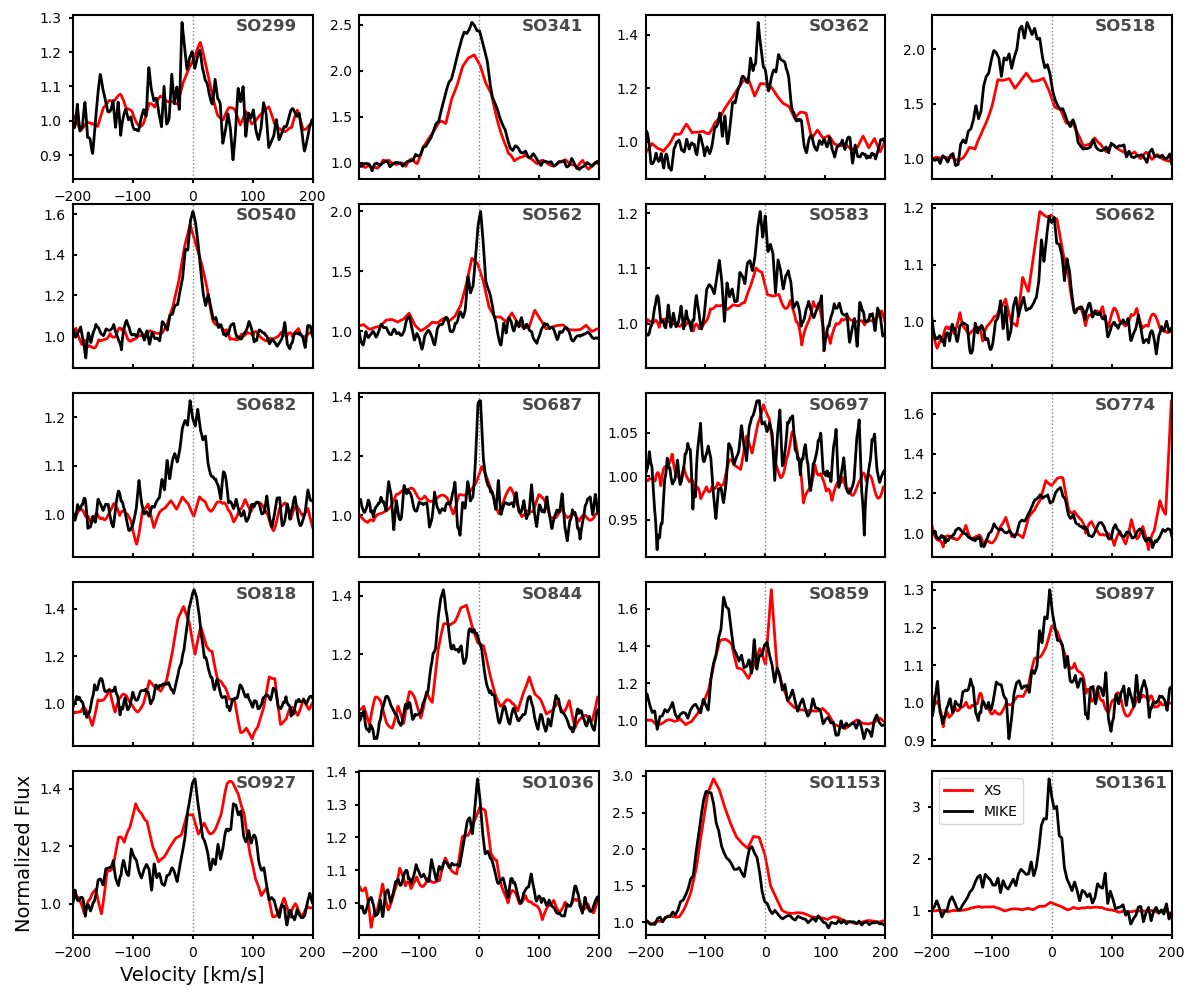}
    \caption{\OI\,LVC line profiles comparison. MIKE spectra is shown in black while X-shooter (XS) spectra is shown in red. SO682 was not detected in X-shooter. SO1361 is detected in X-shooter but with a lower intensity than in MIKE spectra, as shown in Fig.~\ref{fig:OI_Gauss_fit_XS}. The vertical dotted line indicates the systemic stellar velocity (0 km/s).}
    \label{fig:MIKE_vs_XS}
\end{figure*}

Figure~\ref{fig:MIKE_vs_XS} shows the comparison between the X-shooter and the MIKE spectra for the overlapping sample. 

\section{$\rm L_{[OI]}\, -\, L_{acc}$ fit: Corner Plots and Bootstrapping Method}\label{ap:corner}

\begin{figure}[ht]
    \centering
	\includegraphics[width=0.5\textwidth]{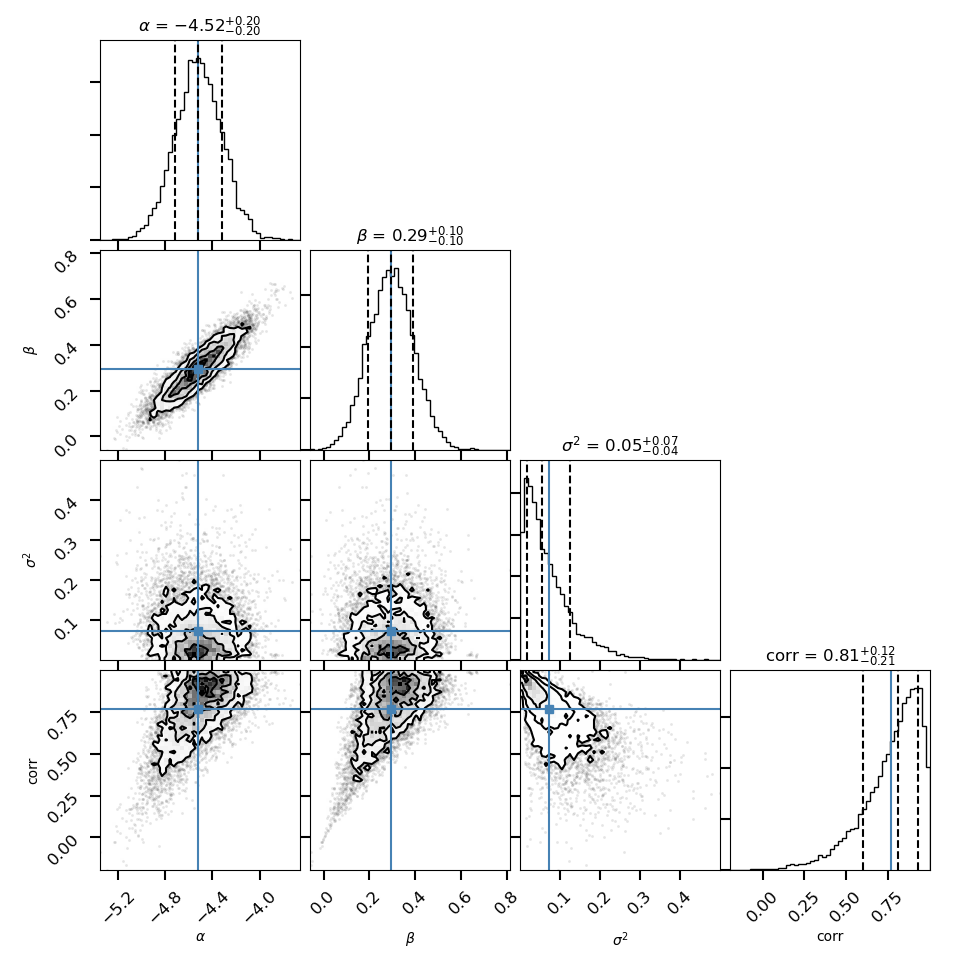}
    \caption{Corner plot of the log$L_{\rm OI, LVC}$ vs. log$L_{\rm acc}$ fit for the complete sample, done using the linmix routine \citep{Kelly2007}.}
    \label{fig:LOI_Lacc_all_fit}
\end{figure}

\begin{figure}[ht]
    \centering
	\includegraphics[width=0.5\textwidth]{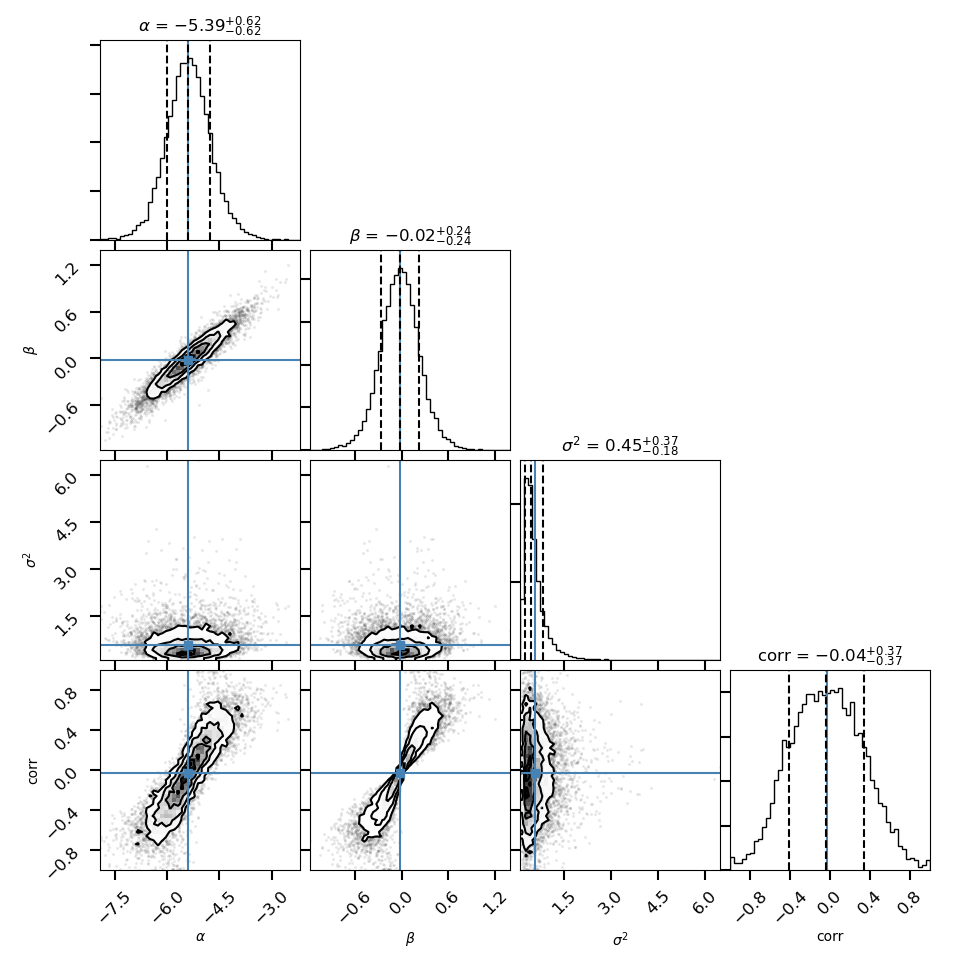}
 \includegraphics[width=0.5\textwidth]{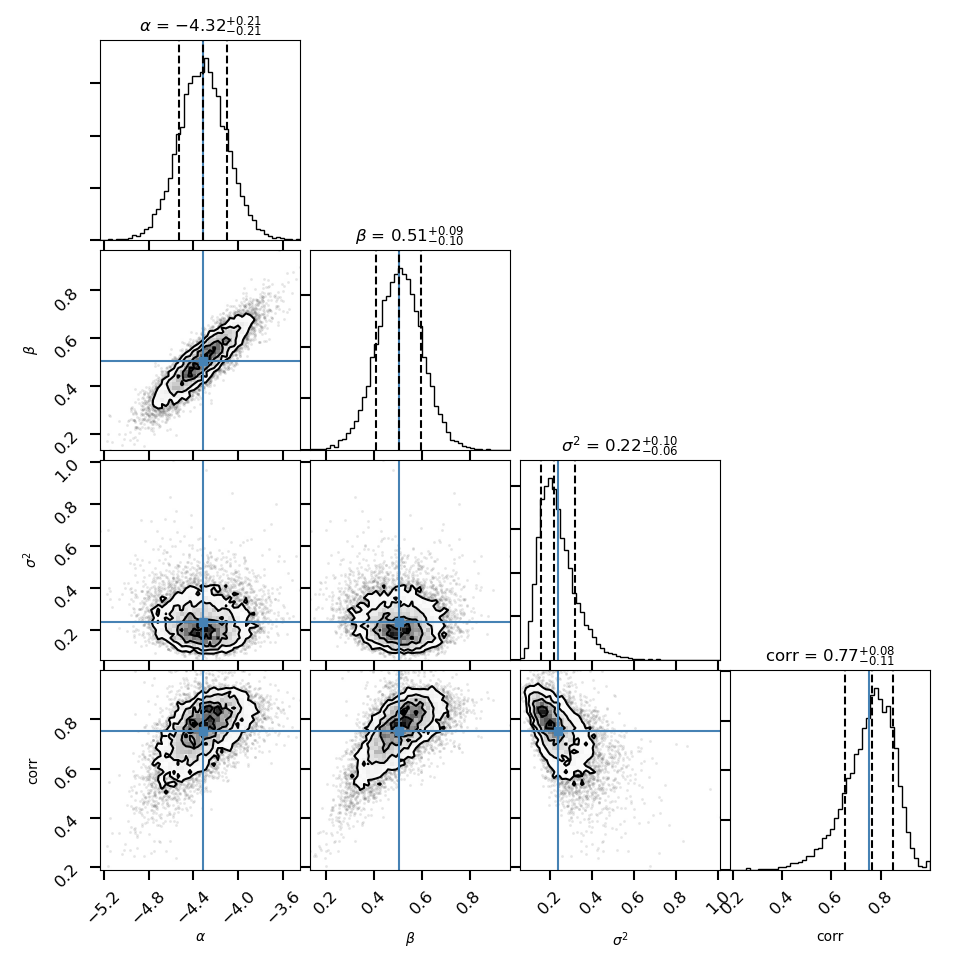}
    \caption{Corner plot of the log$L_{\rm OI, LVC}$ vs. log$L_{\rm acc}$ fit for the near (top) and far (bottom) sample, done using the linmix routine  \citep{Kelly2007}.}
    \label{fig:LOI_Lacc_near_far_fit}
\end{figure}

We show in Figures~\ref{fig:LOI_Lacc_all_fit} and \ref{fig:LOI_Lacc_near_far_fit} the corner plots showing the intercept,
slope, standard deviation, and correlation coefficient for the fit of the log$L_{\rm OI, LVC}$ vs. log$L_{\rm acc}$ relation (see sect.~\ref{sec:OI_lacc}, Fig.~\ref{fig:OI_Lacc}). This is done using \texttt{linmix} \citep{Kelly2007}, considering data with detection on both axes, and including upper limits.

\begin{figure*}[ht]
    \centering
	\includegraphics[width=0.45\textwidth]{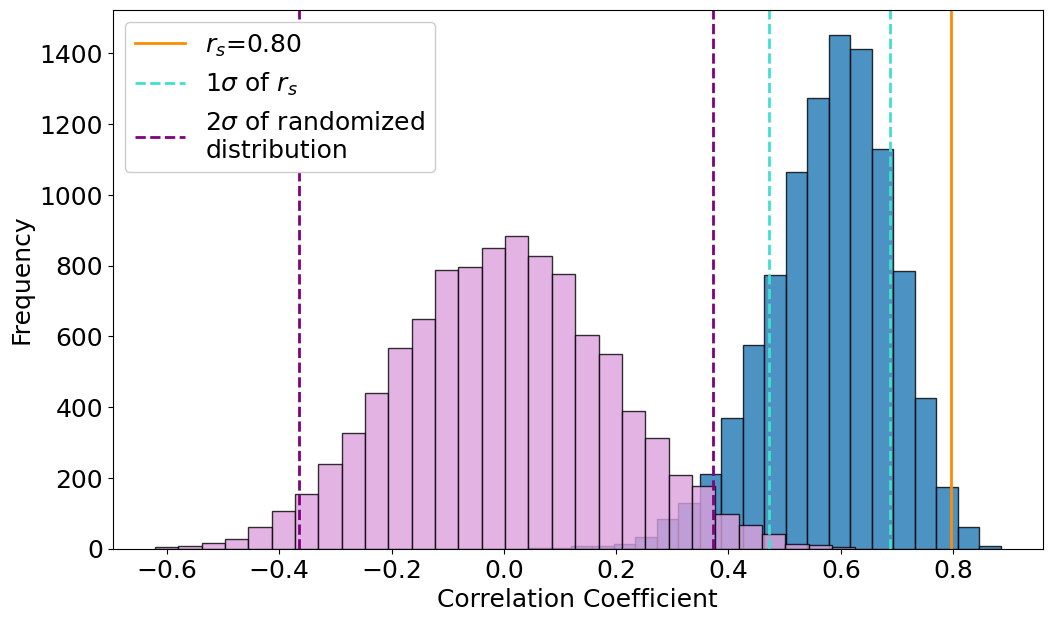}
 \includegraphics[width=0.45\textwidth]{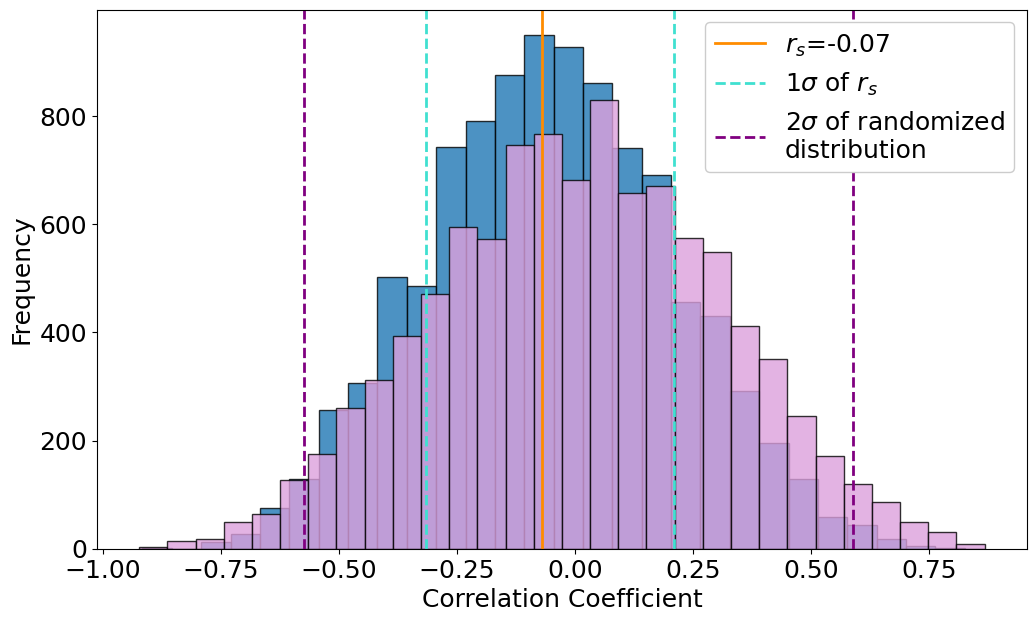}
    \caption{Correlation coefficient analysis for: sources beyond 0.8 pc (\textit{letf}) and within 0.8 pc (\textit{right}). The violet histogram corresponds to the bootstrapped randomized x-axis values; the dashed purple lines mark the 2$\sigma$ ranges of the realizations. The blue histogram results from bootstrapping the y-axis values within the respective error bars, and the 1$\sigma$ ranges of the bootstrap are shown with cyan lines. We show that the Spearman correlation coefficient for the measured data (orange line), lies outside 2$\sigma$ of the randomized distribution for the far sample and within 2$\sigma$ for the near sample.}
    \label{fig:LOI_Lacc_bootstrap}
\end{figure*}

We show the bootstrapping method applied to the log$L_{\rm OI, LVC}$ vs. log$L_{\rm acc}$ relationship in Fig.~\ref{fig:LOI_Lacc_bootstrap}. Following \citet{Aru2024,McLeod2021}, we use a bootstrap sampling method with $10^4$ iterations, where we randomize the x-values (log$L_{\rm acc}$) and compute the Spearman correlation coefficient. Secondly, we assess the uncertainty of the correlation coefficient by bootstrapping with $10^4$ iterations while varying the y-axis values (log$L_{\rm OI, LVC}$) within their respective uncertainties taken as the standard deviation of a Gaussian distribution of random points, for both the far and near samples. These tests are then compared with the actual Spearman correlation coefficient $r_{\rm s}$. 
The left panel of Fig.~\ref{fig:LOI_Lacc_bootstrap} shows the test for the far sample while the right panel shows it for the near sample.  We show that the Spearman correlation coefficient for the measured data (orange line), lies outside 2$\sigma$ of the randomized distribution for the far sample, i.e, log$L_{\rm OI, LVC}$ and log$L_{\rm acc}$ are correlated, while the observed Spearman correlation coefficient as well as both histograms are similar for the near sample, i.e., log$L_{\rm OI, LVC}$ and log$L_{\rm acc}$ are not correlated.

\section{The \NII\,and \SII, \SIIb\,forbidden lines}\label{ap:other_lines}

\subsection{Line profiles fits}

\begin{figure*}[ht]
    \centering
	\includegraphics[width=0.9\textwidth]{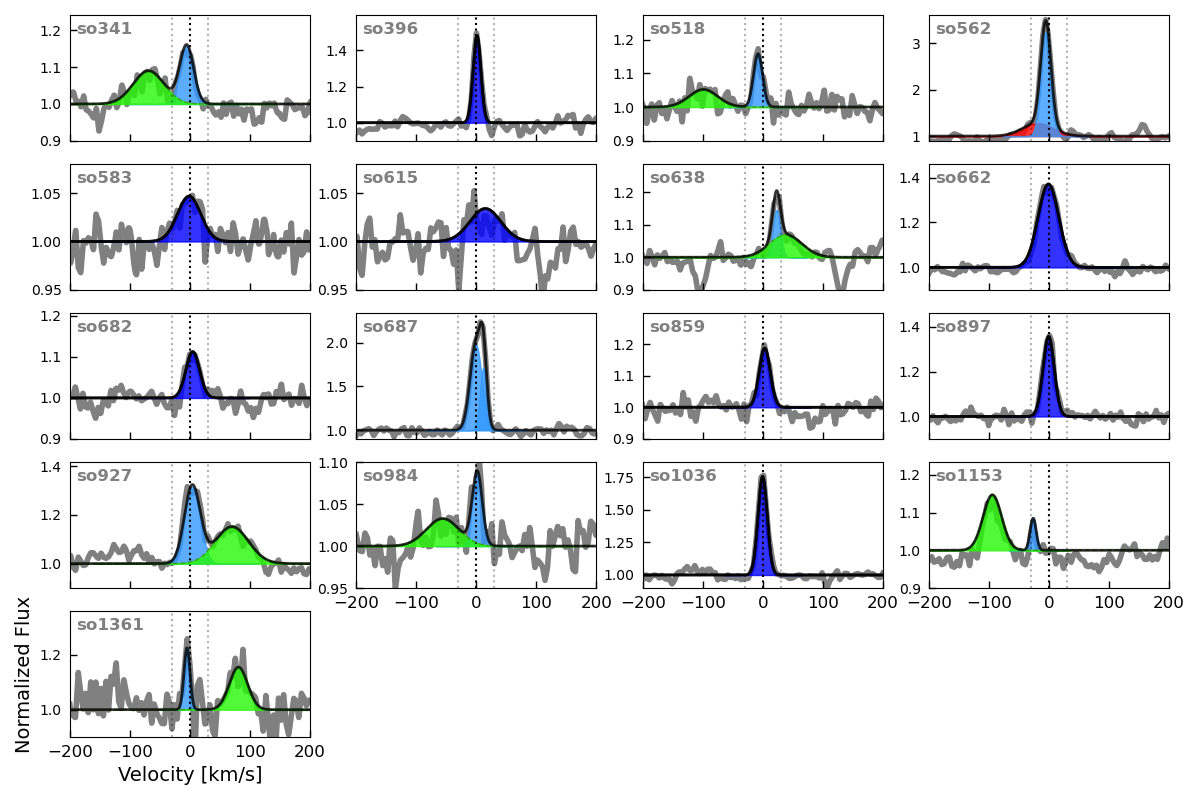}
    \caption{\NII\,line fit profiles for the MIKE sample. The colors indicate the type of Gaussian component as described in sec~\ref{sec:Gauss_fit}. The black dotted line is located at 0 km/s, while the gray dotted lines indicate the velocity threshold for the high/low-velocity component ($\pm$30 km/s).}
    \label{fig:NII6583_fit}
\end{figure*}

\begin{figure*}[ht]
    \centering
	\includegraphics[width=0.9\textwidth]{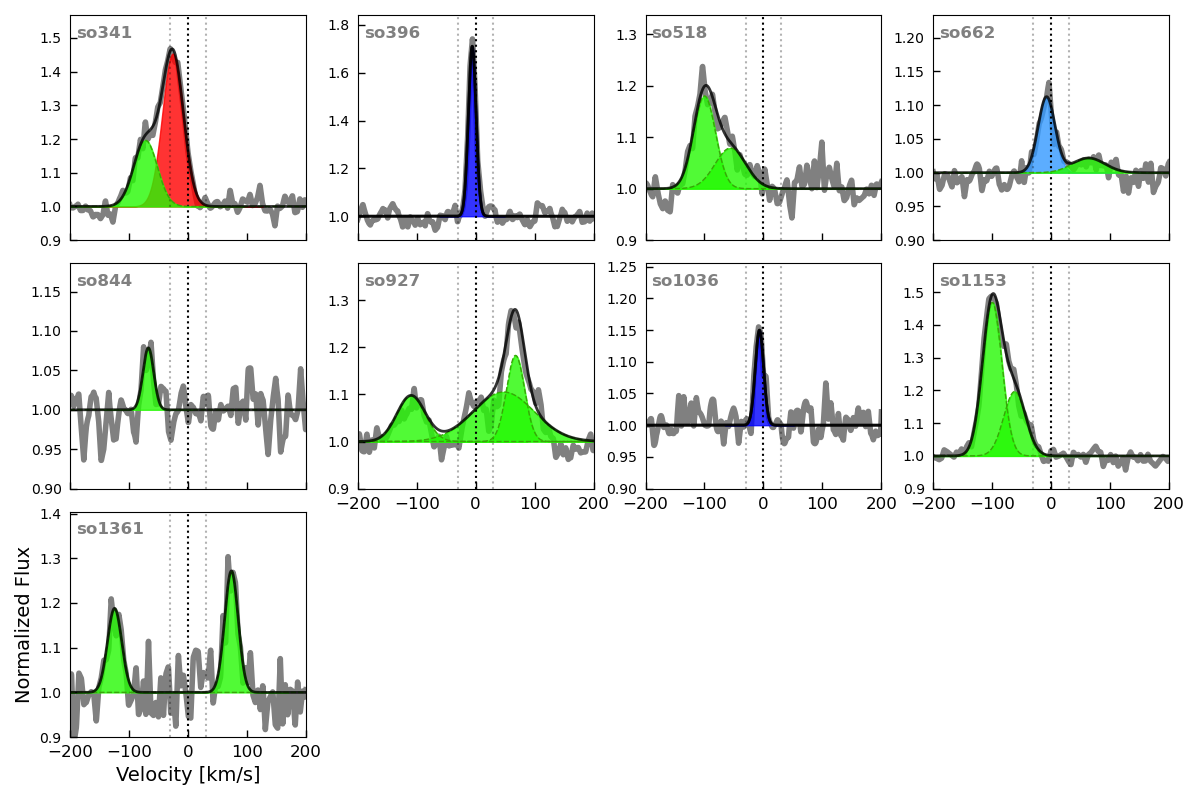}
    \caption{\SII\,line fit profiles for the MIKE sample. The colors indicate the type of Gaussian component as described in sec~\ref{sec:Gauss_fit}. The black dotted line is located at 0 km/s, while the gray dotted lines indicate the velocity threshold for the high/low-velocity component ($\pm$30 km/s).}
    \label{fig:SII6730_fit}
\end{figure*}

\begin{figure*}[ht]
    \centering
	\includegraphics[width=0.9\textwidth]{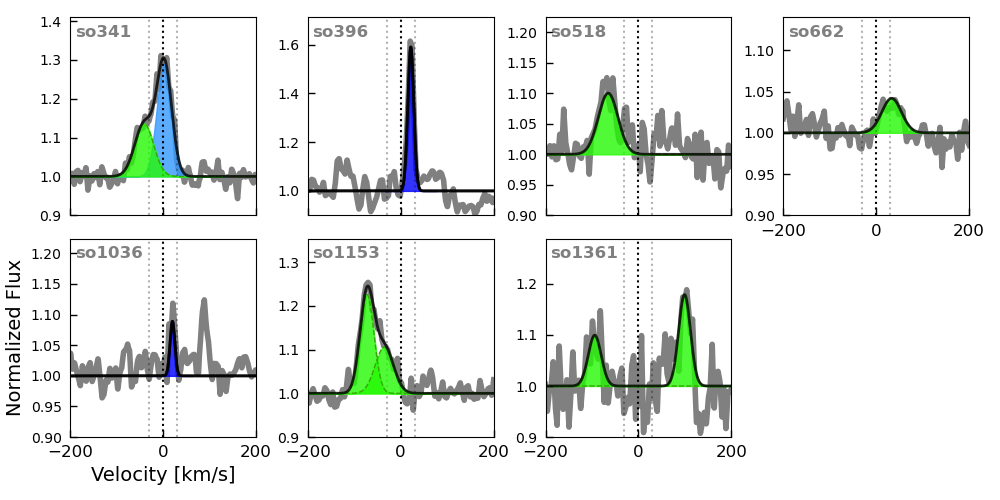}
    \caption{\SIIb\,line fit profiles for the MIKE sample. The colors indicate the type of Gaussian component as described in sec~\ref{sec:Gauss_fit}. The black dotted line is located at 0 km/s, while the gray dotted lines indicate the velocity threshold for the high/low-velocity component ($\pm$30 km/s).}
    \label{fig:SII6716_fit}
\end{figure*}

This section presents the fit results for the \NII, \SIIb, and \SII\,forbidden lines for the MIKE and X-shooter samples. For the MIKE sample, Tables~\ref{tab:NIIfit}, \ref{tab:SIIfit}, and \ref{tab:SIIbfit} list the fit results for the \NII, \SII, and \SIIb\,lines, respectively. The line profiles, with their corresponding Gaussian components fits, are shown in Figures~\ref{fig:NII6583_fit}, \ref{fig:SII6730_fit}, and \ref{fig:SII6716_fit}. 

\begin{figure*}[ht]
    \centering
	\includegraphics[width=0.9\textwidth]{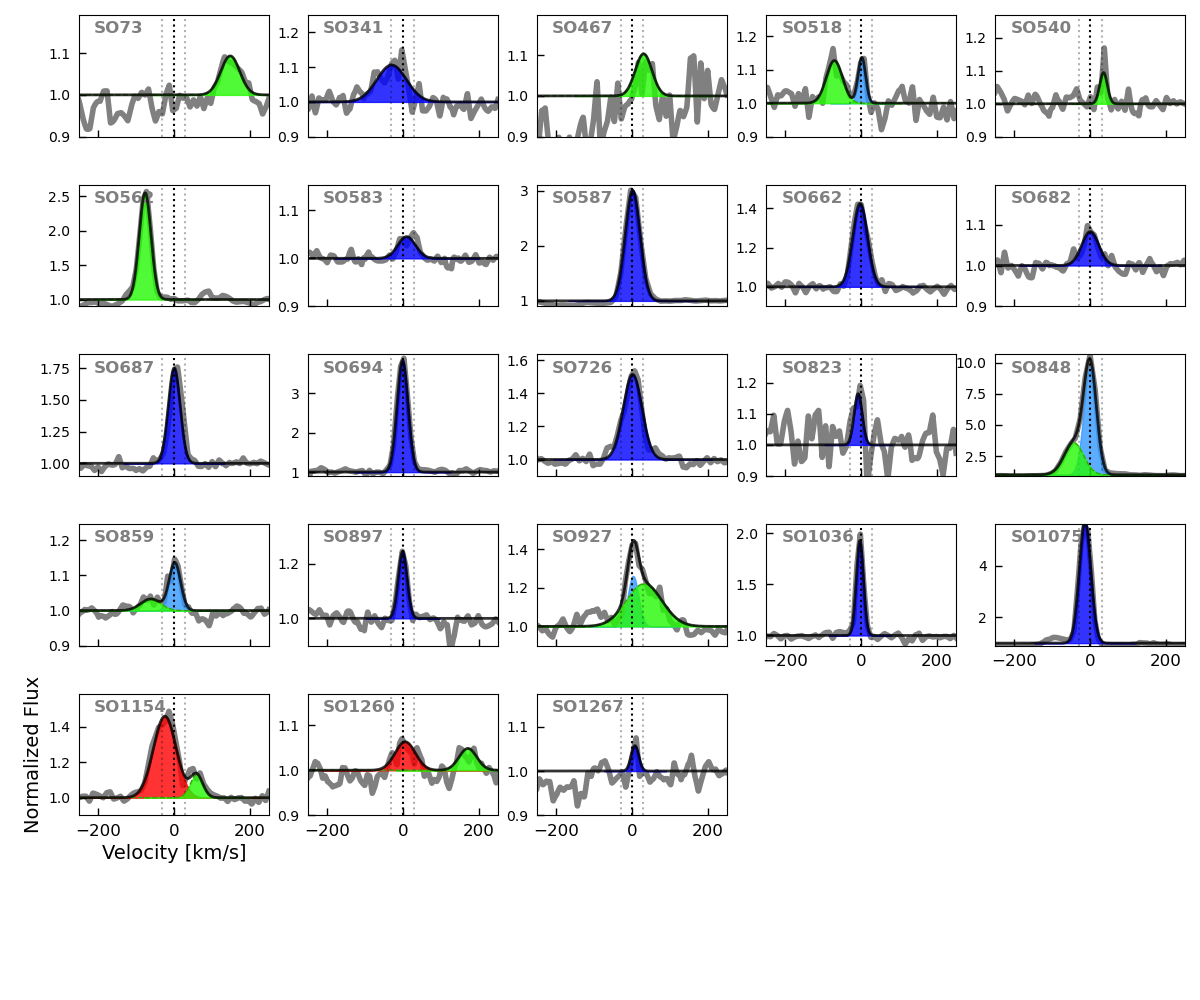}
    \caption{\NII\,line fit profiles for the X-shooter sample. The colors indicate the type of Gaussian component as described in sec~\ref{sec:Gauss_fit}. The black dotted line is located at 0 km/s, while the gray dotted lines indicate the velocity threshold for the high/low-velocity component ($\pm$30 km/s).}
    \label{fig:NII6583_fit_XS}
\end{figure*}

\begin{figure*}[ht]
    \centering
	\includegraphics[width=0.9\textwidth]{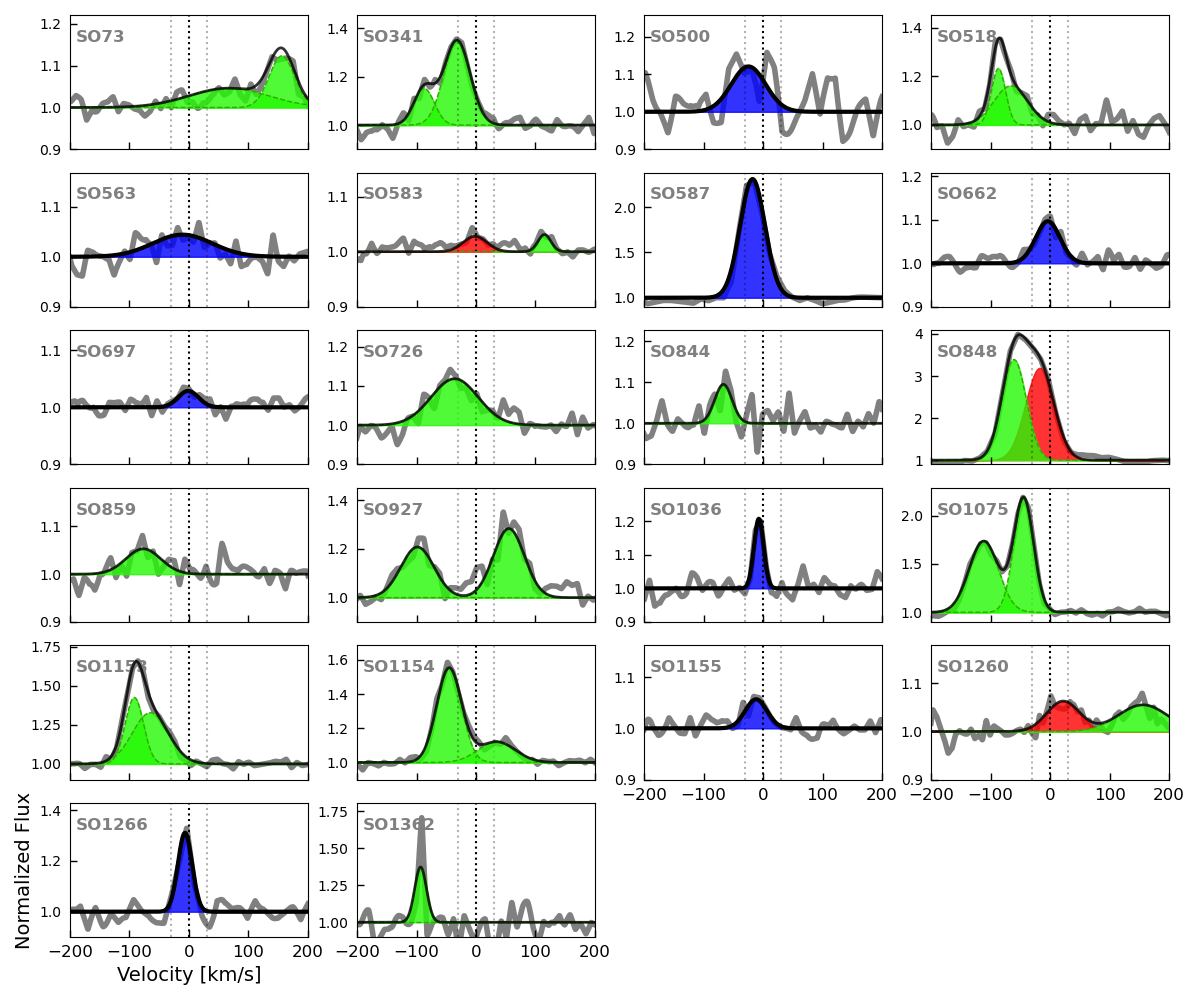}
    \caption{\SII\,line fit profiles for the X-shooter sample. The colors indicate the type of Gaussian component as described in sec~\ref{sec:Gauss_fit}. The black dotted line is located at 0 km/s, while the gray dotted lines indicate the velocity threshold for the high/low-velocity component ($\pm$30 km/s).}
    \label{fig:SII6730_fit_XS}
\end{figure*}

\begin{figure*}[ht]
    \centering
	\includegraphics[width=0.8\textwidth]{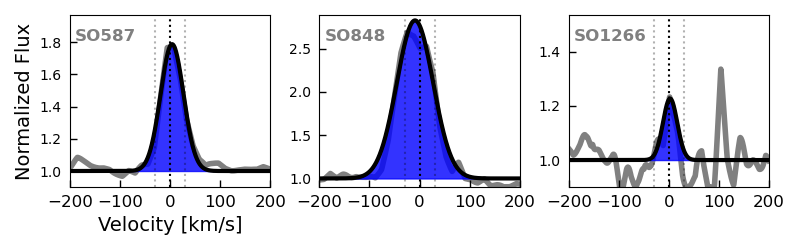}
    \caption{\SIIb\,line fit profiles for the X-shooter sample. The colors indicate the type of Gaussian component as described in sec~\ref{sec:Gauss_fit}. The black dotted line is located at 0 km/s, while the gray dotted lines indicate the velocity threshold for the high/low-velocity component ($\pm$30 km/s).}
    \label{fig:SII6716_fit_XS}
\end{figure*}

For the X-shooter sample, Tables~\ref{tab:NIIfit_XS}, \ref{tab:SIIfit_XS}, and \ref{tab:SIIbfit_XS} list the fit results for the \NII, \SII, and \SIIb\,lines, respectively. The line profiles, with their corresponding Gaussian components fits, are shown in Figures~\ref{fig:NII6583_fit_XS}, \ref{fig:SII6730_fit_XS}, and Figures~\ref{fig:SII6716_fit_XS}.

\begin{table*}
\begin{center}
\begin{threeparttable}[b]
\caption{\label{tab:NIIfit} Parameters for the \NII\,line LVC for the MIKE sample}
\begin{tabular}{lccccccc}
\hline\hline
Name & log$L_{\rm NII, LVC}$\tnote{1} & FWHM\tnote{2}   & $v_{\rm peak}$\tnote{2} & EW    &  $F_{\rm cont}$   & log$L_{\rm NII,GC}$\tnote{3} & Class \\
     & [\lsun]  &  [km/s] & [km/s]         & [\AA] &   [erg $\rm s^{-1}\,nm^{-1}\,cm^{-2}$ ]   &   [\lsun]         &       \\
\hline
SO341 &    -5.34 &     27.1 &    -5.43 &      0.1 & 8.66e-14 &    -5.34 & LVC-NC  \\ [0.5ex]
SO396 &    -5.12 &     15.4 &     1.05 &     0.18 & 8.66e-14 &    -5.12 & LVC-SC  \\ [0.5ex]
SO518 &    -6.66 &     17.8 &    -7.73 &      0.0 & 1.26e-13 &    -6.66 & LVC-NC  \\ [0.5ex]
\hline
SO562 &    -4.83 &     60.1 &   -18.13 &      0.4 & 1.99e-14 &     -5.4 & LVC-BC  \\ [0.5ex]
SO562 &          &     19.6 &    -5.34 &     1.03 & 1.99e-14 &    -4.99 & LVC-NC  \\ [0.5ex]
\hline
SO583 &    -4.65 &     46.1 &    -1.49 &     0.05 & 8.81e-13 &    -4.65 & LVC-SC  \\ [0.5ex]
SO615 &    -4.64 &     60.8 &    14.98 &     0.05 & 8.81e-13 &    -4.64 & LVC-SC  \\ [0.5ex]
SO638 &    -5.82 &     14.9 &    23.03 &      0.0 & 8.81e-13 &    -5.82 & LVC-NC  \\ [0.5ex]
SO662 &    -4.77 &     40.8 &    -1.11 &     0.36 & 9.44e-14 &    -4.77 & LVC-SC  \\ [0.5ex]
SO682 &    -5.44 &     26.9 &     4.59 &     0.07 & 9.61e-14 &    -5.44 & LVC-SC  \\ [0.5ex]
\hline
SO687 &     -4.5 &     22.9 &    -0.75 &     0.52 & 7.89e-14 &    -4.66 & LVC-NC  \\ [0.5ex]
SO687 &          &     13.3 &    11.58 &     0.22 & 7.89e-14 &    -5.03 & LVC-NC  \\ [0.5ex]
\hline
SO859 &    -5.89 &     21.8 &      3.3 &      0.1 & 2.59e-14 &    -5.89 & LVC-SC  \\ [0.5ex]
SO897 &    -4.86 &     22.1 &    -0.75 &     0.18 & 1.49e-13 &    -4.86 & LVC-SC  \\ [0.5ex]
SO927 &     -6.8 &     31.1 &     3.77 &     0.01 & 4.25e-14 &     -6.8 & LVC-NC  \\ [0.5ex]
SO984 &    -7.01 &     19.6 &     1.53 &      0.0 & 9.82e-14 &    -7.01 & LVC-NC  \\ [0.5ex]
SO1036 &    -4.78 &     18.2 &    -0.14 &     0.32 & 1.06e-13 &    -4.78 & LVC-SC  \\ [0.5ex]
SO1153 &    -6.69 &     11.8 &   -26.49 &      0.0 & 2.23e-13 &    -6.69 & LVC-NC  \\ [0.5ex]
SO1361 &    -5.43 &     11.8 &    -4.99 &     0.06 & 1.17e-13 &    -5.43 & LVC-NC  \\ [0.5ex]
\hline
\end{tabular}
\begin{tablenotes}
    \item [1] Total LVC luminosity.
    \item [2] Median uncertainties. SC: $\delta_{\rm v_{peak}}$ = 3.0, $\delta_{\rm FWHM}$ = 4.5. NC: $\delta_{\rm v_{peak}}$ = 1.9 , $\delta_{\rm FWHM}$ = 2.3. BC: $\delta_{\rm v_{peak}}$ = 4.6, $\delta_{\rm FWHM}$ = 7.2.
    \item [3] Luminosity of each Gaussian component.
\end{tablenotes}
\end{threeparttable}
\end{center}
\end{table*}

\begin{table*}
\begin{center}
\begin{threeparttable}[b]
\caption{\label{tab:SIIfit} Parameters for the \SII\,LVC for the MIKE sample}
\begin{tabular}{lccccccc}
\hline\hline
Name & log$L_{\rm SII, LVC}$\tnote{1} & FWHM\tnote{2}   & $v_{\rm peak}$\tnote{2} & EW    &  $F_{\rm cont}$   & log$L_{\rm SII,GC}$\tnote{3} & Class \\
     & [\lsun]  &  [km/s] & [km/s]         & [\AA] &   [erg $\rm s^{-1}\,nm^{-1}\,cm^{-2}$ ]   &   [\lsun]         &       \\
\hline
SO341 &    -4.74 &     42.4 &   -25.52 &     0.46 & 7.61e-14 &    -4.74 & LVC-BC  \\ [0.5ex]
SO396 &    -5.01 &     15.4 &     -5.8 &     0.26 & 7.61e-14 &    -5.01 & LVC-SC  \\ [0.5ex]
SO662 &    -5.41 &     33.7 &    -7.86 &     0.09 & 8.55e-14 &    -5.41 & LVC-NC  \\ [0.5ex]
SO1036 &    -5.58 &     16.7 &     -6.7 &     0.06 & 9.08e-14 &    -5.58 & LVC-SC  \\ [0.5ex]
\hline
\end{tabular}
\begin{tablenotes}
    \item [1] Total LVC luminosity.
    \item [2] Median uncertainties. SC: $\delta_{\rm v_{peak}}$ = 1.3, $\delta_{\rm FWHM}$ = 1.8. NC: $\delta_{\rm v_{peak}}$ = 1.3 , $\delta_{\rm FWHM}$ = 1.9. BC: $\delta_{\rm v_{peak}}$ = 0.8, $\delta_{\rm FWHM}$ = 0.9.    
    \item [3] Luminosity of each Gaussian component.
\end{tablenotes}
\end{threeparttable}
\end{center}
\end{table*}

\begin{table*}
\begin{center}
\begin{threeparttable}[b]
\caption{\label{tab:SIIbfit} Parameters for the \SIIb\,LVC for the MIKE sample}
\begin{tabular}{lccccccc}
\hline\hline
Name & log$L_{\rm SII, LVC}$\tnote{1} & FWHM\tnote{2}   & $v_{\rm peak}$\tnote{2} & EW    &  $F_{\rm cont}$   & log$L_{\rm SII,GC}$\tnote{3} & Class \\
     & [\lsun]  &  [km/s] & [km/s]         & [\AA] &   [erg $\rm s^{-1}\,nm^{-1}\,cm^{-2}$ ]   &   [\lsun]         &       \\
\hline
SO341 &    -4.98  &    37.7 &     2.79 &     0.41  & 7.61e-14 &    -4.98 & LVC-NC  \\ [0.5ex]
SO396 &    -4.73  &    15.4 &     21.79 &    0.21  & 7.61e-14 &    -4.73 & LVC-SC \\ [0.5ex]
SO1036 &   -5.94  &    12.3 &     20.54 &    0.02  & 9.08e-14 &    -5.94 & LVC-SC  \\ [0.5ex]
\hline
\end{tabular}
\begin{tablenotes}
    \item [1] Total LVC luminosity.
    \item [2] Median uncertainties. SC: $\delta_{\rm v_{peak}}$ = 1.3, $\delta_{\rm FWHM}$ = 1.8. NC: $\delta_{\rm v_{peak}}$ = 1.3 , $\delta_{\rm FWHM}$ = 1.9. BC: $\delta_{\rm v_{peak}}$ = 0.8, $\delta_{\rm FWHM}$ = 0.9.    
    \item [3] Luminosity of each Gaussian component.
\end{tablenotes}
\end{threeparttable}
\end{center}
\end{table*}

\begin{table*}
\begin{center}
\begin{threeparttable}[b]
\caption{\label{tab:NIIfit_XS} Parameters for the \NII\,LVC for the X-shooter sample}
\begin{tabular}{lccccccc}
\hline\hline
Name & log$L_{\rm NII, LVC}$\tnote{1} & FWHM\tnote{2}   & $v_{\rm peak}$\tnote{2} & EW    &  $F_{\rm cont}$   & log$L_{\rm NII,GC}$\tnote{3} & Class \\
     & [\lsun]  &  [km/s] & [km/s]         & [\AA] &   [erg $\rm s^{-1}\,nm^{-1}\,cm^{-2}$ ]   &   [\lsun]         &       \\
\hline
SO341 &    -4.99 &     91.6 &   -29.56 &     0.23 & 8.66e-14 &    -4.99 & LVC-SC  \\ [0.5ex]
SO518 &    -5.33 &     23.5 &      2.1 &     0.07 & 1.27e-13 &    -5.33 & LVC-NC  \\ [0.5ex]
SO583 &    -4.59 &     54.9 &     10.0 &     0.06 & 8.81e-13 &    -4.59 & LVC-SC  \\ [0.5ex]
SO587 &    -4.69 &     46.0 &     1.87 &     2.17 & 1.87e-14 &    -4.69 & LVC-SC  \\ [0.5ex]
SO662 &    -4.66 &     45.6 &     -2.4 &     0.45 & 9.44e-14 &    -4.66 & LVC-SC  \\ [0.5ex]
SO682 &    -5.28 &     53.1 &      0.3 &      0.1 & 9.61e-14 &    -5.28 & LVC-SC  \\ [0.5ex]
SO687 &    -4.57 &     36.6 &      1.1 &     0.64 & 7.89e-14 &    -4.57 & LVC-SC  \\ [0.5ex]
SO694 &    -5.25 &     34.3 &    -0.76 &     2.23 &  5.20e-15 &    -5.25 & LVC-SC  \\ [0.5ex]
SO726 &    -4.59 &     58.0 &     1.76 &      0.7 &  7.20e-14 &    -4.59 & LVC-SC  \\ [0.5ex]
SO823 &     -5.7 &     25.1 &    -7.55 &      0.1 & 4.19e-14 &     -5.7 & LVC-SC  \\ [0.5ex]
SO848 &    -5.25 &     39.5 &    -0.06 &      8.0 &  1.80e-15 &    -5.25 & LVC-NC  \\ [0.5ex]
SO859 &    -5.82 &     35.4 &     2.98 &     0.11 & 2.59e-14 &    -5.82 & LVC-NC  \\ [0.5ex]
SO897 &    -4.92 &     27.6 &    -0.63 &     0.16 & 1.49e-13 &    -4.92 & LVC-SC  \\ [0.5ex]
SO927 &    -5.33 &     33.3 &     2.97 &      0.2 & 4.25e-14 &    -5.33 & LVC-NC  \\ [0.5ex]
SO1036 &     -4.6 &     22.4 &    -3.08 &     0.49 & 1.06e-13 &     -4.6 & LVC-SC  \\ [0.5ex]
SO1075 &    -4.43 &     35.9 &   -13.37 &     4.09 & 1.92e-14 &    -4.43 & LVC-SC  \\ [0.5ex]
SO1154 &    -5.04 &     70.6 &   -23.51 &     0.76 & 2.39e-14 &    -5.04 & LVC-BC  \\ [0.5ex]
SO1260 &    -6.28 &     62.6 &     6.19 &     0.09 & 1.22e-14 &    -6.28 & LVC-BC  \\ [0.5ex]
SO1267 &     -6.0 &     23.5 &     7.94 &     0.03 & 6.56e-14 &     -6.0 & LVC-SC  \\ [0.5ex]
\hline
\end{tabular}
\begin{tablenotes}
    \item [1] Total LVC luminosity.
    \item [2] Median uncertainties. SC: $\delta_{\rm v_{peak}}$ = 5.2, $\delta_{\rm FWHM}$ = 3.5. NC: $\delta_{\rm v_{peak}}$ = 3.6 , $\delta_{\rm FWHM}$ = 2.2. BC: $\delta_{\rm v_{peak}}$ = 5.5, $\delta_{\rm FWHM}$ = 6.5.
    \item [3] Luminosity of each Gaussian component.
\end{tablenotes}
\end{threeparttable}
\end{center}
\end{table*}

\begin{table*}
\begin{center}
\begin{threeparttable}[b]
\caption{\label{tab:SIIfit_XS} Parameters for the \SII\,LVC for the X-shooter sample}
\begin{tabular}{lccccccc}
\hline\hline
Name & log$L_{\rm SII, LVC}$\tnote{1} & FWHM\tnote{2}   & $v_{\rm peak}$\tnote{2} & EW    &  $F_{\rm cont}$   & log$L_{\rm SII,GC}$\tnote{3} & Class \\
     & [\lsun]  &  [km/s] & [km/s]         & [\AA] &   [erg $\rm s^{-1}\,nm^{-1}\,cm^{-2}$ ]   &   [\lsun]         &       \\
\hline
SO500 &    -7.48 &     68.0 &   -24.59 &     0.19 & 3.00e-16 &    -7.48 & LVC-SC  \\ [0.5ex]
SO563 &    -5.53 &    117.5 &   -10.92 &     0.12 & 4.72e-14 &    -5.53 & LVC-SC  \\ [0.5ex]
SO583 &    -4.88 &     47.1 &    -1.64 &     0.03 & 8.32e-13 &    -4.88 & LVC-BC  \\ [0.5ex]
SO587 &    -4.96 &     49.6 &   -17.72 &     1.59 & 1.37e-14 &    -4.96 & LVC-SC  \\ [0.5ex]
SO662 &    -5.32 &     48.3 &     -3.8 &     0.11 & 8.55e-14 &    -5.32 & LVC-SC  \\ [0.5ex]
SO697 &    -5.65 &     40.1 &    -1.15 &     0.03 & 1.59e-13 &    -5.65 & LVC-SC  \\ [0.5ex]
SO848 &    -5.82 &     56.9 &   -17.63 &     2.98 &  1.30e-15 &    -5.82 & LVC-BC  \\ [0.5ex]
SO1036 &    -5.37 &     19.3 &    -7.08 &      0.1 & 9.08e-14 &    -5.37 & LVC-SC  \\ [0.5ex]
SO1155 &    -5.05 &     43.1 &   -11.76 &     0.06 &  3.00e-13 &    -5.05 & LVC-SC  \\ [0.5ex]
SO1260 &    -6.38 &     64.6 &    21.75 &      0.1 &  9.10e-15 &    -6.38 & LVC-BC  \\ [0.5ex]
SO1266 &    -6.52 &     28.1 &    -6.64 &     0.21 &  2.80e-15 &    -6.52 & LVC-SC  \\ [0.5ex]
\hline
\end{tabular}
\begin{tablenotes}
    \item [1] Total LVC luminosity.
    \item [2] Median uncertainties. SC: $\delta_{\rm v_{peak}}$ = 6.8, $\delta_{\rm FWHM}$ = 8.7. BC: $\delta_{\rm v_{peak}}$ = 4.5, $\delta_{\rm FWHM}$ = 5.0.
    \item [3] Luminosity of each Gaussian component.
\end{tablenotes}
\end{threeparttable}
\end{center}
\end{table*}

\begin{table*}
\begin{center}
\begin{threeparttable}[b]
\caption{\label{tab:SIIbfit_XS} Parameters for the \SIIb\,LVC for the X-shooter sample}
\begin{tabular}{lccccccc}
\hline\hline
Name & log$L_{\rm SII, LVC}$\tnote{1} & FWHM\tnote{2}   & $v_{\rm peak}$\tnote{2} & EW    &  $F_{\rm cont}$   & log$L_{\rm SII,GC}$\tnote{3} & Class \\
     & [\lsun]  &  [km/s] & [km/s]         & [\AA] &   [erg $\rm s^{-1}\,nm^{-1}\,cm^{-2}$ ]   &   [\lsun]         &       \\
\hline
SO587 &    -5.15 &     53.8 &   3.84   &     1.03 & 1.37e-14 &    -5.15 & LVC-SC  \\ [0.5ex]
SO848 &    -5.70 &     85.39 &   -9.07 &     4.00 &  1.30e-15 &    -5.70 & LVC-SC  \\ [0.5ex]
SO1266 &   -6.56 &     33.67 &    2.32  &     0.19 &  2.80e-15 &    -6.56 & LVC-SC  \\ [0.5ex]
\hline
\end{tabular}
\begin{tablenotes}
    \item [1] Total LVC luminosity.
    \item [2] Median uncertainties. SC: $\delta_{\rm v_{peak}}$ = 6.8, $\delta_{\rm FWHM}$ = 8.7. BC: $\delta_{\rm v_{peak}}$ = 4.5, $\delta_{\rm FWHM}$ = 5.0.
    \item [3] Luminosity of each Gaussian component.
\end{tablenotes}
\end{threeparttable}
\end{center}
\end{table*}

\subsection{Line kinematics with respect to projected distance to \sori}

\begin{figure*}[ht]
    \centering
	\includegraphics[width=0.9\textwidth]{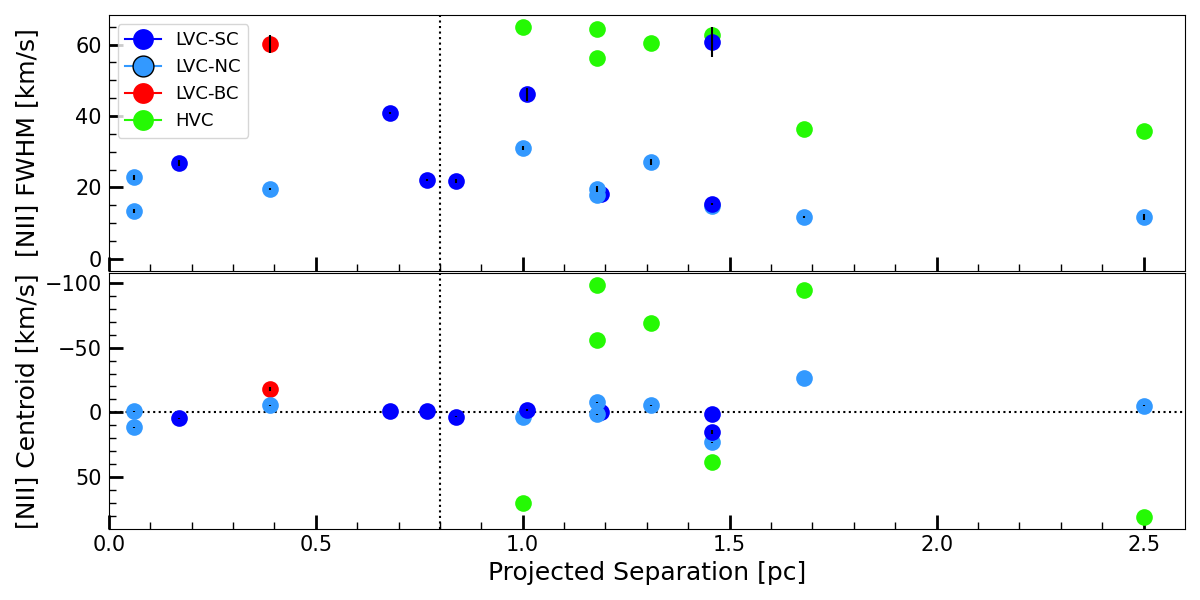}
    \caption{\NII\,lines FWHM (top) and peak velocity (bottom) as a function of projected separation from \sori\,for the MIKE sample. The colors represent the type of Gaussian component as explained in sec~\ref{sec:Gauss_fit}. The vertical dotted line is positioned at 0.8 pc. The horizontal line indicates the systemic stellar velocity (0 km/s).}
    \label{fig:NII6583_dp}
\end{figure*}

\begin{figure*}[ht]
    \centering
	\includegraphics[width=0.9\textwidth]{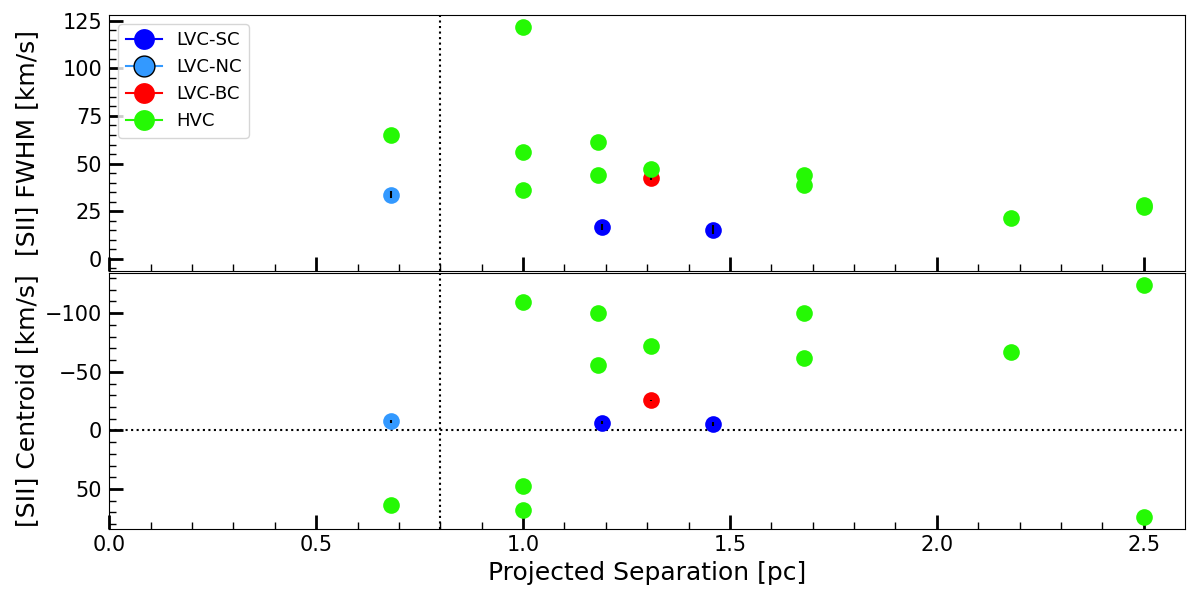}
    \caption{\SII\,lines FWHM (top) and peak velocity (bottom) as a function of projected separation from \sori\, for the MIKE sample. The colors represent the type of Gaussian component as explained in sec~\ref{sec:Gauss_fit}. The vertical dotted line is positioned at 0.8 pc. The horizontal line indicates the systemic stellar velocity (0 km/s).}
    \label{fig:SII6730_dp}
\end{figure*}

\begin{figure*}[ht]
    \centering
	\includegraphics[width=0.9\textwidth]{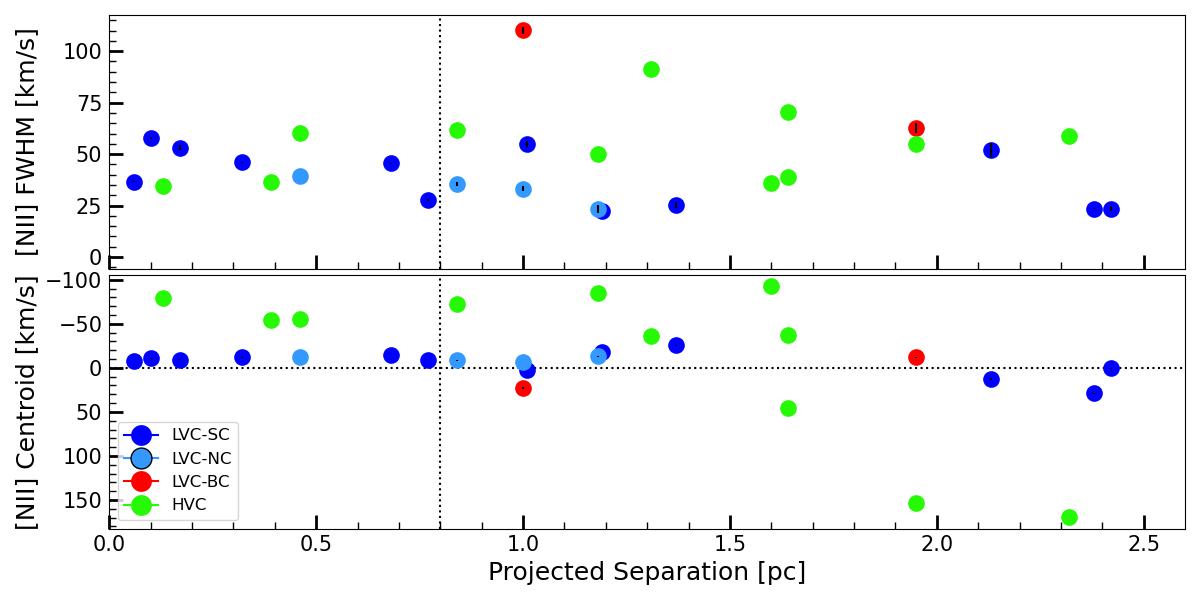}
    \caption{\NII\,lines FWHM (top) and peak velocity (bottom) as a function of projected separation from \sori\,for the X-shooter sample. The colors represent the type of Gaussian component as explained in sec~\ref{sec:Gauss_fit}. The vertical dotted line is positioned at 0.8 pc. The horizontal line indicates the systemic stellar velocity (0 km/s).}
    \label{fig:NII6583_dp_XS}
\end{figure*}

\begin{figure*}[ht]
    \centering
	\includegraphics[width=0.9\textwidth]{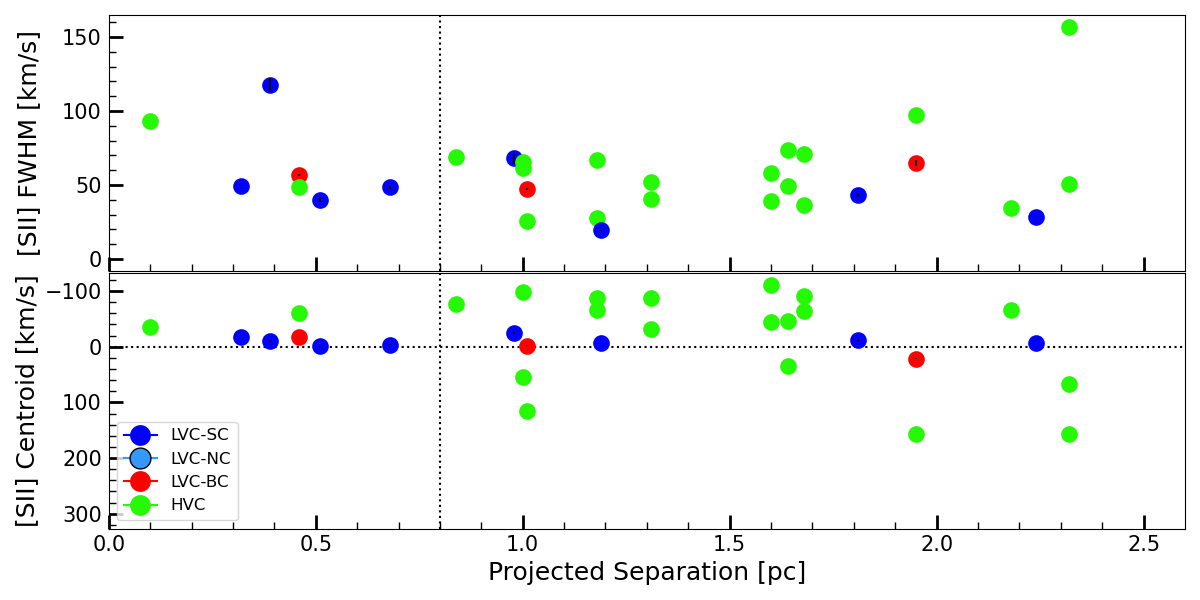}
    \caption{\SII\,lines FWHM (top) and peak velocity (bottom) as a function of projected separation from \sori\,for the X-shooter sample. The colors represent the type of Gaussian component as explained in sec~\ref{sec:Gauss_fit}. The vertical dotted line is positioned at 0.8 pc. The horizontal line indicates the systemic stellar velocity (0 km/s).}
    \label{fig:SII6730_dp_XS}
\end{figure*}

The kinematic properties of the lines ($v_{\rm peak}$, FWHM) as a function of the projected separation to the massive system \sori\,are shown in Figures~\ref{fig:NII6583_dp} and \ref{fig:SII6730_dp} for \NII\,and \SII\,lines respectively, as observed in our MIKE sample, and in Figures~\ref{fig:NII6583_dp_XS} and \ref{fig:SII6730_dp_XS} for \NII\,and \SII\,lines respectively, in the X-shooter sample. 

\section{High-velocity components Fits}\label{sec:ap_HVC}
This section presents the fit results for the HVC of all the forbidden lines studied here (\OI, \NII, \SIIb, \SII). These are listed in tables~\ref{tab:HVCfit}, \ref{tab:HVCfit_XS}, \ref{tab:HVCfit_NII}, \ref{tab:HVCfit_NII_XS}, \ref{tab:HVCfit_SII}, \ref{tab:HVCfit_SII_XS}, \ref{tab:HVCfit_SIIb}.

\begin{table}
\begin{center}
\caption{\label{tab:HVCfit} Kinematics Properties of the \OI\,HVC for the MIKE sample}
\begin{tabular}{lcc}
\hline\hline
Name  & $v_{\rm peak}$  &   FWHM  \\
      &  [km/s]          & [km/s]  \\
\hline
SO362  & 32.1   $\pm$ 1.1 & 26.6   $\pm$  0.1  \\[0.5ex]
\hline
SO518  & -102.1 $\pm$ 0.8 & 49.5   $\pm$  0.1  \\[0.5ex]
SO518  & -37.4  $\pm$ 0.8 & 74.7   $\pm$  0.0  \\[0.5ex]
\hline
SO844  & -60.6  $\pm$ 1.2 & 33.4   $\pm$  1.5  \\[0.5ex]
SO859  & -69.2  $\pm$ 1.8 & 28.0   $\pm$  0.1  \\[0.5ex]
\hline
SO927  & -93.9  $\pm$ 0.8 & 121.9  $\pm$  0.1  \\[0.5ex]
SO927  & 75.1   $\pm$ 0.8 & 69.3   $\pm$  0.0  \\[0.5ex]
\hline
SO1036 & -91.2  $\pm$ 12.2& 92.4   $\pm$  20.3 \\[0.5ex]
\hline
SO1153 & -97.3  $\pm$ 0.8 & 33.1   $\pm$  0.0  \\[0.5ex]
SO1153 & -69.3  $\pm$ 0.8 & 106.3  $\pm$  0.0  \\[0.5ex]
\hline
SO1361 & -119.0 $\pm$ 3.2 & 30.2   $\pm$  5.2  \\[0.5ex]
SO1361 & -31.1  $\pm$ 9.1 & 157.1  $\pm$  15.1 \\[0.5ex]
SO1361 & 86.7   $\pm$ 2.5 & 22.6   $\pm$  4.0  \\[0.5ex]
\hline
\end{tabular}
\end{center}
\end{table}

\begin{table}
\begin{center}
\caption{\label{tab:HVCfit_XS} Kinematics Properties of the \OI\,HVC for the X-shooter sample}
\begin{tabular}{lcc}
\hline\hline
Name  & $v_{\rm peak}$  &   FWHM  \\
      &  [km/s]          & [km/s]  \\
\hline
SO518  &  -32.5   $\pm$  3.9  & 83.6  $\pm$  3.0  \\[0.5ex]
SO518  &  -88.3   $\pm$  3.7  & 45.0  $\pm$  2.2  \\[0.5ex]
\hline
SO687  &  -44.6   $\pm$  24.5 & 188.4 $\pm$  40.9 \\[0.5ex]
SO859  &  -65.7   $\pm$  4.0  & 52.6  $\pm$  4.5  \\[0.5ex]
\hline
SO927  &  -95.4   $\pm$  3.0  & 65.2  $\pm$  2.0  \\[0.5ex]
SO927  &  67.8    $\pm$  2.8  & 50.6  $\pm$  1.2  \\[0.5ex]
\hline
SO1036 &  -66.5   $\pm$  10.9 & 117.7 $\pm$  17.6 \\[0.5ex]
\hline
SO1075 &  -96.9   $\pm$  6.6  & 66.1  $\pm$  1.0  \\[0.5ex]
SO1075 &  -42.8   $\pm$  6.6  & 42.6  $\pm$  0.8  \\[0.5ex]
\hline
SO1153 &  -84.6   $\pm$  3.1  & 54.6  $\pm$  1.2  \\[0.5ex]
SO1154 &  -30.4   $\pm$  3.3  & 79.7  $\pm$  1.7  \\[0.5ex]
SO1156 &  -33.6   $\pm$  6.1  & 61.8  $\pm$  9.1  \\[0.5ex]
SO1260 &  100.0   $\pm$  31.7 & 152.6 $\pm$  52.5 \\[0.5ex]
\hline
SO1361 &  -110.1  $\pm$  5.4  & 56.3  $\pm$  7.6  \\[0.5ex]
SO1361 &  67.4    $\pm$  4.7  & 46.4  $\pm$  6.2  \\[0.5ex]
\hline
\end{tabular}
\end{center}
\end{table}

\begin{table}
\begin{center}
\caption{\label{tab:HVCfit_NII} Kinematics Properties of the \NII\,HVC for the MIKE sample}
\begin{tabular}{lcc}
\hline\hline
Name  & $v_{\rm peak}$  &   FWHM  \\
      &  [km/s]          & [km/s]  \\
\hline
SO341  &  -69.0  $\pm$  3.6  & 60.4  $\pm$   6.0   \\[0.5ex]
SO518  &  -98.8  $\pm$  17.8 & 56.2  $\pm$   29.6  \\[0.5ex]
SO638  &   38.3  $\pm$  19.0 & 62.7  $\pm$   31.6  \\[0.5ex]
SO927  &   69.9  $\pm$  16.1 & 65.0  $\pm$   26.8  \\[0.5ex]
SO984  &  -56.0  $\pm$  11.1 & 64.3  $\pm$   18.3  \\[0.5ex]
SO1153 &  -94.5  $\pm$  4.5  & 36.5  $\pm$   7.4   \\[0.5ex]
SO1361 &   80.8  $\pm$  4.1  & 35.9  $\pm$   6.8   \\[0.5ex]
\hline
\end{tabular}
\end{center}
\end{table}

\begin{table}
\begin{center}
\caption{\label{tab:HVCfit_NII_XS} Kinematics Properties of the \NII\,HVC for the X-shooter sample}
\begin{tabular}{lcc}
\hline\hline
Name  & $v_{\rm peak}$  &   FWHM  \\
      &  [km/s]          & [km/s]  \\
\hline
SO73   &   148.9 $\pm$   4.8  & 58.7   $\pm$  8.0  \\[0.5ex]
SO467  &   30.7  $\pm$   7.7  & 51.9   $\pm$  11.1 \\[0.5ex]
SO518  &  -70.2  $\pm$   5.2  & 50.1   $\pm$  6.3  \\[0.5ex]
SO540  &   35.3  $\pm$   3.4  & 23.5   $\pm$  3.4  \\[0.5ex]
SO562  &  -75.4  $\pm$   0.5  & 36.5   $\pm$  0.8  \\[0.5ex]
SO848  &  -43.0  $\pm$   3.5  & 60.1   $\pm$  1.8  \\[0.5ex]
SO859  &  -60.7  $\pm$   11.6 & 61.8   $\pm$  18.6 \\[0.5ex]
SO927  &   32.1  $\pm$   8.0  & 110.1  $\pm$  12.3 \\[0.5ex]
SO1154 &   60.0  $\pm$   4.9  & 38.8   $\pm$  5.9  \\[0.5ex]
SO1260 &   171.4 $\pm$   7.8  & 54.8   $\pm$  11.3 \\[0.5ex]
\hline
\end{tabular}
\end{center}
\end{table}

\begin{table}
\begin{center}
\caption{\label{tab:HVCfit_SII} Kinematics Properties of the \SII\,HVC for the MIKE sample}
\begin{tabular}{lcc}
\hline\hline
Name  & $v_{\rm peak}$  &   FWHM  \\
      &  [km/s]          & [km/s]  \\
\hline
SO341   &  -72.0 $\pm$  1.5  & 46.9 $\pm$  2.4  \\[0.5ex]
\hline
SO518   &  -55.9 $\pm$  19.1 & 61.5 $\pm$  31.8 \\[0.5ex]
SO518   &  -100.0$\pm$  6.0  & 44.1 $\pm$  10.0 \\[0.5ex]
\hline
SO662   &   63.4 $\pm$ 11.3  & 64.8 $\pm$  18.8 \\[0.5ex]
SO844   &  -67.1 $\pm$  2.0  & 21.4 $\pm$  3.1  \\[0.5ex]
\hline
SO927   &  -109.8$\pm$ 19.8  & 56.2 $\pm$  33.0 \\[0.5ex]
SO927   &   67.6$\pm$  6.8   & 36.1 $\pm$  11.2 \\[0.5ex]
SO927   &   47.4$\pm$  39.6  & 121.9 $\pm$ 66.0 \\[0.5ex]
\hline
SO1153  &  -61.8$\pm$  13.5 & 44.1 $\pm$  22.4 \\[0.5ex]
SO1153  &  -100.0 $\pm$  5.0  & 38.9 $\pm$  8.3  \\[0.5ex]
\hline
SO1361  &  73.8  $\pm$  1.9  & 27.2 $\pm$  2.9  \\[0.5ex]
SO1361  &  -124.4 $\pm$  2.7  & 28.4 $\pm$  4.4  \\[0.5ex]
\hline
\end{tabular}
\end{center}
\end{table}

\begin{table}
\begin{center}
\caption{\label{tab:HVCfit_SII_XS} Kinematics Properties of the \SII\,HVC for the X-shooter sample}
\begin{tabular}{lcc}
\hline\hline
Name  & $v_{\rm peak}$  &   FWHM  \\
      &  [km/s]          & [km/s]  \\
\hline
SO73    &  67.0   $\pm$  20.0 & 157.0  $\pm$  32.8 \\[0.5ex]
SO73    &  156.0  $\pm$  4.1  & 50.5   $\pm$  3.9  \\[0.5ex]
\hline
SO341   &  -87.3  $\pm$  3.3  & 40.4   $\pm$  2.6  \\[0.5ex]
SO341   &  -32.0  $\pm$  3.0  & 52.3   $\pm$  1.4  \\[0.5ex]
\hline
SO518   &  -86.9  $\pm$  3.7  & 27.9   $\pm$  2.0  \\[0.5ex]
SO518   &  -66.1  $\pm$  5.4  & 66.6   $\pm$  7.0  \\[0.5ex]
\hline
SO583   &  115.1  $\pm$  2.1  & 25.4   $\pm$  3.6  \\[0.5ex]
SO726   &  -36.0  $\pm$  7.3  & 93.0   $\pm$  11.0 \\[0.5ex]
SO844   &  -66.7  $\pm$  5.1  & 34.3   $\pm$  6.9  \\[0.5ex]
SO848   &  -60.8  $\pm$  3.6  & 48.4   $\pm$  0.8  \\[0.5ex]
SO859   &  -76.8  $\pm$  10.4 & 69.1   $\pm$  16.6 \\[0.5ex]
\hline
SO927   &  -98.5  $\pm$  5.5  & 65.8   $\pm$  7.9  \\[0.5ex]
SO927   &  55.3   $\pm$  4.2  & 61.4   $\pm$  5.4  \\[0.5ex]
\hline
SO1075  &  -44.4  $\pm$  6.6  & 39.2   $\pm$  0.4  \\[0.5ex]
SO1075  &  -111.6 $\pm$  6.6  & 57.9   $\pm$  0.9  \\[0.5ex]
\hline
SO1153  &  -65.0  $\pm$  3.4  & 71.1   $\pm$  2.5  \\[0.5ex]
SO1153  &  -91.9  $\pm$  3.1  & 36.2   $\pm$  1.0  \\[0.5ex]
\hline
SO1154  &  34.7   $\pm$  5.6  & 73.6   $\pm$  7.6  \\[0.5ex]
SO1154  &  -45.3  $\pm$  3.2  & 49.2   $\pm$  1.1  \\[0.5ex]
\hline
SO1260  &  156.0  $\pm$  7.8  & 97.6   $\pm$  11.6 \\[0.5ex]
\hline
SO1362  &  305.8  $\pm$  5.0  & 26.1   $\pm$  2.4  \\[0.5ex]
SO1362  &  -93.2  $\pm$  5.5  & 23.5   $\pm$  4.6  \\[0.5ex]
\hline
\end{tabular}
\end{center}
\end{table}

\begin{table}
\begin{center}
\caption{\label{tab:HVCfit_SIIb} Kinematics Properties of the \SIIb\,HVC for the MIKE sample}
\begin{tabular}{lcc}
\hline\hline
Name  & $v_{\rm peak}$  &   FWHM  \\
      &  [km/s]          & [km/s]  \\
\hline
SO341   &  -40.4 $\pm$  1.2  & 47.6 $\pm$  2.2  \\[0.5ex]
SO518   &  -65.1 $\pm$  0.8  & 49.9 $\pm$  1.8 \\[0.5ex]
SO662   &   33.3 $\pm$  0.7 & 49.6 $\pm$  1.9 \\[0.5ex]
\hline
SO1153  &  -35.2 $\pm$  2.2  & 46.4 $\pm$  3.4 \\[0.5ex]
SO1153  &  -72.5 $\pm$  0.7  & 35.3 $\pm$  1.0  \\[0.5ex]
\hline
SO1361  &  99.3  $\pm$  0.5  & 30.1 $\pm$  1.3  \\[0.5ex]
SO1361  &  -94.1 $\pm$  1.0  & 32.9 $\pm$  2.5  \\[0.5ex]
\hline
\end{tabular}
\end{center}
\end{table}

\section{Additional information on targets}\label{sec:ap_additional_info}

Table~\ref{tab:targets_info} lists the additional information on targets used in this work \citep{Mauco2023,Mauco2016}. Typical uncertainties on $A_{\rm V}$, $L_*$, $M_*$, $\rm L_{acc}$, $\dot{M}_{\rm acc}$ are 0.1 mag, 0.2 dex, 0.1 dex, 0.25 dex, 0.45 dex, respectively \citep{Manara2021}.

\begin{table*}
\begin{center}
\caption{\label{tab:targets_info} Additional information on targets}
\begin{tabular}{lccccccccc}
\hline\hline
Name & A$_{\rm v}$ &  L$_{*}$  & log$L_{\rm acc}$   & M$_{*}$ &  log$\dot{M}_{\rm acc}$  & Distance & dp     & logG$_0$  & RV \\
     & [mag]       &  [\lsun]  & [\lsun]            & [\msun] & [\msun/yr]               & [pc]     & [pc]   &           & [km/s] \\
\hline
SO73   &  1.0  & 0.20  &  -1.13   &  0.29  &  -7.89   &  359.2 $_{-4.4 }^{ +4.2 }$ & 2.32  & 2.34  &  33.1   \\[0.5ex]
SO299  &  0.2  & 0.22  &  -2.62   &  0.24  &  -9.26   &  355.5 $_{-4.4 }^{ +4.3 }$ & 1.52  & 2.7   &  12.91  \\[0.5ex]
SO341  &  0.8  & 0.55  &  -1.18   &  0.59  &  -8.14   &  409.0 $_{-4.4 }^{ +4.3 }$ & 1.31  & 2.83  &  27.75  \\[0.5ex]
SO362  &  1.4  & 0.60  &  -0.7    &  0.3   &  -7.23   &  402.3 $_{-4.8 }^{ +4.6 }$ & 1.07  & 3.01  &  32.92  \\[0.5ex]
SO396  &  0.64 & 0.40  &  -1.20   &  0.42  &  -8.10   &  397.7 $_{-4.2 }^{ +4.2 }$ & 1.32  & 2.76  &  28.70  \\[0.5ex]
SO397  &  0.0  & 0.24  &  -2.62   &  0.19  &  -9.06   &  401.0                     & 1.47  & 2.73  &  29.95  \\[0.5ex]
SO467  &  0.3  & 0.07  &  -3.18   &  0.1   &  -9.57   &  383.3 $_{-9.0 }^{ +8.6 }$ & 2.13  & 2.41  &  39.0   \\[0.5ex]
SO490  &  0.0  & 0.1   &  -3.01   &  0.13  &  -9.41   &  401.0                     & 1.88  & 2.52  &  32.3   \\[0.5ex]
SO500  &  0.0  & 0.02  &  -3.84   &  0.06  &  -10.22  &  409.2 $_{-45.4}^{ +37.2}$ & 0.98  & 3.09  &  26.15  \\[0.5ex]
SO518  &  1.6  & 0.48  &  -0.69   &  0.8   &  -7.86   &  399.0 $_{-4.0 }^{ +3.9 }$ & 1.18  & 2.93  &  36.32  \\[0.5ex]
SO520  &  0.1  & 0.23  &  -2.01   &  0.18  &  -8.45   &  402.6 $_{-6.5 }^{ +6.3 }$ & 0.52  & 3.64  &  33.95  \\[0.5ex]
SO540  &  0.5  & 0.57  &  -1.84   &  0.77  &  -8.96   &  406.0 $_{-3.6 }^{ +3.5 }$ & 2.38  & 2.32  &  22.85  \\[0.5ex]
SO562  &  0.3  & 0.26  &  -1.44   &  0.15  &  -7.7    &  401.0                     & 0.39  & 3.88  &  25.05  \\[0.5ex]
SO563  &  0.6  & 0.36  &  -1.27   &  0.64  &  -8.36   &  401.0                     & 0.39  & 3.88  &  33.3   \\[0.5ex]
SO583  &  1.0  & 4.06  &  -0.69   &  1.18  &  -7.62   &  401.0                     & 1.01  & 3.06  &  20.35  \\[0.5ex]
SO587  &  0.0  & 0.35  &  -3.91   &  0.21  &  -10.31  &  401.0                     & 0.32  & 4.06  &  34.65  \\[0.5ex]
SO615  &  1.9  & 3.27  &  -1.92   &  1.49  &  -9.11   &  397.9 $_{-6.2 }^{ +6.2 }$ & 0.85  & 3.14  &  30.55  \\[0.5ex]
SO646  &  0.0  & 0.12  &  -2.9    &  0.25  &  -9.66   &  404.6 $_{-6.8 }^{ +6.6 }$ & 1.13  & 2.96  &  32.1   \\[0.5ex]
SO662  &  0.3  & 0.68  &  -3.79   &  0.64  &  -10.77  &  401.2 $_{-3.4 }^{ +3.3 }$ & 0.68  & 3.41  &  27.55  \\[0.5ex]
SO682  &  0.7  & 0.76  &  -2.02   &  0.57  &  -8.89   &  409.8 $_{-4.8 }^{ +4.7 }$ & 0.17  & 4.63  &  28.67  \\[0.5ex]
SO687  &  0.8  & 0.73  &  -1.21   &  0.44  &  -7.94   &  412.8 $_{-4.3 }^{ +4.2 }$ & 0.06  & 5.52  &  25.22  \\[0.5ex]
SO694  &  0.1  & 0.16  &  -2.51   &  0.12  &  -8.82   &  392.3 $_{-9.6 }^{ +9.2 }$ & 0.13  & 4.85  &  40.2   \\[0.5ex]
SO697  &  0.2  & 0.97  &  -3.11   &  0.67  &  -10.05  &  404.5 $_{-2.4 }^{ +2.4 }$ & 0.51  & 3.66  &  30.15  \\[0.5ex]
SO726  &  0.6  & 0.56  &  -2.19   &  0.59  &  -9.15   &  403.9 $_{-7.0 }^{ +6.8 }$ & 0.1   & 5.03  &  --     \\[0.5ex]
SO736  &  0.1  & 1.49  &  -1.48   &  0.55  &  -8.23   &  401.0                     & 1.03  & 3.05  &  29.9   \\[0.5ex]
SO739  &  0.1  & 0.1   &  -3.06   &  0.1   &  -9.35   &  433.3 $_{-22.3}^{ +20.3}$ & 1.02  & 3.06  &  --     \\[0.5ex]
SO774  &  0.0  & 0.49  &  -2.75   &  0.7   &  -9.84   &  403.3 $_{-3.4 }^{ +3.3 }$ & 1.28  & 2.86  &  31.17  \\[0.5ex]
SO818  &  0.4  & 0.29  &  -2.11   &  0.78  &  -9.36   &  405.4 $_{-4.2 }^{ +4.1 }$ & 2.37  & 2.32  &  18.11  \\[0.5ex]
SO823  &  1.5  & 0.32  &  -2.43   &  0.77  &  -9.66   &  401.0                     & 1.37  & 2.79  &  36.7   \\[0.5ex]
SO844  &  0.7  & 0.62  &  -1.37   &  0.44  &  -8.14   &  415.5 $_{-3.8 }^{ +3.7 }$ & 2.18  & 2.39  &  29.22  \\[0.5ex]
SO848  &  0.0  & 0.02  &  -3.51   &  0.17  &  -10.47  &  356.3 $_{-18.0}^{ +16.3}$ & 0.46  & 3.75  &  --     \\[0.5ex]
SO859  &  0.6  & 0.41  &  -1.72   &  0.29  &  -8.31   &  407.9 $_{-6.6 }^{ +6.4 }$ & 0.84  & 3.22  &  25.15  \\[0.5ex]
SO897  &  0.6  & 0.85  &  -1.34   &  0.7   &  -8.33   &  401.0                     & 0.77  & 3.29  &  23.56  \\[0.5ex]
SO927  &  0.6  & 0.33  &  -1.92   &  0.65  &  -9.03   &  413.6 $_{-4.8 }^{ +4.7 }$ & 1.0   & 3.07  &  20.72  \\[0.5ex]
SO984  &  0.1  & 0.72  &  -3.5    &  0.64  &  -10.46  &  409.6 $_{-3.2 }^{ +3.1 }$ & 1.18  & 2.92  &  27.4   \\[0.5ex]
SO1036 &  0.7  & 0.53  &  -0.89   &  0.59  &  -7.86   &  395.0 $_{-3.5 }^{ +3.4 }$ & 1.19  & 2.92  &  31.23  \\[0.5ex]
SO1075 &  0.6  & 0.14  &  -1.38   &  0.3   &  -8.22   &  390.0 $_{-8.6 }^{ +8.2 }$ & 1.6   & 2.66  &  --     \\[0.5ex]
SO1152 &  0.8  & 0.61  &  -1.26   &  0.58  &  -8.19   &  398.6 $_{-3.9 }^{ +3.8 }$ & 2.71  & 2.21  &  30.1   \\[0.5ex]
SO1153 &  1.5  & 0.33  &  0.02    &  0.9   &  -7.3    &  396.6 $_{-4.3 }^{ +4.2 }$ & 1.68  & 2.62  &  30.25  \\[0.5ex]
SO1154 &  1.8  & 0.08  &  -0.78   &  0.62  &  -8.19   &  401.0                     & 1.64  & 2.64  &  31.6   \\[0.5ex]
SO1155 &  0.6  & 1.45  &  -0.92   &  0.86  &  -7.94   &  401.0                     & 1.81  & 2.55  &  29.6   \\[0.5ex]
SO1156 &  0.4  & 0.66  &  -1.26   &  0.74  &  -8.33   &  403.8 $_{-2.6 }^{ +2.6 }$ & 2.42  & 2.3   &  28.15  \\[0.5ex]
SO1248 &  0.0  & 0.18  &  -3.3    &  0.13  &  -9.6    &  398.4 $_{-7.9 }^{ +7.6 }$ & 2.47  & 2.28  &  --     \\[0.5ex]
SO1260 &  0.0  & 0.15  &  -1.94   &  0.19  &  -8.53   &  386.3 $_{-6.4 }^{ +6.2 }$ & 1.95  & 2.49  &  35.6   \\[0.5ex]
SO1266 &  0.0  & 0.07  &  -4.76   &  0.15  &  -11.36  &  399.1 $_{-11.0}^{ +10.4}$ & 2.24  & 2.37  &  42.6   \\[0.5ex]
SO1267 &  0.6  & 0.76  &  -1.85   &  0.43  &  -8.57   &  400.5 $_{-5.3 }^{ +5.2 }$ & 2.42  & 2.3   &  27.65  \\[0.5ex]
SO1274 &  0.0  & 0.68  &  -0.98   &  0.64  &  -7.95   &  407.3 $_{-2.7 }^{ +2.7 }$ & 2.42  & 2.3   &  28.05  \\[0.5ex]
SO1327 &  0.1  & 0.33  &  -1.91   &  0.21  &  -8.32   &  397.7 $_{-5.8 }^{ +5.7 }$ & 2.79  & 2.18  &  33.15  \\[0.5ex]
SO1361 &  0.5  & 0.47  &  -0.61   &  0.46  &  -7.46   &  406.0 $_{-4.0 }^{ +3.9 }$ & 2.5   & 2.27  &  21.01  \\[0.5ex]
SO1362 &  0.0  & 0.1   &  -2.96   &  0.13  &  -9.37   &  399.4 $_{-10.7}^{ +10.2}$ & 2.76  & 2.19  &  35.85  \\[0.5ex]
SO1369 &  0.0  & 1.26  &  -1.45   &  0.57  &  -8.24   &  402.5 $_{-2.5 }^{ +2.5 }$ & 3.05  & 2.1   &  30.3   \\[0.5ex]
\hline
\end{tabular}
\end{center}
\end{table*}

\end{appendix}

\end{document}